\documentclass[%
superscriptaddress,
reprint,
nofootinbib,
 amsmath,amssymb,
 aps,
 pre,
floatfix,
twocolumn
]{revtex4-1}

\usepackage{csquotes}
\usepackage{graphicx}% Include figure files
\usepackage{dcolumn}% Align table columns on decimal point
\usepackage{bm}% bold math
\usepackage[utf8x]{inputenc}
\usepackage[usenames,dvipsnames,svgnames,table]{xcolor}
\usepackage{selinput}
\usepackage{soul}
\usepackage[normalem]{ulem}
\usepackage{cancel}
\newcommand{\lowt}[1]{_{\text{#1}}}
\newcommand{\hight}[1]{^{\text{\tiny{#1}}}}

\begin{document}

\preprint{APS/123-QED}

\title{Thin-Film Modeling of Resting and Moving Active Droplets}

\author{Sarah Trinschek}
\thanks{These two authors contributed equally}
\affiliation{Institut für Theoretische Physik, Westfälische Wilhelms-Universität Münster, Wilhelm-Klemm-Str. 9, 48149 Münster, Germany}
\affiliation{Universit\'{e} Grenoble-Alpes, CNRS, Laboratoire Interdisciplinaire de Physique 38000 Grenoble, France}

\author{Fenna Stegemerten}
\thanks{These two authors contributed equally}
\affiliation{Institut für Theoretische Physik, Westfälische Wilhelms-Universität Münster, Wilhelm-Klemm-Str. 9, 48149 Münster, Germany}

 \author{Karin John}
\affiliation{Universit\'{e} Grenoble-Alpes, CNRS, Laboratoire Interdisciplinaire de Physique 38000 Grenoble, France}

\author{Uwe Thiele}
\email{u.thiele@uni-muenster.de}
\homepage{http://www.uwethiele.de}
\thanks{ORCID ID: 0000-0001-7989-9271}
\affiliation{Institut f\"ur Theoretische Physik, Westf\"alische Wilhelms-Universit\"at M\"unster, Wilhelm-Klemm-Str.\ 9, 48149 M\"unster, Germany}
\affiliation{Center for Nonlinear Science (CeNoS), Westf{\"a}lische Wilhelms-Universit\"at M\"unster, Corrensstr.\ 2, 48149 M\"unster, Germany}

\date{\today}

\begin{abstract}
We propose a generic model for thin films and shallow drops of a polar active liquid that have a free surface and are in contact with a solid substrate.  The model couples evolution equations for the film height and the local polarization in the form of a gradient dynamics supplemented with active stresses and fluxes.  A wetting energy for a partially wetting liquid is incorporated allowing for motion of the liquid-solid-gas contact line. This gives a consistent basis for the description of drops of dense bacterial suspensions or compact aggregates of living cells on solid substrates.  As example, we analyze the dynamics of two-dimensional active drops (i.e., ridges) and demonstrate how active forces compete with passive surface forces to shape droplets and drive their motion. In our simple two-dimensional scenario we find that defect structures within the polarization profile drastically influence the shape and motility of active droplets. Thus, we can observe a transition from resting to motile droplets via the elimination of defects in the polarization profile. Furthermore, droplet motility is modulated by strong active stresses.
Contractile stresses even lead to topological changes, i.e., drop splitting, which is naturally encoded in the evolution equations.

\end{abstract}

\pacs{Valid PACS appear here}
\maketitle

\section{Introduction \label{sec:level1}}
Active media far from thermodynamic equilibrium display a rich spectrum of bulk phenomena. Meso-scale turbulence in bacterial suspensions \cite{WDH+2012pnasusa}, the emergence of large-scale structures in microtubule-motor assemblies \cite{NSM+1997n, SNL+2001s, SNS+2012n}, and dynamical clustering in bacterial colonies \cite{PSJ+2012prl, ZBF+2010potnaos} or suspensions of artificial Janus particles \cite{BBK+2013prl} are some examples of intriguing reported observations.  In these systems the nonequilibrium character manifests itself via the generation of active stresses and/or the self-propulsion of active particles.  When active matter features a free surface, motility-induced active forces compete with passive interfacial forces. This results in novel features, e.g., vortex flows in bacterial suspensions confined into an oil-immersed drop \cite{WWD+2013prl}, spontaneous symmetry breaking in the actin cortex at the interface of water-in-oil emulsions induced by myosin activity \cite{SK2014e} and the autonomous self-sustained motion of freely suspended droplets containing microtubule-motor assemblies \cite{SCD+2012n}.\\
Swarming bacterial colonies or compact aggregates and thin layers of living cells with free edges form a special class of soft active media where a free surface is in contact with a solid substrate. In some cases, the concept of passive wetting can be employed to gain insight into the dynamics of these systems. When a drop of \textit{passive} liquid is deposited on a solid substrate, the shape of the drop is determined solely by the interfacial tensions of the involved interfaces and its equilibrium three-phase contact angle can directly be obtained from Young's law \cite{DeGennes1985}. In the embryogenesis of zebrafish, the collective cell migration follows the laws of wetting \cite{WTY+2018bj}  and the observed shapes can roughly be explained by variations of the interfacial tensions. Also the spreading of cell aggregates at long times has been successfully studied as a wetting problem \cite{DGN+2011potnaos, DDB2012sm}.
However, the ability of the active liquids' constituents to polarize and generate active stresses can drastically affect the dynamics. Recently, it has been shown that a wetting transition in a thin layer of epithelial tissue on a collagen surface can be explained by the competition between traction forces and contractile intercellular stresses \cite{PAB+2019np}. In the epiboli of zebrafish, tissue contraction results in anisotropic stresses that affect the shape of the egg \cite{MGB+2017dc}. These examples show that the interplay of passive interface forces, i.e., capillarity and wettability, and of activity is a crucial determinant of the dynamics of droplets of living matter on surfaces. However the consistent theoretical description of the droplet's dynamical properties constitutes a challenge and shall be the objective of the present work.\\
In a coarse-grained modeling approach, active bulk liquids can be described by a small number of macroscopic fields, such as the particle density and a macroscopic polarization. Usually, the polarization is hereby defined as the local average over the orientation of the individual constituents which at high densities typically tend to orientationaly order (for reviews see, for example, \cite{MJR+2013rmp, Menzel2015pr, Ramaswamy2010}). One important class of coarse-grained models for active media is based on liquid crystal hydrodynamics \cite{Chandrasekhar1992, Prost1995}. Activity is introduced into this passive theory by endowing the constituents of the liquid with self-generated active stresses. The resulting evolution equations for the macroscopic fields are either derived from microscopic theory \cite{BM2009potnaos, LM2006prl, RKBH2018pre} or are phenomenologically derived based on symmetry arguments \cite{HRR+2004prl}. In the context of the cytoskeleton of living cells, a description of active polar gels \cite{KJJ+2005tepje, JKP+2007pr, JP2009hj, PJJ2015np} is developed and successfully applied to study, e.g., the effect of defect structures \cite{KJJ+2004prl}, the transition to spontaneous flow \cite{TCM2011sm}, concentration banding \cite{GML2008prl}, multi-component \cite{JJK+2007NJoP} and compressible \cite{VJP2006prl} active polar films.\\
Thin layers of a suspension of active particles in the gap between parallel solid plates are considered in Refs.~\cite{CoEL2016el} and \cite{LoEL2018prl,LTEL2019jcp}, for resting and sheared plates respectively. Inspired by cellular motility, several studies consider active liquids with free boundaries suspended in a passive fluid using phase-field models \cite{TMC2012potnaos, WMV+2014tepje, MWP2015jotrsi, WhHa2016njp}. Thereby evolution equations for the active matter are coupled to a description of the surrounding passive fluid, i.e., to the Navier-Stokes or Stokes equations. Activity is found to lead to spontaneous symmetry breaking accompanied by deformation and self-propulsion of the droplet.\\
Active droplets in contact with flat solid surfaces are studied within the context of cell crawling \cite{ZiSA2012jrsi,GD2014prl,ZiAr2016ncm}. However, the employed models are two-dimensional (2D) and only consider the dimensions parallel to the substrate. The direction perpendicular to the substrate is neglected, i.e., height profiles are not considered. Interfacial forces are incorporated via a line tension (or its equivalent in a diffuse interface description) between the active and passive phase. The presence of the solid substrate is incorporated via solid friction terms.\\
Alternatively, recent direct numerical simulations of three-dimensional (3D) drops of an active liquid with contractility and treadmilling find motile (stationary moving) states of biologically relevant shapes \cite{TTMC2015nc}. They use an advanced phase-field model, namely, an active version of model-H \cite{AnMW1998arfm}, but seemingly do not implement a parameter controlling the contact angle, i.e., the physics of the contact of the active fluid drop with the solid substrate is not explicitly considered. Their figures indicate a fixed 90~degree microscopic contact angle, implicitly enforced via boundary conditions. A treadmilling speed is imposed in a finite thickness layer near the substrate.\\
Since simulations of 3D active droplets on substrates are computationally expensive, some studies employ a long-wave approximation \cite{ODB1997rmp,Thiele2007} 
to derive thin-film models of passive nematic liquid crystals \cite{BeCu2001pf,LCA+2013pf,LKT+2013jofm} and active polar gels \cite{SR2009prl,JoRa2012jfm,KA2015pre,KMW2018potrsa}. In particular, Ref.~\cite{KMW2018potrsa} derives a thin-film theory for an active liquid crystal based on the Beris-Edwards theory that uses a tensorial order parameter (instead of a polarization field). Thin-film models for active polar liquids are employed to study wave-forming linear instabilities of free-surface films \cite{SR2009prl}\footnote{As they introduce their model as a generalization of the passive model in \cite{BeCu2001pf}, and that paper was shown to have an arguable elastic contribution \cite{LCA+2013pf} the status of the model is not clear.} and the effect of a highly symmetric polarization field on steady drop shapes and the scaling law for drop spreading in the limit of dominant active stress \cite{JoRa2012jfm}. 
A recently proposed model for droplets of active nematics derives an effective thin-film model for the evolution of the film height profile \cite{LoEL2019prl} and describes transformations in drop shape and drop motion with increasing active stress. It employs slip at the substrate, directly imposes a static microscopic contact angle and assumes instantaneous adaptation of the polarization profile to changes in the height profile. It is further analyzed in \cite{AuET2020sm} where also the case of drop motion driven by self-propulsion is considered.
Further, the self-propulsion of active drops has been associated with topological defects in the polarization field in a model that prescribes static polarization patterns and drop profiles and employs a long-wave approximation to determine the induced instantaneous velocity field and instantaneous propulsion velocity, without specifying a fully dynamical model \cite{KA2015pre}. Steady shapes of resting drops are also obtained there.
Note, that none of the mentioned thin-film models of active media provides a closed form of fully nonlinear coupled evolution equations for film height profile \textit{and} polarization field. Neither are dynamic wetting effects captured.
However, one striking result seems to emerge from both, thin-film and fully three-dimensional active liquid approaches: macroscopic motion does not require active self-propulsion in polar liquids. Active contractile stresses related to nematic order are sufficient to induce waves \cite{SR2009prl} and droplet motion \cite{TMC2012potnaos,KA2015pre}. In general, the literature only scarcely addresses the interplay between active stresses and self-propulsion on the one hand and passive wetting and interfacial forces on the other hand. 
Here, we present a thin-film model that allows for a systematic study of the interplay between activity (self-propulsion and active stresses) and passive wetting forces for partially wetting liquids that form droplets with a finite contact angle on solid substrates.  In a first model analysis of 2D active droplets we are investigating how wetting and active forces combine to shape the droplet and to naturally induce droplet motion.  In section~\ref{sec:model}, we construct a generic phenomenological model that couples evolution equations for the film height profile of the liquid and its local height-integrated polarization.  The passive part of the model is written as a gradient dynamics on an underlying free energy functional that explicitly includes wettability.  The passive model is then supplemented by self-propulsion and active stresses that enter in the form of additional non-variational terms. We discuss the individual contributions to the energy functional and reduce the model further for 2D droplets (i.e., liquid ridges on 1D substrates).  The next section~\ref{sec:flat-film} analyzes the linear stability and the dewetting of flat films of active liquids while the subsequent section~\ref{sec:res-drop} focuses on the dynamics of 2D (active) drops. In particular, we investigate drop shapes and motion in the weak and strong activity regime, depending on the defect structure in the polarization field within the droplet.
We conclude with a summary and outlook in section~\ref{sec:conc} that includes a discussion of literature models in the context of our obtained results.
\section{Model for active polar drops} 
\label{sec:model}
We develop a generic model that couples an evolution equation for the height profile of the droplet to the dynamics of a polarization field. 
In the following, we first introduce the general modeling framework before discussing the specific choices for the energetic contributions and presenting the model equations in a one-dimensional geometry.
\begin{figure}[htbp]
\includegraphics[width=0.5\textwidth]{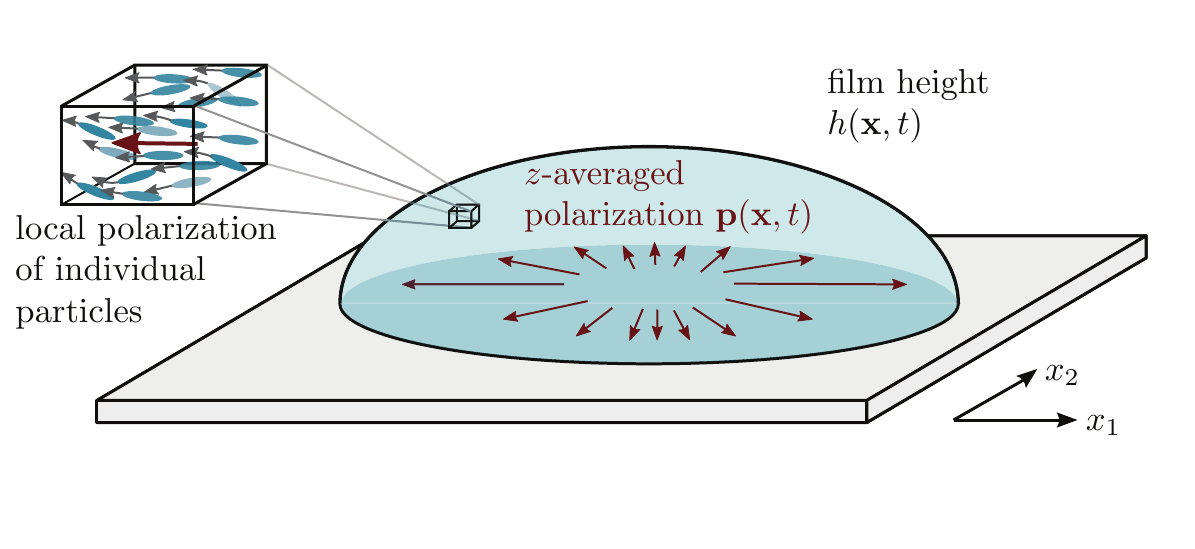}
\caption{Droplet of active polar liquid on a solid substrate. 
The polarization $\mathbf{p}(\mathbf{x},t)$ (red arrows) represents the local height-averaged value of the polarization of the individual particles (gray arrows in the inset).
Its dynamics is coupled to the dynamics of the film height $h(\mathbf{x},t)$. The local height-integrated polarization is given by $\mathbf{P}(\mathbf{x},t) = h(\mathbf{x},t) \,  \mathbf{p}(\mathbf{x},t) $. \label{Sketch:Polarization} }
\end{figure}
\subsection{Model framework and structure}
\noindent
We consider an active polar liquid film of height $h(x_1,x_2,t)$ and introduce the polarization field $\mathbf{p}(x_1,x_2,t)$  as the height-averaged value of the local microscopic $z$-dependent polarization of the individual particles as sketched in Fig.~\ref{Sketch:Polarization}. 
We assume ad hoc that the component of the local polarization perpendicular to the substrate is small as compared to the components parallel to the substrate. In consequence we only consider the latter components and write
\begin{equation}
\mathbf{p} = \begin{pmatrix} p_1(x_1,x_2,t) \\ p_2(x_1,x_2,t)  \end{pmatrix} \, .
\end{equation}
In other words, we assume that polarization is nearly parallel to the substrate and replace its local strength by its vertically averaged value which forms a basic variable in our phenomenological model. 
Although we study an \textit{active} polar liquid, we construct the passive core of our model as a gradient dynamics on a free energy functional. This guarantees that  in the absence of activity it describes the approach to and the characteristics of well-defined steady equilibrium states.
In particular, we introduce the free energy functional $\mathcal{F}[h,\mathbf{p}]$,
that accounts for various effects that may influence the dynamics of the polar liquid droplet. Namely, we consider capillarity, wettability, spontaneous polarization, the elastic energy of the polarization and a coupling between the polarization vector and the shape of the free surface of the drop (see section~\ref{sec:en} below).
The dynamics of a \textit{passive} polar liquid close to 
equilibrium is modeled by constructing a gradient dynamics based on the energy functional $\mathcal{F}[h,\mathbf{p}]$.
However, $\mathcal{F}[h,\mathbf{p}]$ needs to be expressed in independent variables. 
We therefore introduce the local height-integrated amount of polarization $\mathbf{P} = h \mathbf{p}$ and perform the transformation
\begin{equation}
F[h,\mathbf{P}]=\mathcal{F}[h,\mathbf{p}(h,\mathbf{P})] \, . \label{trafo}
\end{equation}
\textit{Activity} is introduced into the model by two non-variational terms that force the system out of equilibrium and which break the Onsager symmetry of the gradient dynamics.
The first contribution is the active stress $\boldsymbol{\sigma}\hight{a}$ 
with the components \cite{Ramaswamy2010}
\begin{equation}
\sigma\hight{a}_{kj} =- c_\mathrm{a} p_k p_j \,,
\end{equation}
where $j,k=1,2$.
The active stress is extensile for $c_\mathrm{a}>0$ (describing, e.g., bacterial suspensions) and contractile for $c_\mathrm{a}<0 $ (describing, e.g., actomyosin solutions). 
The second active contribution is the self-propulsion of the particles in the direction of their polarization $\mathbf{p}$. It gives rise to an active force of the form
\begin{equation}
\boldsymbol{\alpha}=\alpha_0 {3\eta\over h^2}\mathbf{p} =\alpha_0 {3\eta\over h^3}\mathbf{P}=\sum_k \alpha_k\mathbf{e}_k
\end{equation}
where $\alpha_0$ is a constant and $\eta$ denotes the viscosity. 
The self-propulsion breaks the $\mathbf{P} \rightarrow -\mathbf{P}$ symmetry of the model.
By combining the passive and the active contributions, we obtain the general form of the coupled evolution equations for film height $h$ and polarization $\mathbf{P}$
\begin{eqnarray}
\partial_t h&  = & \sum_{k} \partial_{x_k} \bigg[ Q_{hh}\Big(\partial_{x_k}{\delta F\over \delta h} -\alpha_k - \sum_{j} \partial_{x_j} \sigma\hight{a}_{kj} \Big) +\nonumber \\
  & & \sum_{j} Q_{hP_j}\partial_{x_k}{\delta F\over \delta P_j}  \bigg] \label{hP:eqs:1}\\
\partial_tP_i &= & \sum_{k} \partial_{x_k} \bigg[  Q_{h P_i}\Big(\partial_{x_k}{\delta F\over  \delta h} -\alpha_k - \sum_{j}  \partial_{x_j}\sigma\hight{a}_{kj}\Big) +  \nonumber  \\
 & & \sum_{j} Q_{P_i P_j}\partial_{x_k}{\delta F\over \delta P_j} \bigg] -Q\lowt{NC}{\delta F\over \delta P_i}\,.  \label{hP:eqs:2}
 \end{eqnarray}
Here $\partial_{x_k}$ refers to the partial derivative with respect to coordinate $x_k$.
In contrast to film thickness, polarization is not a conserved quantity. It describes a certain order that may occur spontaneously and can also be created by the surface profile. The respective mobility is $Q\lowt{NC}$.
The remaining mobilities 
\begin{align}
 &Q_{hh}=\frac{h^3}{3\eta} \notag \\
 &Q_{hP_i}=\frac{h^2P_i}{3\eta}=\frac{h^3 p_i}{3\eta} \\
 &Q_{P_i P_j}=h\Big(\frac{P_iP_j}{3\eta}+M\delta_{ij}\Big)=\frac{h^3p_ip_j}{3\eta}+hM\delta_{ij} \notag 
 \end{align}
correspond to scalar, vector and tensor quantities, respectively, and can be understood
in analogy to Ref.~\cite{XTQ2015JPCM} where an analogous thin-film model for a mixture of scalar quantities is discussed.
The evolution equations \eqref{hP:eqs:1}-\eqref{hP:eqs:2} for film height and polarization can be expressed in the hydrodynamic form 
\begin{eqnarray}
\partial_th &= & -\nabla\cdot\mathbf{j}^\mathrm{C} \label{Eq:conth} \\
\partial_tP_i=\partial_t(hp_i) &= &-\nabla\cdot\left(p_i \mathbf{j}^\mathrm{C}+\mathbf{j}^{\mathrm{D}P_i}\right)+j^{\mathrm{R}}_i\label{Eq:contP}\nonumber
\end{eqnarray}
by introducing the convective flux $\mathbf{j}^\mathrm{C}$, the diffusive fluxes
$\mathbf{j}^{\mathrm{D} P_i}$ and the reactive (i.e. rotational) flux $j^{\mathrm{R}}_i$ as
\begin{eqnarray}
\mathbf{j}^\mathrm{C} & = & -{h^3\over 3\eta}\Big(\nabla {\delta
    F\over \delta h}- \sum_j {P_j\over h} \nabla{\delta
    F\over \delta
    P_j}- \nabla \cdot \boldsymbol{\sigma}\hight{a}\Big)+ \alpha_0 \mathbf{P} \, , \notag  \\
\mathbf{j}^{\mathrm{D} P_i} & = & -hM\nabla {\delta F\over \delta
  P_i}\, , \label{fluxes}\\
j^{\mathrm{R}}_i & = & -Q\lowt{NC} {\delta F\over \delta P_i} \notag  \,  ,
\end{eqnarray}
respectively. The mass conservation for the liquid implies that Eq.~(\ref{Eq:conth}) has the form of a continuity equation. 
The polarization equation \eqref{Eq:contP} combines a conserved dynamics representing the transport of polarization by diffusion and with the liquid flow by convection with a non-conserved reactive flux describing reorientation, e.g., due to spontaneous polarization and rotational diffusion.

\subsection{Specific choices for the energies}
\label{sec:en}
\noindent 
Now, we specify the individual contributions to the free energy functional that underlies the dynamics of the active droplet.
We employ
\begin{align}
\mathcal{F}[h,\mathbf{p}]  
=& \mathcal{F}\lowt{cap} + \mathcal{F}\lowt{w}+ \mathcal{F}\lowt{spo}+ \mathcal{F}\lowt{el}+ \mathcal{F}\lowt{coupl} \nonumber \\
   =& \int  \Big[ \tfrac{ \gamma }{2} (\nabla h )^2 + f\lowt{w}(h)+
h f\lowt{spo}\left(h, \mathbf{p}^2\right) \nonumber  \\
& +h f\lowt{el}(\nabla\mathbf{p}) 
+f\lowt{coupl}(\nabla h, \mathbf{p}) \Big] \mathrm{d}\mathbf{x}.  \label{en:hpspecific} 
\end{align}
Namely, $\mathcal{F}\lowt{cap}$ models capillarity, i.e., it consists of the energy of the liquid-air interface (in long-wave approximation) where $\gamma$ denotes the interfacial tension.
The second contribution, $\mathcal{F}\lowt{w}$, represents wettability and accounts for interactions between the liquid and the underlying substrate. 
For a partially wetting simple liquid that forms drops of finite equilibrium contact angle coexisting with a thin adsorption layer of height $h\lowt{a}$ \cite{DeGennes1985,BEI+2009rmp}, we employ the wetting energy 
\begin{align}
 f\lowt{w}(h)= A \left( - \frac{1}{2 h^2} + \frac{h\lowt{a}^3}{5 h^5} \right) 
\end{align}
where  $A$ denotes the Hamaker constant \cite{Thiele2010jpcm}.

\begin{figure}[htbp]
\begin{center}
\includegraphics[width=0.5\textwidth]{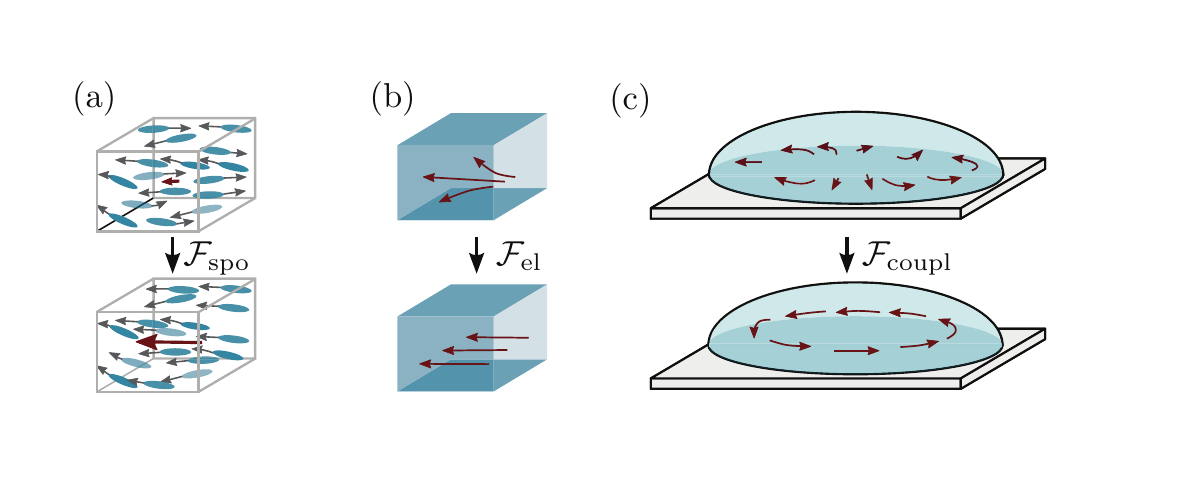}
\caption{Schematic illustration of the effect of the energetic contributions occurring in droplets of polar active liquids. (a) $\mathcal{F}\lowt{spo}$ describes a spontaneous transition between an isotropic, 
microscopically disordered state with $|\mathbf{p}| \approx 0$ to an ordered state with $|\mathbf{p}| \approx 1$. (b) $\mathcal{F}\lowt{el}$ is a liquid crystal elastic energy that represents the energetic cost of horizontal gradients in the polarization. 
(c) $\mathcal{F}\lowt{coupl}$ couples the polarization to the gradient of the free interface. Shown is an example for $c\lowt{hp}>0$ where a polarization along the liquid-solid-gas contact line is energetically favored. Note, that we assume $\mathbf{p}$ to be always parallel to the substrate. \label{Sketch:ActiveEnergies} }
\end{center}
\end{figure}
The contribution $\mathcal{F}\lowt{spo}$ accounts for spontaneous polarization of the liquid and e.g.\ drives a transition between an isotropic, microscopically disordered and 
a polarized state as illustrated in Fig.~\ref{Sketch:ActiveEnergies}~(a). We employ the double-well energy 
\begin{eqnarray}
f\lowt{spo}(\mathbf{p}^2)  &=& -{c\lowt{sp2}\over 2} \left[ 1-2\beta\kappa(h)\right]\mathbf{p}\cdot\mathbf{p}+\nonumber \\
& &{c\lowt{sp4} \over 4}   (\mathbf{p}\cdot\mathbf{p})^2  \label{eq:fspo}
\label{eq:fhspo}
\end{eqnarray}
with $c\lowt{sp2},c\lowt{sp4}>0$, $\beta>{1\over 2}$ and 
\begin{eqnarray}
\kappa(h) = \frac{h\lowt{a} f\lowt{w}(h) }{ h f\lowt{w}(h\lowt{a})}. \label{kappa:h}
\end{eqnarray}
Depending on the film height $h$, Eq.~\eqref{eq:fspo} allows for the existence of a disordered ($|\mathbf{p}|=0$) and an ordered state $\left(|\mathbf{p}|=\sqrt{\frac{c\lowt{sp2}}{c\lowt{sp4}}\left[1-2\beta\kappa(h)\right]}\right)$. 
Note, that $|\mathbf{p}|$ is the strength of the polarization, i.e., 
it measures the amount of aligned particles. 
In the adsorption layer one has $\lim_{h\rightarrow h\lowt{a}} \kappa(h) = 1$, i.e., the disordered state $|\mathbf{p}|=0$ is the only possible (stable) state. 
For large film heights $\kappa\to0$, the disordered state looses stability and the energetically favored ordered state $|\mathbf{p}|=\sqrt{\frac{c\lowt{sp2}}{c\lowt{sp4}}}$ is adopted. 
Unless otherwise specified we chose $c\lowt{sp2}=c\lowt{sp4}=c\lowt{sp}$, i.e., $|\mathbf{p}|=1$. 
The choice of the parameter $\beta>{1\over 2}$ controls the film height above which the ordered polarization state exists. Here we restrict ourselves to the choice $\beta=1$.\\
The contribution $\mathcal{F}\lowt{el} $ accounts for a liquid crystal elastic energy with
 \begin{eqnarray}  
f\lowt{el}(\nabla\mathbf{p}) = {c\lowt{p}\over  2} \nabla\mathbf{p}:\nabla\mathbf{p} \label{eq:fel}
\end{eqnarray}
representing the energetic cost of gradients in the polarization along the substrate as illustrated in Fig.~\ref{Sketch:ActiveEnergies}~(b). For simplicity, we assume the same value of stiffness associated with splay and bend deformations, i.e., we use the single elastic constant $c\lowt{p}$ \cite{Prost1995}.
The final contribution, $\mathcal{F}\lowt{coupl}$, couples the
polarization and the gradient of the free surface via the energy
\begin{equation}
f\lowt{coupl}(\nabla h, \mathbf{p}) = {c\lowt{hp}\over 2} (\mathbf{p} \cdot \nabla h)^2 \, . 
\end{equation}
The constant $c\lowt{hp}$ can be chosen positive for an alignment of the polarization field parallel to the interface as shown in Fig.~\ref{Sketch:ActiveEnergies}~(c, bottom) or negative (for an alignment of $\mathbf{p}$ parallel to $\nabla h$). 
Note that alternatively, coupling terms $\sim \mathbf{p} \cdot \nabla h$ may be applied to energetically favor an outward or inward pointing polarization.

\subsection{Liquid ridge (2D) geometry}
To investigate the basic behavior of the developed model, we consider the case of a 1D substrate. With other words we assume that the system is translation-invariant in the $x_2$-direction, i.e., all gradients and the polarization component in $x_2$-direction vanish. 
Then, the evolution equations strongly simplify as polarization and all mobilities become scalar quantities. In the following, we use the notations $x=x_1$, $p = p_1$ and $P=hp_1$.
In consequence, we neglect the coupling between polarization and interface slope, since the polarization cannot minimize anymore the interaction with the interface by rotating in the substrate plane.
However, still the coupling between film height and polarization guarantees that the polarization decays to zero in the contact line region.
The variations of $F$ [Eqs.~\eqref{trafo} and \eqref{en:hpspecific}] with respect to film height and polarization then read 
\begin{align}
 \frac{\delta F}{\delta h} =&-\gamma\partial_{xx} h+\partial_h f\lowt{w}
+f\lowt{spo} + h \partial_h f\lowt{spo}   \nonumber  \\
& - p \partial_{p} f\lowt{spo}+ {c\lowt{p}\over2}(\partial_x p)^2 + {c\lowt{p} p\over h}\partial_x (h\partial_xp)\\
 \frac{\delta F}{\delta P} =& \,\partial_{p} f\lowt{spo}-{c\lowt{p}\over h}\partial_x\left(h\partial_x p\right)\,. \label{var:h:P}
\end{align}
The time evolution is given by
\begin{eqnarray}
\partial_t h & =  &-\partial_x j^\mathrm{C}\label{eq:activeEE1D_1} \\
\partial_t (hp) &= & -\partial_x\left(p j^\mathrm{C}+j^\mathrm{D}\right)+j^\mathrm{R} \label{eq:activeEE1D_2}
\end{eqnarray}
with the $x_1$-component of the fluxes \eqref{fluxes}
\begin{eqnarray}
j^\mathrm{C} & = &-{h^3\over 3\eta}\left[\partial_x \left({\delta F \over \delta h}\right)+p \, \partial_x  \left({\delta F \over \delta P}\right)+  c_\mathrm{a} \partial_x(p^2) \right]\nonumber\\
&&+ \alpha_0 hp \nonumber \\
j^{\mathrm{D}}& = &-hM\partial_x  \left({\delta F \over \delta  P}\right) \label{eq:fluxes}\\
j^{\mathrm{R}} & = & -Q\lowt{NC} \, {\delta F\over \delta P} \nonumber \, .
\end{eqnarray}
Note, that $f\lowt{spo}$ does not contribute to the convective flux $j^\mathrm{C}$ as the respective terms cancel out. This is analogous to the fact that for a thin film of a liquid mixture or suspension the osmotic pressure does not contribute to the convective flux \cite{TTL2013prl}.
In the following, we analyze the developed model for active polar liquids in the 2D case. 
On the one hand the film and drop dynamics is studied by time simulations employing finite element schemes provided by the modular toolbox DUNE-PDELAB \cite{BBD+2008c,BBD+2008c} and the open source library \texttt{oomph-lib} \cite{HH2006}. On the other hand, we employ pseudo-arclength path continuation techniques
\cite{DoKK1991ijbc,DWCD2014ccp,EGUW2019springer} to efficiently study the effect of parameter changes on the properties of steady sitting and steadily moving drops. To do so we transform the evolution equations \eqref{eq:activeEE1D_1}-\eqref{eq:activeEE1D_2} 
into a frame moving with a constant velocity $v$ and use the continuation package 
\textsc{pde2path} \cite{UeWR2014nmma,UeWe2014sjads}.
First, we consider flat homogeneous films and discuss the instabilities introduced by the passive and active components of the model. Next, we show that the model describes resting and moving drops of active liquids and study the influence of the model parameters on their shape and velocity.
\section{Films of passive and active polar fluids}
\label{sec:flat-film}
\subsection{Linear stability analysis of the flat film}
\label{sec:res-linstab}
We begin the analysis of flat homogeneous films of active polar liquid with a linear stability analysis.
The model possesses flat film solutions of arbitrary thickness $h=h_0$ with up to three distinct homogeneous polarization states $P_0$ that can be determined from the condition of a vanishing reactive flux $j^\mathrm{R}$, i.e., 
${\partial f_{spo}\over\partial p}=0$ [see Eqs. \eqref{var:h:P} and \eqref{eq:fluxes}].
The solutions correspond to unpolarized films with $P_0=0$ and to polarized films with $P_0=B\,h_0$ with $B=\pm \sqrt{1-2\kappa(h_0)}$ 
for film heights with $\kappa(h_0)\le 1/2$ and approach a mean polarization of $P_0\approx \pm h_0$ for thick films with $h_0 \gg h\lowt{a}$.
\begin{figure}[htbp]
\includegraphics[width=0.5\textwidth]{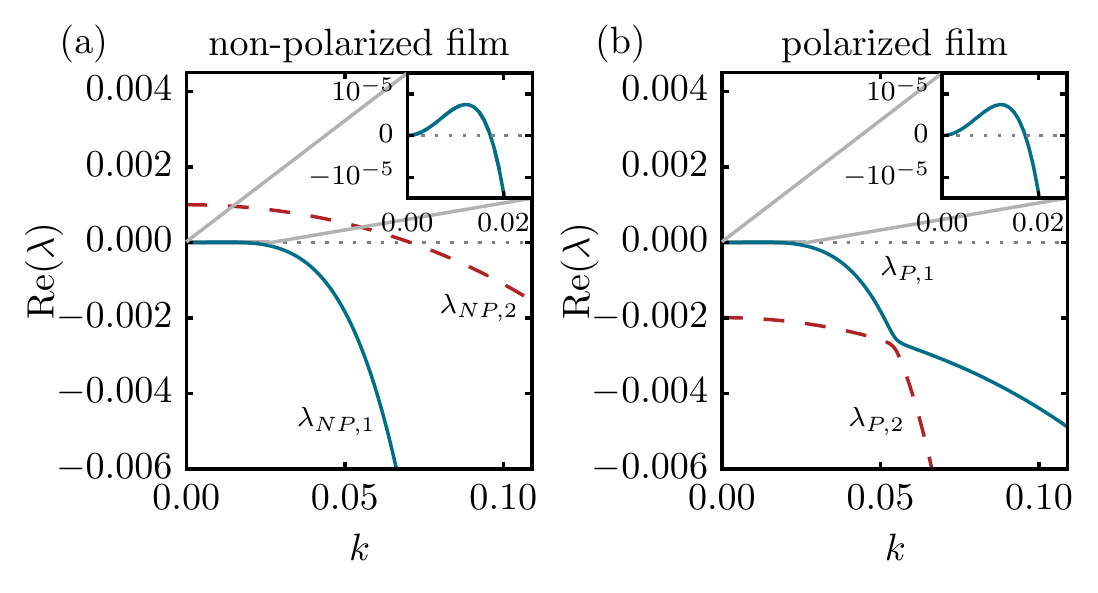} 
\caption{Dispersion relations for homogeneous (a) unpolarized $(h_0,P_0=0)$ and (b) polarized $(h_0,P_0=B h_0)$ flat films of height $h_0 = 10 $.
Note that the eigenvalues $\lambda\lowt{P,i}$ are complex for the polarized film with imaginary part $- i \alpha_0 k$. The remaining parameters are $c\lowt{sp2}=c\lowt{sp} = 0.01$, $A = 1$, $M = 1$, $h\lowt{a} = 1$, $\eta = 1$, $\gamma = 1$, $c\lowt{p} = 2$, $Q\lowt{NC} = 1$, $c_\mathrm{a} = 0.01$ and $\alpha_0 =0.001$. \label{Fig:EVs}} 
\end{figure}
The linear stability of the flat homogeneous films is determined by inserting the harmonic ansatz
\begin{align}
h(x) &= h_0 + \epsilon h_1 e^{i k x + \lambda t}\\
P(x) &= P_0 + \epsilon P_1 e^{i k x + \lambda t}
\end{align}
into the evolution equations (\ref{eq:activeEE1D_1})-(\ref{eq:activeEE1D_2}), linearizing in $\epsilon \ll 1$, and solving the resulting eigenvalue problem.
The two branches of the dispersion relation for the \textit{unpolarized flat film} with $(h, P) = (h_0, 0)$ are given by
\begin{align}
 \lambda_{\mathrm{NP},1}(k) =& - \frac{h_0^3 \gamma}{3 \eta} k^4 + \left[ 1 - 2 \left(\frac{h\lowt{a}}{h_0}\right)^3 \right] \frac{A}{\eta h_0} k^2 \notag  \\
 \lambda_{\mathrm{NP},2}(k) =&- M c\lowt{p} k^4 - \frac{1}{h_0} \left[  Q\lowt{NC} c\lowt{p} - M h_0 c\lowt{sp} B^2 \right] k^2  \label{eq:EV_NP} \\
 & + Q\lowt{NC} \frac{c\lowt{sp}}{h_0} B^2 \notag \, .
  \end{align}
The two eigenvalues are shown in Fig.~\ref{Fig:EVs}~(a) for a flat film of height $h_0=10$ and can easily be interpreted because the effects of film height and polarization decouple: 
The eigenvalue $\lambda_{NP,1}(k)$ corresponds to the dispersion relation of a thin film of a simple, partially wetting liquid \cite{Thiele2007}. It exhibits a typical long-wave instability of a conserved quantity with $\lambda_\mathrm{NP,1}(0)=0$. 
For film heights $h_0>\sqrt[3]{2} h\lowt{a}$, there exists an unstable band of wavenumbers $0 \leq k \leq k\lowt{c}$ where $k\lowt{c}= \sqrt{- \partial_{hh} f\lowt{w}(h_0)}$, 
and the fastest growing mode is at $k\lowt{h} = k\lowt{c}/ \sqrt{2}$. 
The film tends to dewet, leading to the formation of droplets \cite{BEI+2009rmp}. 
In contrast, the eigenvalue $\lambda_{NP,2}(k)$ captures the influence of spontaneous polarization, a non-conserved quantity, which uniformly destabilizes the unpolarized state of the film.  
Above onset, i.e., for $h_0^2 B^2 < (Q\lowt{NC} c\lowt{p})/(M c\lowt{sp})$, there exists an unstable band of wavenumbers $0 \leq k \leq k\lowt{c}$ whereby the fastest growing mode is always at $k=0$.
The energetic costs of gradients in the polarization [$c\lowt{p}>0$ in Eq.~\eqref{eq:fel}] thus  result in a spatially homogeneous polarization for flat films. 
Note that for $c\lowt{p}\to0$, the eigenvalue $\lambda_{NP,2}(k)$ diverges for large wavenumbers $k$. The elastic energy \eqref{eq:fel} is therefore a crucial ingredient to provide a small scale cut-off of instabilities triggered 
by $\lambda_\mathrm{NP,2}$ and is needed to ensure a physical behavior.
The eigenvalues of the \textit{polarized flat film} with $(h,P)=(h_0,B h_0)$ shown in Fig.~\ref{Fig:EVs}~(b) are also  obtained analytically, however,
we do not print the rather lengthy fully coupled expressions [plotted in Fig.~\ref{Fig:EVs}~(b) for $h_0=10$]. In the limit of thick films,  $\left( h\lowt{a} / h_0 \right)^3 \ll 1$ and they reduce to 
\begin{align}
\begin{split}
 \lambda_{P,1}(k) =& -  \frac{\gamma\,{{  h_0}}^{3}}{3 \eta}  k^4   + \frac{A }{ \eta h_0} k^2 -\,i{  \alpha_0} k \\
   \lambda_{P,2}(k) =&  -{  c\lowt{p}}\,M{k}^{4}-2\, \left( M  c\lowt{sp} +\frac{Q\lowt{NC} }{h_0} \,  {  c\lowt{p}}  \right) \,{k}^{2} \\
 & -  \frac{Q\lowt{NC}}{h_0} 2\,{  c\lowt{sp}} -i{  \alpha_0} k\,. \label{eq:EV_P}
\end{split}
 \end{align}
Both eigenvalues are now complex for $\alpha_0 \neq 0$. The dewetting instability is still present in $\lambda_{P,1}(k)$ and is independent of the polarization state of the film. The complex eigenvalues lead here to exponentially growing dewetting waves. 
In contrast, the eigenvalue $\lambda_{P,2}(k)$ connected to the influence of polarization has always a negative real part for the polarized flat film. This reflects that the polarized state is already the one favored by the spontaneous polarization energy $f_\mathrm{spo}$ \eqref{eq:fhspo}.  
The analysis so far has shown that both, unpolarized and polarized homogeneous flat films are linearly unstable for all film heights $h_0>\sqrt[3]{2} h\lowt{a}$.
\subsection{Dewetting dynamics}
\label{subsec:dewett:dyn}
\begin{figure*}[htbp]
passive film $c_\mathrm{a}=0$, $\alpha_{0}=0$\\
\begin{minipage}[t]{0.48\hsize}
(a) initially polarized\\
\includegraphics[width=1.\textwidth]{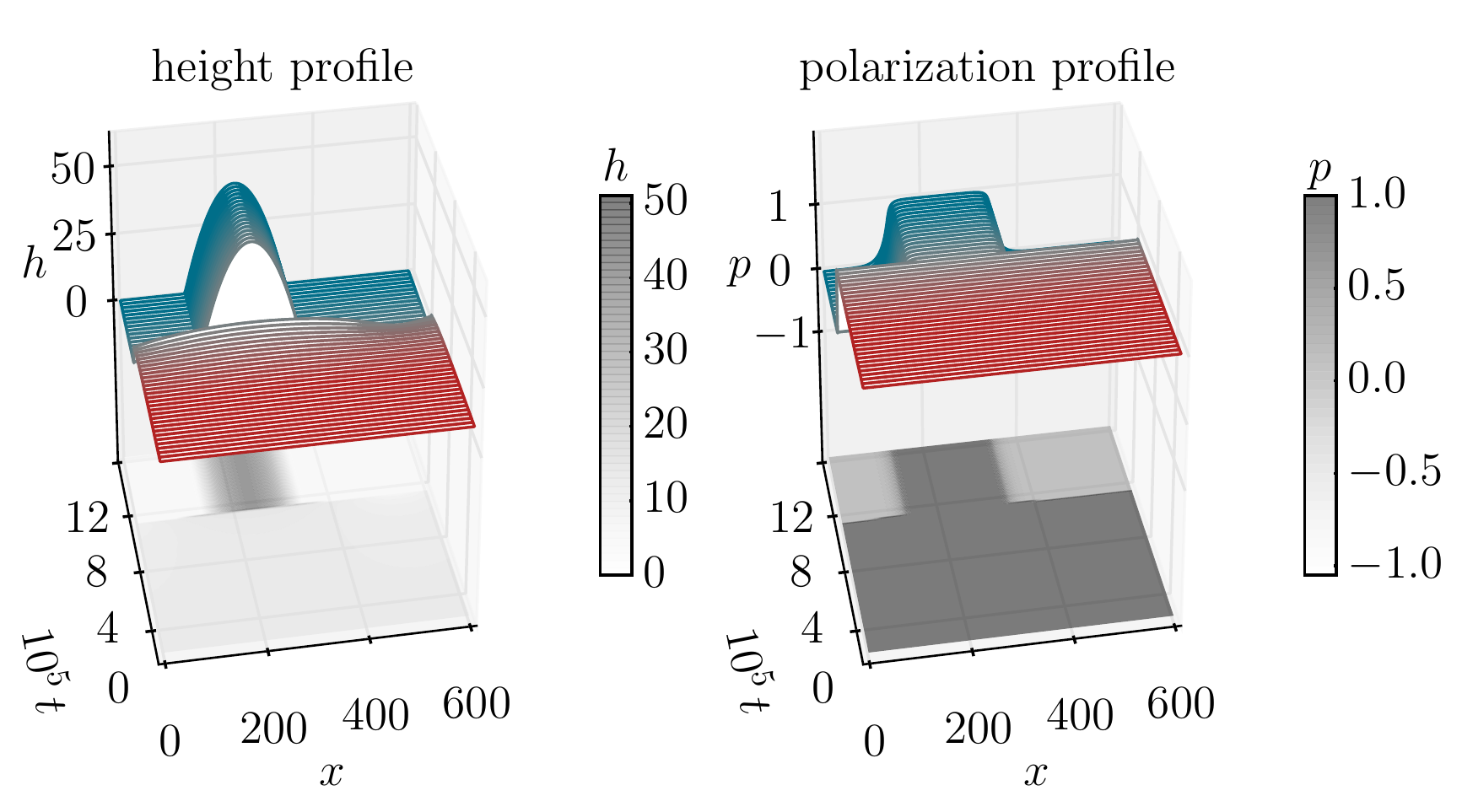}
\end{minipage}\hfill
\begin{minipage}[t]{0.48\hsize}
(b) initially unpolarized\\
\includegraphics[width=1.\textwidth]{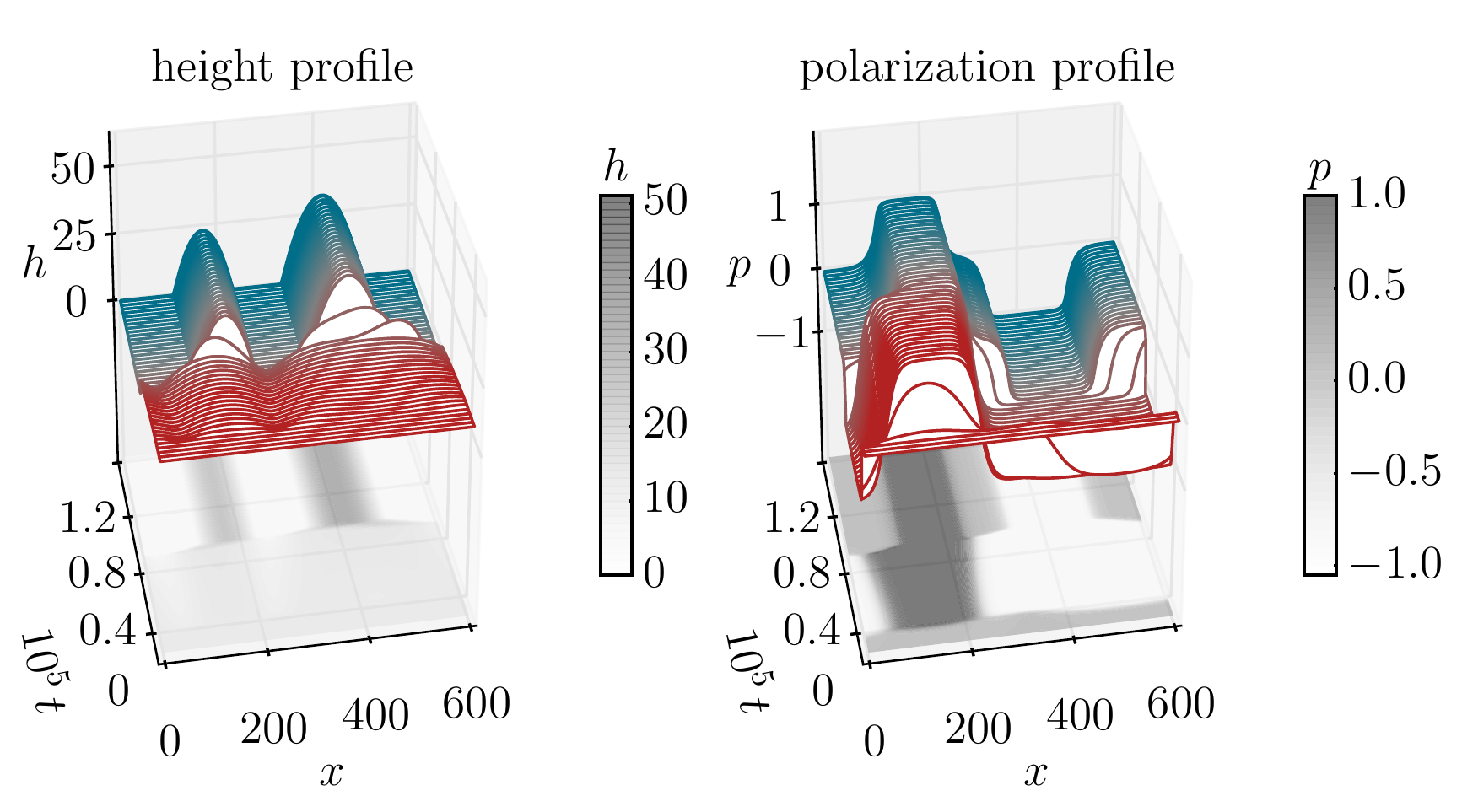}
\end{minipage}
\vspace*{4ex}
active film $c_\mathrm{a}=-0.01$, $\alpha_{0}=0.001$\\
\begin{minipage}[t]{0.48\hsize}
(c) initially polarized\\
\includegraphics[width=1.0\textwidth]{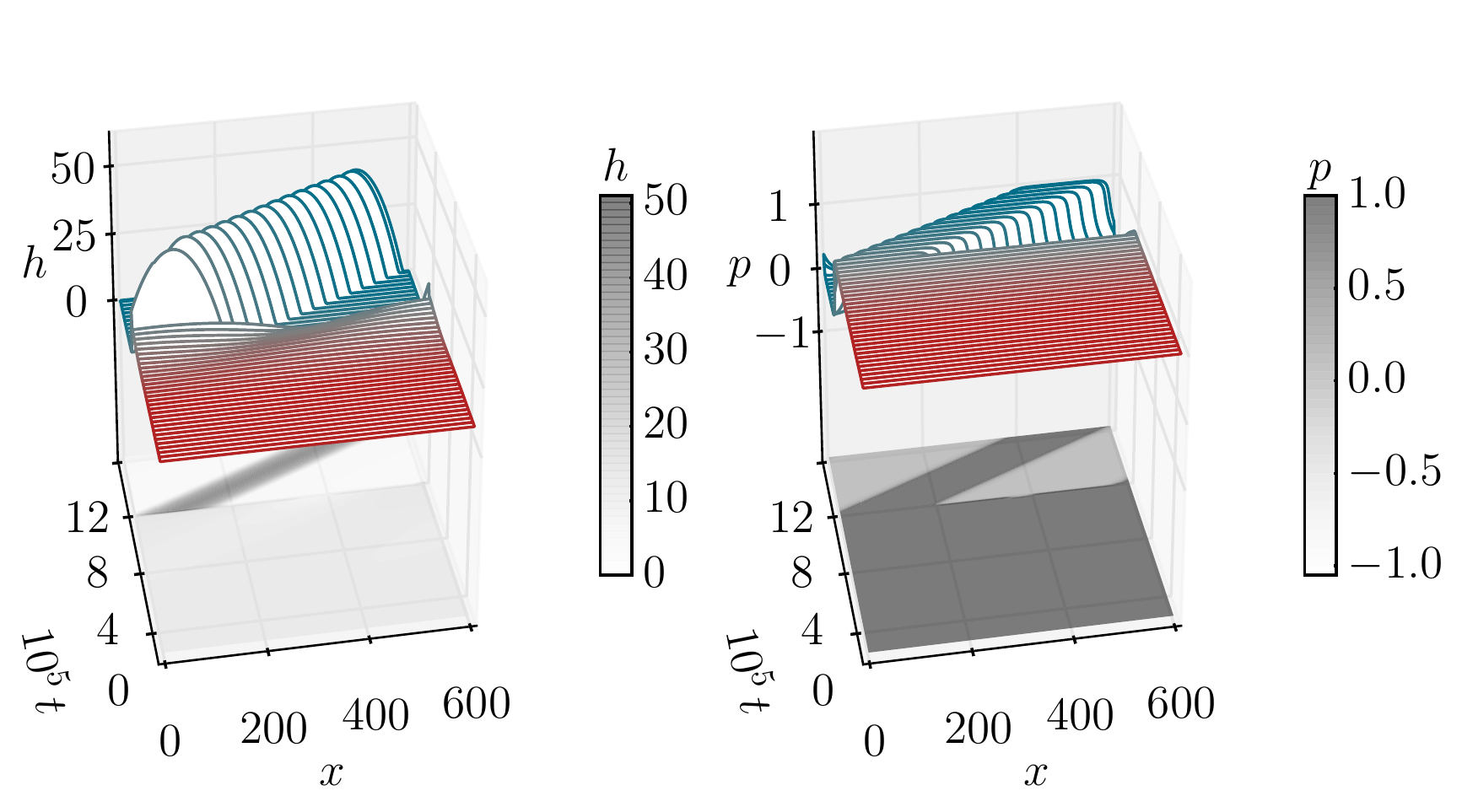}
\end{minipage}\hfill
\begin{minipage}[t]{0.48\hsize}
(d) initially unpolarized\\
\includegraphics[width=1.0\textwidth]{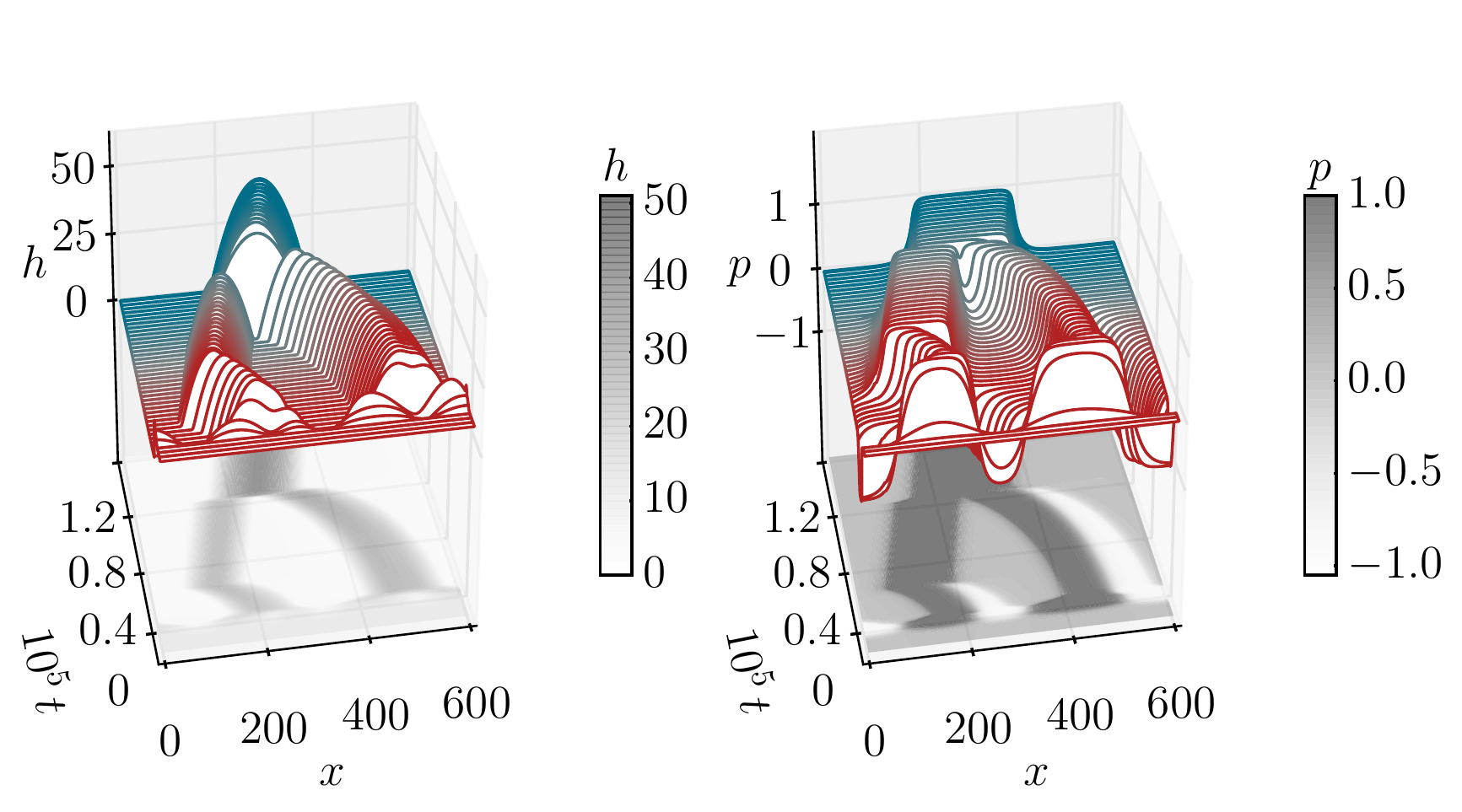}
\end{minipage}
\caption{Direct time simulations of the dewetting dynamics of initially flat homogeneous films of height $h_0=10$. 
Panels (a) and (b) give the dynamics of a passive film ($c_\mathrm{a}=0$, $\alpha_0=0$) for initially homogeneous polarized and unpolarized films, respectively.
Panels (c) and (d) show the cases with active stress and self-propulsion ($c_\mathrm{a}=-0.01$, $\alpha_0=0.001$) for initially homogeneous polarized and unpolarized films, respectively.
Shown are space-time plots of height profiles $h(x,t)$ (left) and polarization $p(x,t)$ (right) at equidistant times with colors varying from red at early times to blue at late times. 
Note the different time scales of the dynamics in the four cases: The dewetting accelerates for films which are already polarized, compared to unpolarized films.
The gray-scale shading below the profiles are contour plots giving an alternative visualization of the dynamics. The remaining parameters are as in Fig.~\ref{Fig:EVs}.\label{Fig:waterfall}}
\end{figure*}

We analyze the dewetting dynamics of flat homogeneous films by performing direct numerical simulations for a system of size $\Omega=[0,600]$ 
discretized on an equidistant mesh with $N_x = 256$ grid points and periodic boundary conditions employing the finite element-based modular toolbox DUNE-PDELAB \cite{BBD+2008c,BBD+2008c}.  Figure~\ref{Fig:waterfall} shows examples of the dewetting dynamics of initially flat films of height $h_0=10$ with a small random noise of amplitude $a=0.2$.  
Space-time plots consist of snapshots of film height and polarization profiles at equidistant times. 
First we consider the passive case, i.e., without active stress ($c_\mathrm{a}=0$) and self-propulsion ($\alpha_0=0$).  Figures~\ref{Fig:waterfall}~(a) and~(b) show the evolution of initially polarized and unpolarized films, 
respectively. 
For the passive polarized film in Fig.~\ref{Fig:waterfall}~(a), as expected, a sinusoidal modulation becomes visible at early times consistent with a spatially periodic instability triggered by the eigenvalue $\lambda\lowt{P,1}$.  
It grows in amplitude, becomes less harmonic beyond the linear regime until, eventually, a steady equilibrium droplet with uniform polarization is approached.  In the precursor film outside the droplet the polarization vanishes.  
If the passive film is initially unpolarized [Fig.~\ref{Fig:waterfall}~(b)], we observe the formation of domains of different orientation of the polarization, i.e., in the present 1D case of different sign. 
Domains of opposite polarization are separated by a domain wall (sometimes also referred to as \enquote{kink}/\enquote{anti-kink} or \enquote{defect}).  
Due to the coupling between polarization and film height, the strong polarization gradient across a domain wall can induce a strong flow that contributes to the modulation of the film height.  
This coupling drastically accelerates the dewetting dynamics as compared to the passive polarized case (by about one order of magnitude). Here, at the end of the simulation of the dewetting process, two steady droplets of opposite polarization are formed.
The two droplets in Fig.~\ref{Fig:waterfall}~(b) coarsen into one droplet on a time scale of order $10^9$ (not shown).
For other initial noise realizations, the early self-polarization process can lead immediately to a homogeneously polarized film (data not shown). 
In that case, dewetting takes place on the same time scale as for the initially polarized passive film shown in Fig.~\ref{Fig:waterfall}~(a).
The presence of active stress ($c_\mathrm{a}=0.01$) and self-propulsion ($\alpha_0=0.001$) modifies the dewetting and coarsening dynamics as shown for initially polarized and unpolarized films in Figs.~\ref{Fig:waterfall}~(c) and (d), respectively. 
For the active polarized film [Fig.~\ref{Fig:waterfall}~(c)], due to the self-propulsion the developing harmonic modulations in the film height travel along the film surface consistent with the complex eigenvalues determined in the linear analysis. 
The modulations grow, become fully nonlinear and the dynamics converges to an active droplet with a uniform polarization moving with constant shape and speed.  For the initially unpolarized active film shown in Fig.~\ref{Fig:waterfall}~(d), the dynamics is more involved as 
it reflects the existence of two unstable eigenvalues $\lambda\lowt{NP,1}(k)$ and $\lambda\lowt{NP,2}(k)$.  At first, on a short timescale the flat film polarizes and domains of positive and negative polarization form. This gives rise to counteracting flows in the film due to self-propulsion.  
In consequence, dewetting is further accelerated as compared to the corresponding passive case in Fig.~\ref{Fig:waterfall}~(b). Droplets with different orientation of the polarization form and move in opposing directions.  This drastically affects and accelerates the coarsening process: 
The droplets coalesce until only one large steadily moving drop of uniform polarization remains.
Interestingly, starting from other realizations of the initial random noise, the same dewetting process may as well result in droplets of non-uniform polarization, i.e., droplets that contain domain walls. 
We investigate this phenomenon in more depth in the next section. Note that the real parts of the eigenvalues $\lambda\lowt{P,1}(k)$ and $\lambda\lowt{NP,1}(k)$ [Eqs.~(\ref{eq:EV_NP}) and (\ref{eq:EV_P}), respectively] 
are identical for thick films $h_\mathrm{a}/h_0\ll 1$, i.e., the linear height mode is decoupled from the polarization and the active stress has no influence. The self-propulsion strength only affects the imaginary part of $\lambda\lowt{P,1}(k)$ for polarized films. For the parameters used in Fig.~\ref{Fig:waterfall}, the fastest growing instability mode connected to spatial modulations in the film height has in all considered cases a wave number of $k_\mathrm{max}\approx0.012$, i.e., a wavelength  $L_\mathrm{max}\approx513$.
Therefore, the difference in the dewetting dynamics and resulting drop number on the time scale of the simulations is not an effect related to system size but results from a nonlinear coupling between film height and polarization.
In the following section, we analyze moving and resting droplets in more detail and study the effect of the active stress parameter $c_\mathrm{a}$ and the self-propulsion speed $\alpha_0$ on the droplet shape and dynamics.
\section{Drops of passive and active polar fluid}
\label{sec:res-drop}
For droplets consisting of active polar particles, self-propulsion of the particles and active stresses modify the fluxes within the droplet as compared to droplets consisting of passive liquids. 
In consequence, the coupling between the film height profile and the polarization field can be expected to give rise to modified steady shapes and may induce active motion of the droplets along the substrate. 
Here, we investigate the drop behavior with direct time simulations employing the finite-element based library \texttt{oomph-lib} \cite{HH2006} and with continuation methods using the \textsc{Matlab} toolbox \texttt{pde2path} \cite{UeWR2014nmma, EGUW2019springer}.
\subsection{Stationary states and dynamics of passive drops}
\label{sec:res-drop-ini-unpol}
The previous section has shown that depending on the initial conditions, the dewetting process can result in drops with a single or several polarization domains. To better understand this phenomenon, we first analyze the existence and stability of stationary states of passive polarized droplets depending on the occurrence of domain walls.
\begin{figure}[htbp]
\begin{center}
\includegraphics[width=0.5\textwidth]{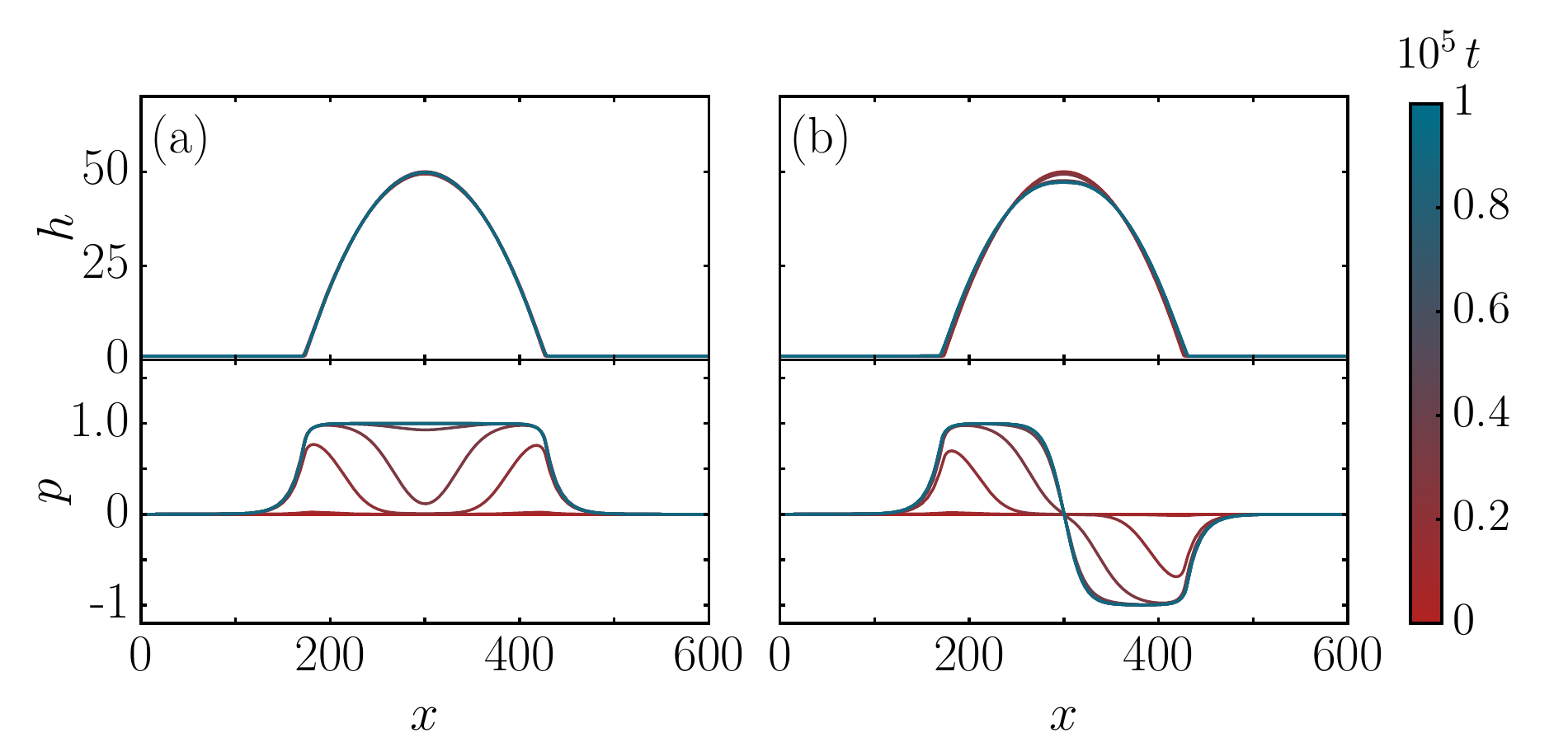}  
\caption{Full time simulations of passive droplets ($c_\mathrm{a} = 0$ and $\alpha_0 =0$). Shown are simulation snapshots of height profiles $h$ (top)  and mean polarization profiles $p$ (bottom) at equidistant times of (a) uniformly polarized and (b) 
non-uniformly polarized droplets with identical parameter values. 
The simulations in (a) and (b) are initiated with slightly different polarization patterns within the droplet, given in Appendix \ref{ap:init:cond}. 
(a) If the polarization profile resulting from self-polarization is uniform within the droplet, the droplet's height profile does not change compared to its initial state. (b) 
If a non-uniform polarization profile with two domains of opposite polarization results, the drop profile becomes wider and lower. Due to symmetry, the height profile is not affected by the transformation $p\rightarrow -p$. The remaining parameters are as in Fig.~\ref{Fig:EVs}. }
\label{Fig:Simulation_active}
\end{center}
\end{figure}
We initiate the time simulations with the equilibrium drop shape of the corresponding passive case (parabolic drop with a contact angle corresponding to the equilibrium contact angle of the non-polar fluid) 
with an added small random polarization field within the droplet. The specific initial conditions for the polarization field are detailed in Appendix \ref{ap:init:cond}.   
The resulting evolution toward two qualitatively different types of passive droplets is shown in Fig.~\ref{Fig:Simulation_active}. 
A uniformly polarized drop develops in Fig.~\ref{Fig:Simulation_active}~(a). An initial self-polarization stage starts by developing positive polarization in both contact line regions, then extends into the entire drop, and reaches a rather uniform plateau with $p=1$ at the drop center. 
Across the contact line region, the polarization decreases to $|p|\approx 0$ smoothed by the elastic energy $f\lowt{el}$ \eqref{eq:fel} that penalizes strong gradients in $p$.
In the course of the process, the total polarization within the droplet monotonically increases.
In contrast, Fig.~\ref{Fig:Simulation_active}~(b) shows a scenario where the initial self-polarization stage starts with the development of opposite polarization in the two contact line regions. 
These then extend towards the drop center where a domain wall develops that separates short plateaus with $p=1$ and $p=-1$. The resulting state after $t=10^5$ is a non-uniformly polarized drop with, in this case inward pointing, counteracting polarization. It is slightly lower and has a slightly smaller contact angle than the initial drop, as depicted in Fig.~\ref{Fig:Simulation_active}~(b).
Due to symmetry, the height profile is not affected by the transformation $p\rightarrow -p$, 
i.e. the additional possible solutions with left pointing and outward pointing polarization profiles, respectively, feature identical droplet profiles as in  Fig.~\ref{Fig:Simulation_active}~(a) and (b).
Note, that these simulations do not imply the long-time stability of the depicted polarized drop states, which will be investigated in the next step.
To understand the connection in parameter space of the different states observed in the previous section and to analyze their stability, we apply continuation methods using the \textsc{Matlab} toolbox \texttt{pde2path} \cite{UeWR2014nmma, EGUW2019springer}. 
\begin{figure*}[htbp]
\begin{center}
\includegraphics[width=1.0\textwidth]{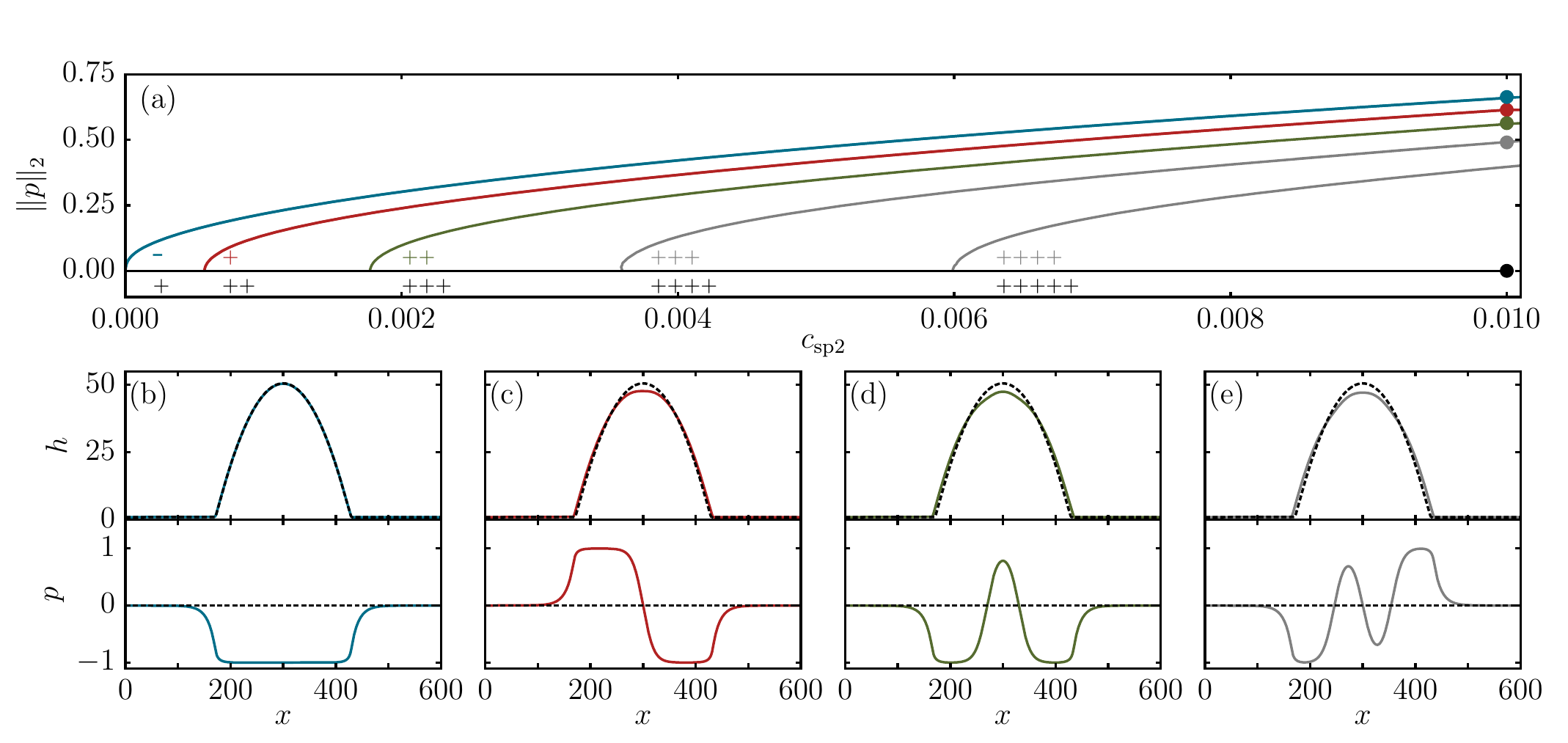}  
\caption{Emergence of various passive states of polarized droplets from unpolarized passive droplet states. (a) shows the bifurcation diagram (characterized by the $L^{2}$-norm ${\lVert p \rVert}_2=\sqrt{\frac{1}{L}\int_{0}^{L} p^2\textrm{d}x}$ of the mean polarization field $p$) depending on the parameter
$c\lowt{sp2}$ [cf. Eqs. \eqref{eq:fspo}]. The stability of each solution branch is indicated by $+$ (unstable) and $-$ (stable) signs in the respective colors. The black branch corresponds to the unpolarized droplet.
The blue branch corresponds to the uniformly polarized droplet and the red branch corresponds to the droplet with two polarization domains of opposite polarization, i.e.,  with one domain wall.  The the green and gray branches indicate non-uniformly polarized droplets with two or more defects, respectively.
Panels (b-e) show the film height and mean polarization profiles respectively, for the specific value $c\lowt{sp2}=0.01$ as indicated by the filled circles in (a). 
For comparison, the black dashed solutions show the height profile of the unpolarized drop solution. Note, that by symmetry the polarization profiles with $p(x)\rightarrow -p(x)$ give identical height profiles. 
Remaining parameters are as in Fig.~\ref{Fig:Simulation_active}. \label{Fig:bif_csp2}}
\end{center}
\end{figure*}
\begin{figure}[htbp]
\begin{center}
\includegraphics[width=0.5\textwidth]{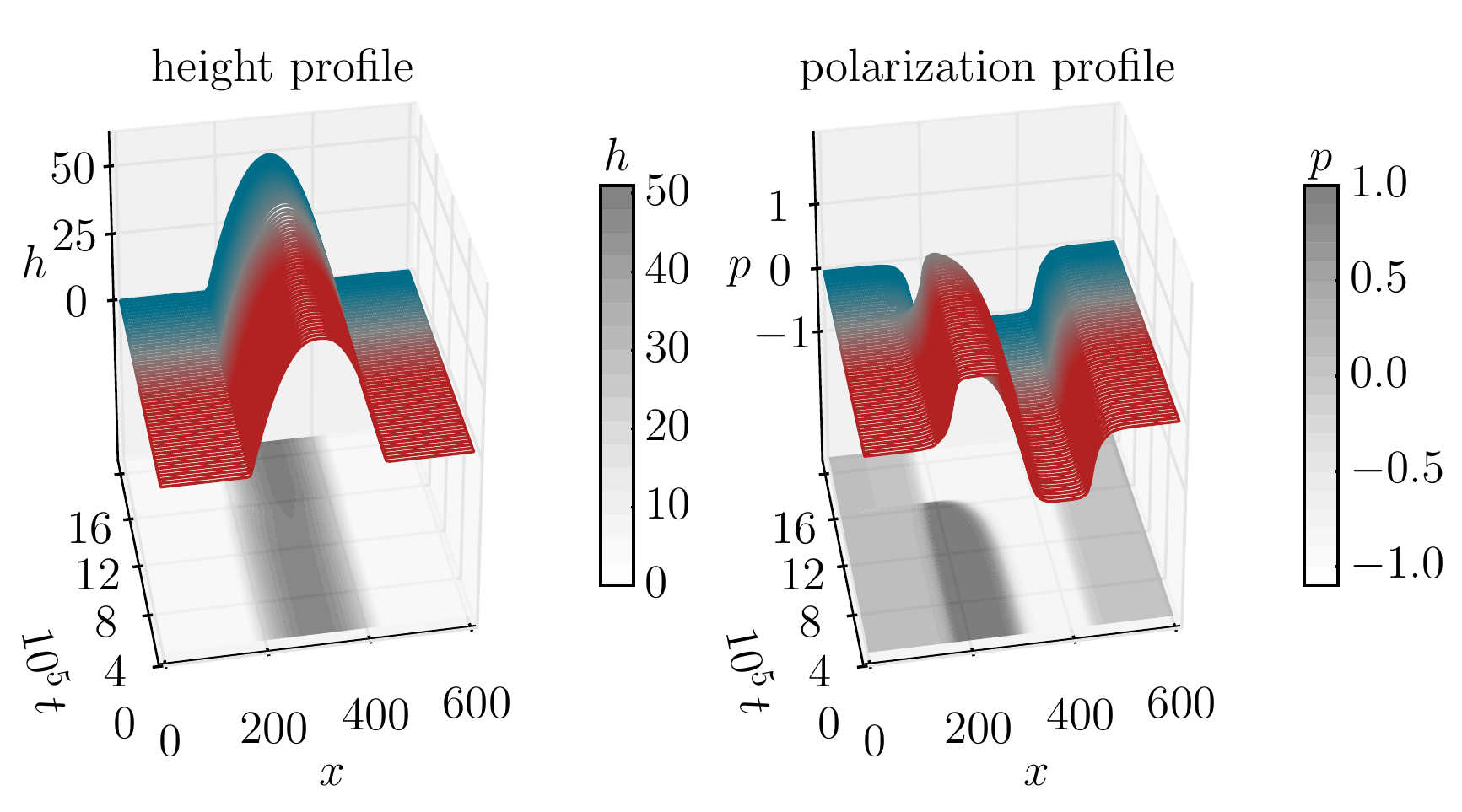} 
\caption{Long time simulation of a passive drop that initially contains two domains of opposite polarization (inward pointing). 
At $t\approx10^6$ the polarization within the drop has become uniform.
Parameters are as in Fig.~\ref{Fig:Simulation_active}.}
\label{Fig:waterfall-appendix}
\end{center}
\end{figure}
Fig.~\ref{Fig:bif_csp2}~(a) shows a bifurcation diagram in terms of the L$^{2}$-norm (independent of system size $L$) and the spontaneous polarization parameter $c_\mathrm{sp2}$, which determines the amplitude of the height averaged spontaneous polarization in the droplet [see Eq.~\eqref{eq:fspo}]. 
Here, we chose $c\lowt{sp2}=0$ as a starting point, i.e., the disordered state [black solid branch in Fig. ~\ref{Fig:bif_csp2}~(a)] is energetically favored and stable. When increasing $c\lowt{sp2}$ towards $c\lowt{sp2}=c\lowt{sp4}=c\lowt{sp}=0.01$ [see Eq.~\eqref{eq:fspo}], first the polarized state $p=1$ (by symmetry also $p=-1$)
arise and gain stability whereas the disordered state $p=0$ becomes unstable as indicated by the $+$ signs corresponding to the number of unstable eigenmodes. As $c\lowt{sp2}$ increases, other unstable branches bifurcate from the unpolarized droplet state.
All polarized states bifurcate supercritically form the unpolarized branch, whereby the number of defects increases with increasing $c\lowt{sp2}$. The uniformly polarized drops, depicted for $c\lowt{sp2}=0.01$ in Fig.~\ref{Fig:bif_csp2}~(b) are stable solutions.
Due to the successive destabilization of the unpolarized drops, the next bifurcating branch, e.g., the red solid branch in Fig.~\ref{Fig:bif_csp2}~(a) is unstable and the steady states shown in Fig.~\ref{Fig:bif_csp2}~(c) transform over time into a uniformly polarized drop as proven by direct time simulation. 
Figure~\ref{Fig:waterfall-appendix} shows a simulation initiated using the same conditions as for Fig.~\ref{Fig:Simulation_active}~(c). However, the simulation was carried out much longer. The non-uniformly polarized droplet emerges and forms a long-time transient state. 
The polarization pattern only changes into the stable uniformly polarized drop with $p=-1$ after $t\approx10^6$. In comparison, the transition from a disordered to a non-uniformly polarized state  takes place on a time scale of about $t\approx10^5$ [cf.~Fig.~\ref{Fig:Simulation_active}~(a, b)].
The additional branches bifurcating from the black branch in Fig.~\ref{Fig:bif_csp2}~(a) correspond to non-uniformly polarized drops with more than one domain wall. As examples, Fig.~\ref{Fig:bif_csp2}~(d) and (e) show droplets containing two and three domain walls, respectively. 
However, these branches feature an increasing number of positive eigenvalues, thus are increasingly unstable. Here, we do not consider them further and focus on the behavior of uniformly polarized droplets and droplets containing one domain wall.
For comparison, the unpolarized drop, i.e., $p=0$, is represented in Fig.~\ref{Fig:bif_csp2}~(b)-(e) as the black dashed lines, to indicate the changes in the height profile.
\subsection{Stability and dynamics of active polar drops}
\label{sec:continuation}
So far we have investigated the polarization states and stability of passive droplets.
\begin{figure*}[htbp]
\begin{center}
\includegraphics[width=1.0\textwidth]{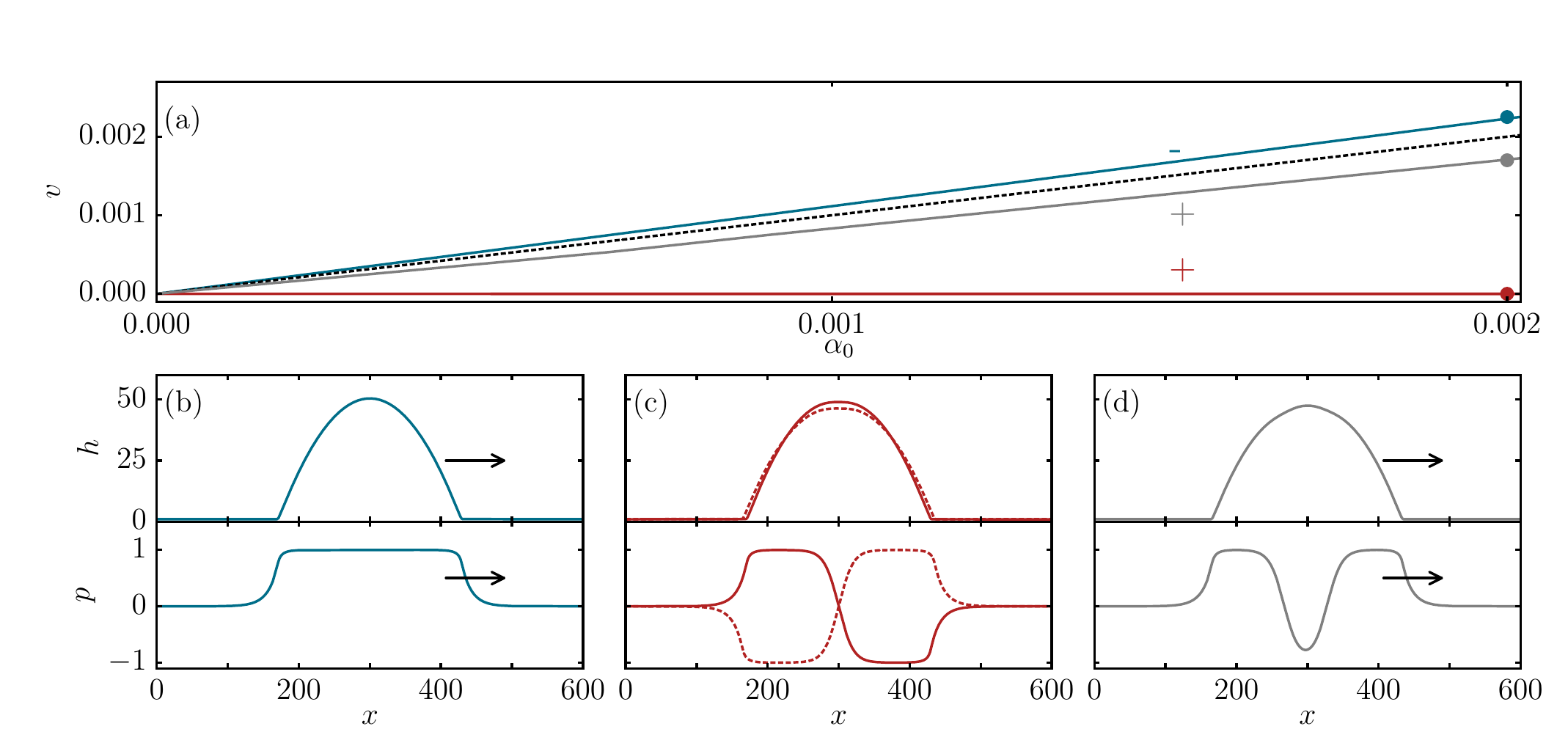}  
\caption{Velocity of uniformly and non-uniformly polarized active droplets in dependence of the self-propulsion strength $\alpha_0>0$ in the case without active stress ($c_\mathrm{a}=0$).The bifurcation diagram in (a) depicts branches of moving and resting droplets, depending on the polarization profiles within the droplet. Profiles are shown in (b)-(d) in corresponding colors. Stability is indicated by '$+$' (unstable) and '$-$' (stable) signs in the respective colors. Note, that neither the uniformly nor the non-uniformly (two defects, panel~(d)) droplets move with exactly $v=\alpha_{0}$  as indicated by the black dashed line in (a). Panels (b)-(d) show height (top) and polarization (bottom) profiles for $\alpha_0 =0.002$ indicated by filled circles in (a). For steady droplets (one defect, panel~(c)), the self-propulsion $\alpha_{0}$ breaks the symmetry w.r.t.\ the transformation $p\rightarrow -p$, resulting in a (small) differences in the droplet height profiles as exemplarily shown in (c) (solid vs.\ dashed line). For moving droplets, the asymmetry of the height profile is most prominent in the contact line region and is not visible on the macroscopic scale represented in (b), (d). Remaining parameters are as in Fig.~\ref{Fig:Simulation_active}.}
\label{Fig:bif_alpha}
\end{center}
\end{figure*}
\begin{figure}[htbp]
\begin{center}
\includegraphics[width=0.5\textwidth]{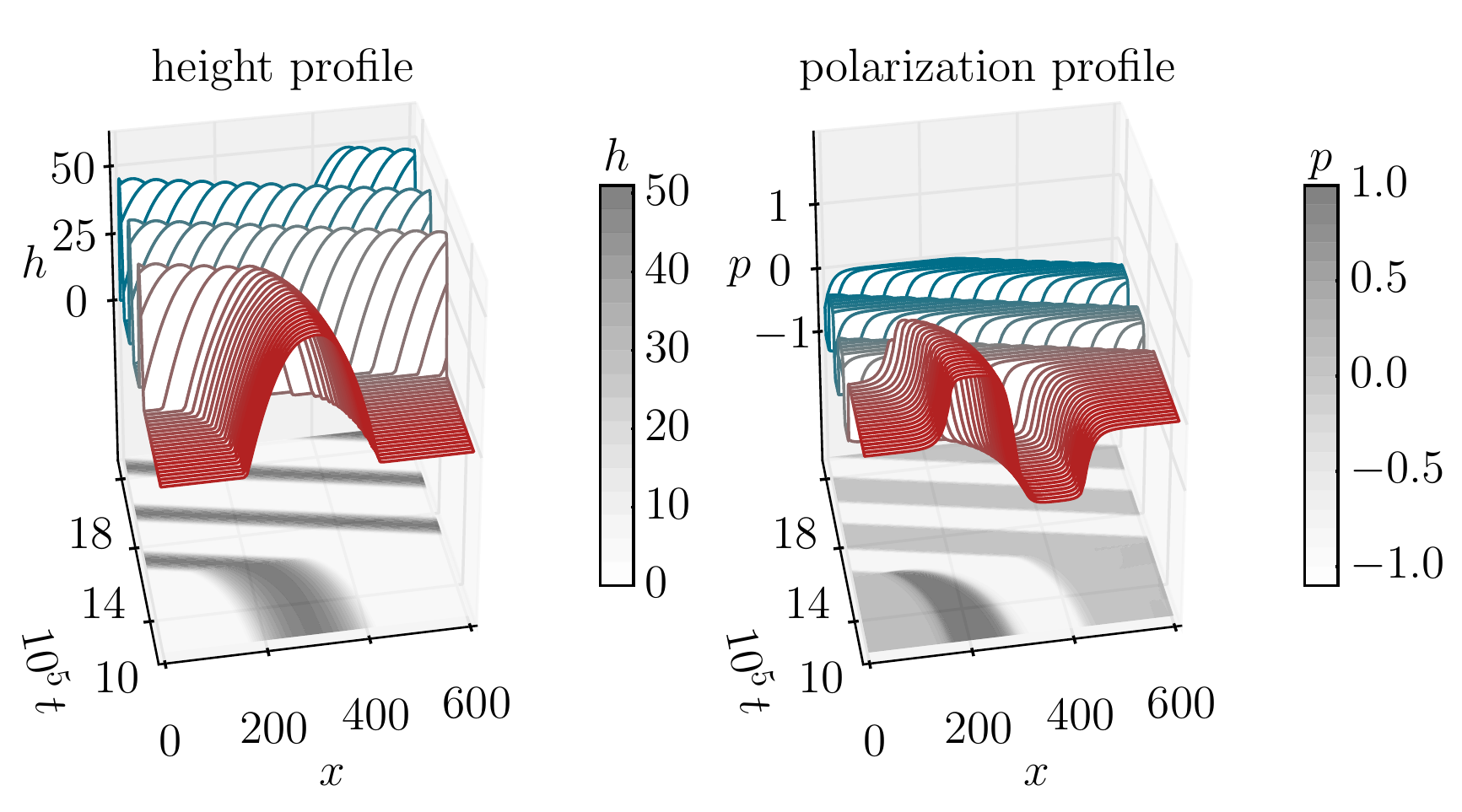} 
\caption{Long time simulation of an initially non-uniformly polarized droplet at self-propulsion strength $\alpha_{0}=0.002$ in the absence of active stress ($c_\mathrm{a}=0$). Shown are the evolution of (left) the droplet height profile and (right) the polarization profile. 
Initially the droplet is stationary and contains two polarization domains of opposite polarization (inward pointing, zero net polarization).
The droplet starts to move to the left after a long transition time when the net polarization is non zero (negative).
Remaining parameters are as in Fig.~\ref{Fig:bif_alpha}.}
\label{Fig:waterfall-appendix-alpha0}
\end{center}
\end{figure}
Next, we focus on the influence of self-propulsion $\alpha_{0}$ on uniformly and non-uniformly polarized droplets. Thereby, the polarization $p$ becomes a polar order parameter, which breaks the parity symmetry.
Starting from the parameter settings indicated by the filled circles in Fig.~\ref{Fig:bif_csp2}~(a) we perform a parameter continuation in $\alpha_{0}$. 
Figure \ref{Fig:bif_alpha}~(a) depicts the dependence of drop velocity on self-propulsion strength $\alpha_0$ for droplet with various polarization states. We find that the uniformly polarized droplets move for any $\alpha_0\neq0$ into their polarization direction, as expected. The speed increases linearly with $\alpha_0$ and the height profile barely changes, for the small values of $\alpha_0$ investigated.
\begin{figure*}[htbp]
\begin{center}
\includegraphics[width=1.0\textwidth]{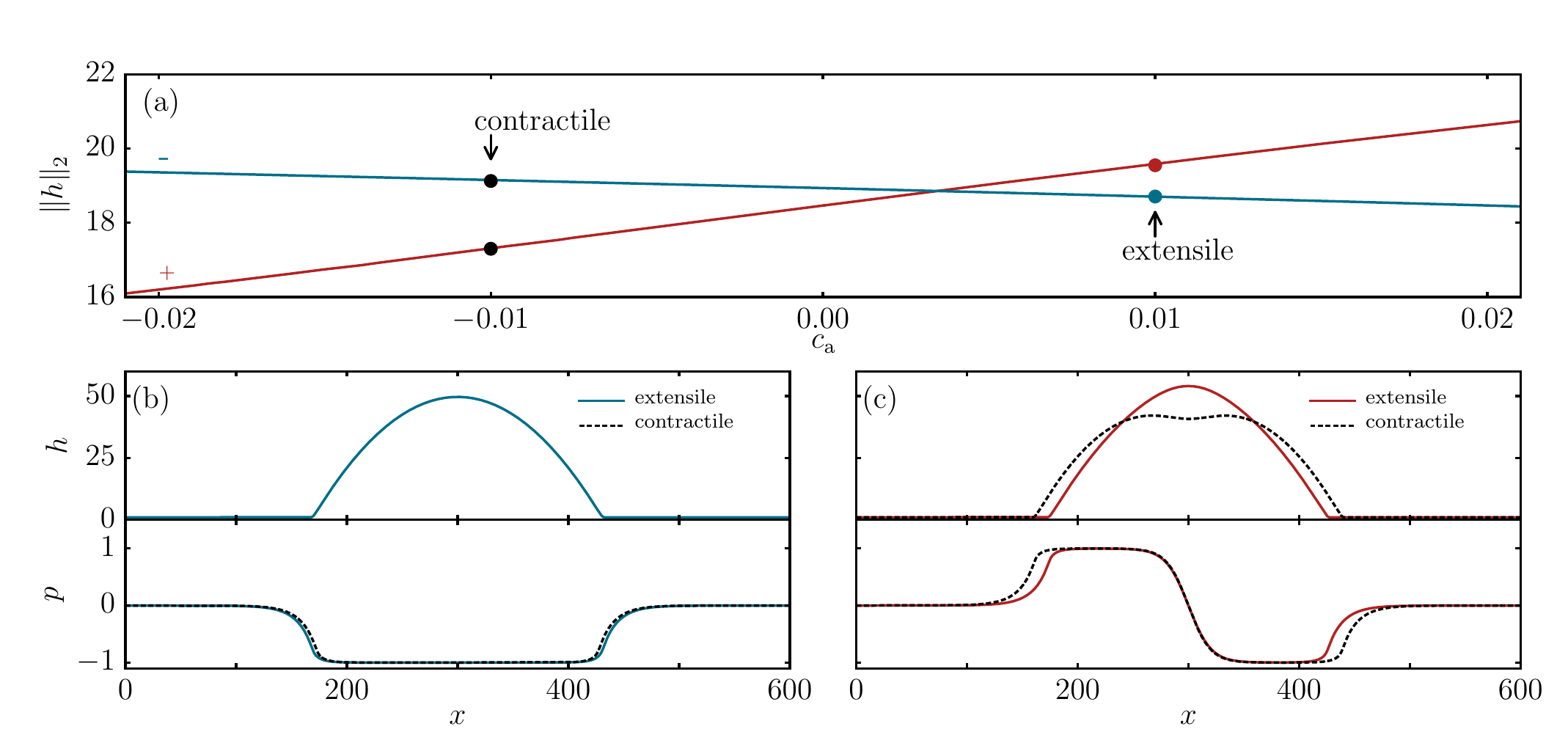}  
\caption{Dependence of droplet height and polarization profiles on the active stress parameter $c_\mathrm{a}$ in the absence of self-propulsion ($\alpha_0 =0$) for uniformly and non-uniformly polarized droplets. All states are stationary. (a) shows the $L^2$-norm of $h$, i.e., the droplet height deviating from its mean value $h\lowt{m}=\frac{1}{L}\int_{0}^{L} h\,\textrm{d} x$, depending on the active stress $c\lowt{a}$. 
The blue (red) solution branch corresponds to uniformly (non-uniformly) polarized drops, and their stability is denoted by '$+$' and '$-$' signs, respectively. 
(b) and (c) show the height (top) and polarization (bottom) profiles of uniformly and non-uniformly polarized droplets respectively, corresponding to the parameters indicated by the respectively colored circles in (a). 
Due to symmetry with respect to the transformation $p\rightarrow -p$, we only present one possible solution for each solution branch.
Remaining parameters are as in Fig.~\ref{Fig:Simulation_active}.}
\label{Fig:bif_ca}
\end{center}
\end{figure*}
\begin{figure}[!h]
\begin{center}
\includegraphics[width=0.5\textwidth]{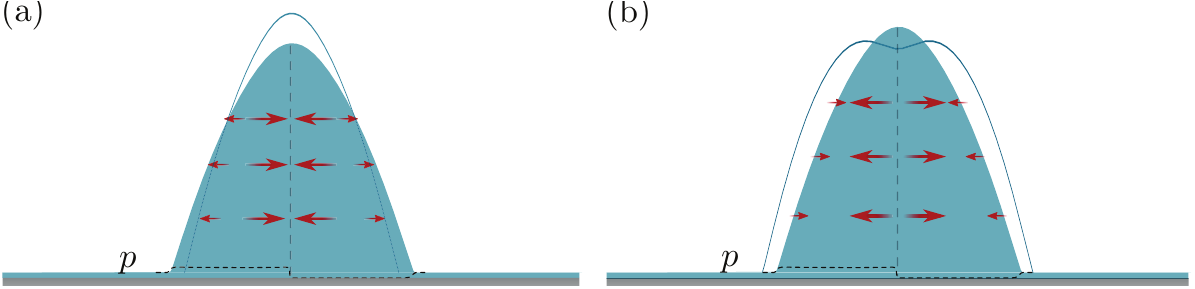}
\caption{Schematic illustration of the effect of active stress on non-uniformly polarized droplets. The shaded droplet represents the droplet shape in the absence of active stress, whereas the blue solid line represents the droplet shape with active stress. The polarization profile is indicated as black dashed line close to the solid-liquid interface.
(a) Extensile stress: Both domains are pushing the fluid outwards. However, due to the scaling $\sim -h^3(\partial_x p^2)$ of the active flux, fluxes are stronger in the center of the droplet, than at its periphery, which is denoted by the different sizes of the red arrows. 
Due to mass conservation the droplet becomes narrower. 
(b) Contractile stress: Each polarization domain attracts fluid. The strong fluxes directed away from the droplet center cause a dip in the height profile. Due to mass conservation the droplet becomes wider. 
Hence, the competition between active stress and mass conservation plays a crucial role for the drop shape.
\label{Sketch:active-stress} }
\end{center}
\end{figure}
For droplets containing 2 polarization domains, although being stationary, increasing self-propulsion breaks the symmetry between the inward pointing (red solid line in Fig.~\ref{Fig:bif_alpha}~(c)) and outward pointing polarization profiles (red dashed line in Fig.~\ref{Fig:bif_alpha}~(c)), visible in the difference in the height profiles. 
Due to their complete anti-symmetric polarization profiles the integral $\int ph \,\textrm{d}x$ vanishes and the 'net' polarization of the droplet is zero: The droplets remain therefore at rest. Droplets containing three polarization domains are moving (Fig.~\ref{Fig:bif_alpha}~(d)), albeit at lower speed than uniformly polarized drops. For the shown polarization profile, the net polarization is positive. Note, that the velocities of the moving states are not exactly equal to the self-propulsion strength $v=\alpha_{0}$ and depend on the polarization profile.
Again, the uniformly polarized states are stable and the non-uniformly polarized states are unstable, which is confirmed by direct time simulations as shown in Fig.~\ref{Fig:waterfall-appendix-alpha0}.
Starting from non-uniformly polarized droplets with one defect, the polarization pattern changes after $t\approx 10^6$ such that a uniformly polarized drop evolves which eventually moves to the left with constant shape and velocity.
For clarity, we only show the dynamics for $t>8\cdot10^5$, as before height and polarization profiles do nearly not change.\\
In a second line of investigation, we analyze the behavior of polarized droplets when varying the active stress $c\lowt{a}\neq 0$ without self-propulsion ($\alpha_{0}=0$).
To that end we perform parameter continuations in $c\lowt{a}$ taking the states from Fig.~\ref{Fig:bif_csp2}~(b) and (c) as starting points. We find that all polarized droplets with active stresses are stationary.
In Fig.~\ref{Fig:bif_ca} we present the dependence of the drop and polarization profiles on the magnitude of the active stress (contractile: $c_\mathrm{a}<0$, extensile: $c_\mathrm{a}>0$) for different polarization states. 
Note, that the active stress is only sensitive to the magnitude of the polarization, but not the direction, hence the polarization takes here the role of a nematic order parameter and we can restrict ourselves to the analysis of positively uniformly polarized droplets [Fig.~\ref{Fig:bif_ca}~(b)] and inward pointing non-uniformly polarized droplets [Fig.~\ref{Fig:bif_ca}~(c)].  
The oppositely polarized states are identical due to the nematic symmetry of the polarization field vis-\`a-vis the active stress tensor.  
As expected, for extensile stresses uniformly polarized drops become lower and wider whereas for contractile active stresses they become higher and narrower.  
However, non-uniformly polarized droplets show a more interesting behavior due to the strong gradient across the domain wall. The presence of two polarization domains within one droplet of a conserved volume leads to the somewhat paradoxical behavior, that droplets are wider in the presence of contractile stresses than in the presence of 
extensile stresses as shown in Fig.~\ref{Fig:bif_ca}~(c).
On the one hand when the active stress is extensile, the oppositely polarized domains push the fluid out, such that the droplet becomes higher at the center [red solid line in Fig.~\ref{Fig:bif_ca}~(c)].
\begin{figure}[htbp]
\begin{minipage}[t]{1.0\hsize}
(a) extensile stress\\
\includegraphics[width=1.0\textwidth]{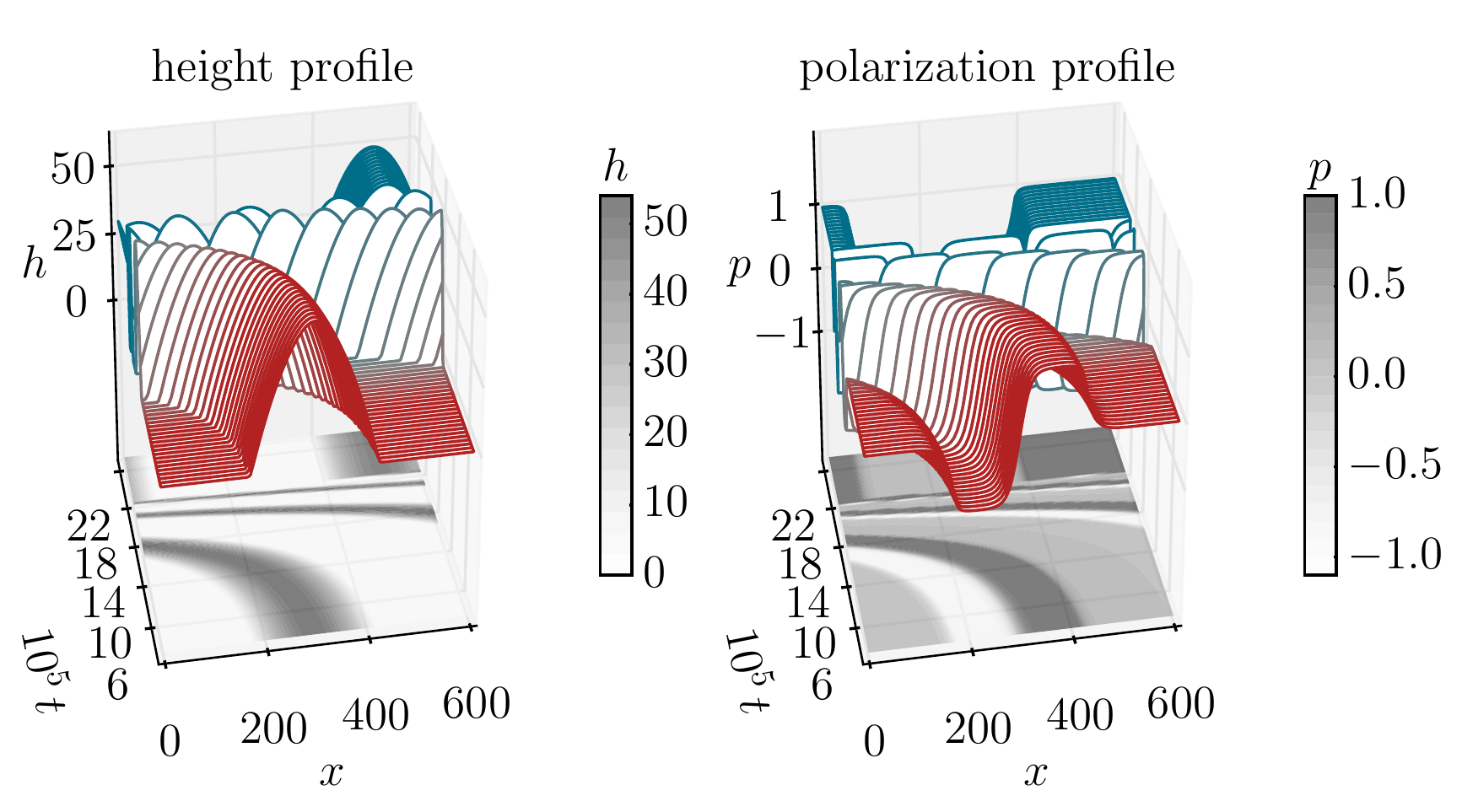} \\
\end{minipage}\hfill
\begin{minipage}[t]{1.0\hsize}
(b) contractile stress\\
\includegraphics[width=1.0\textwidth]{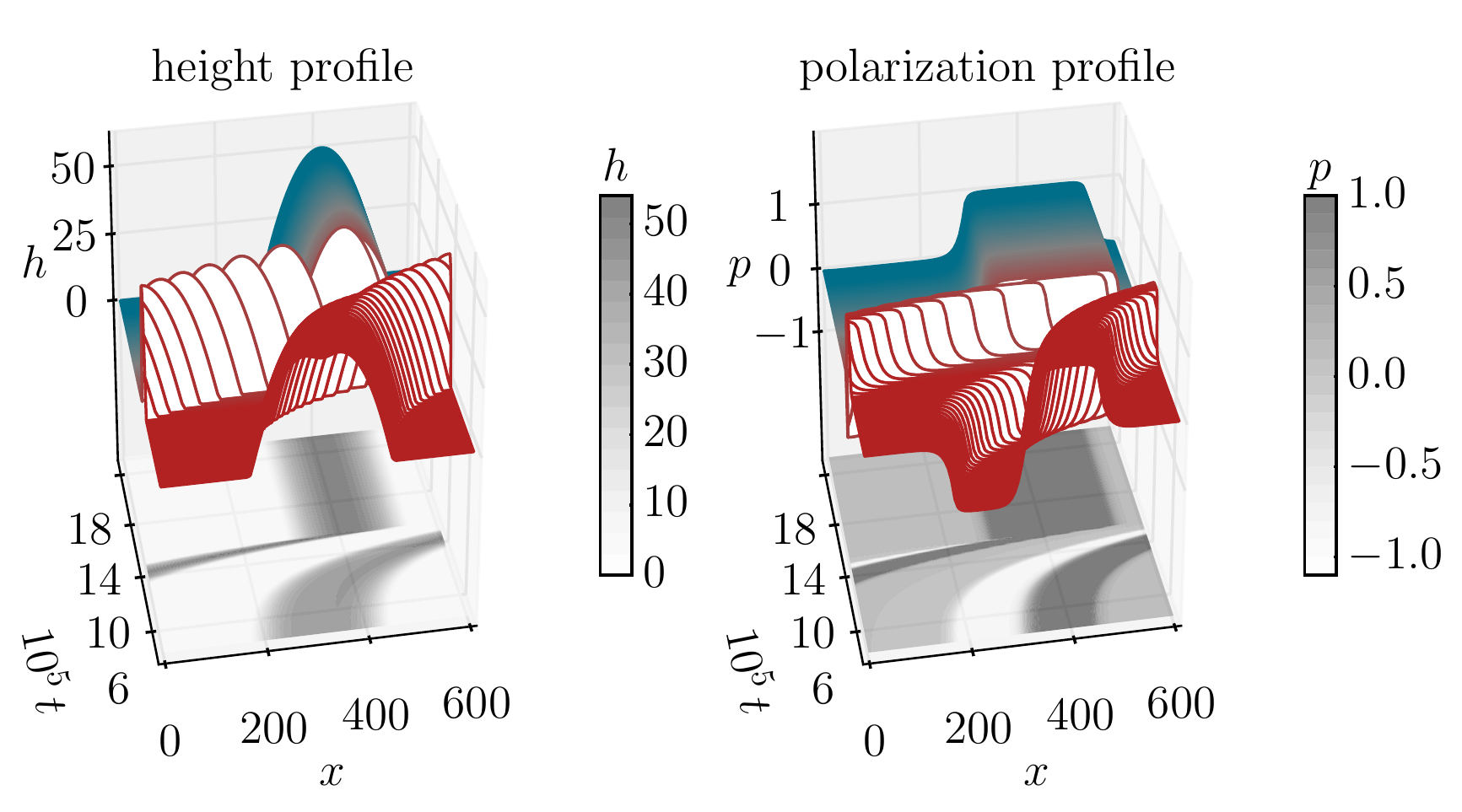} \\
\end{minipage}\hfill
\caption{Long time simulation for an initially non-uniformly polarized droplet with active extensile ($ c_\mathrm{a}=0.01$) (a) and contractile ($ c_\mathrm{a}=-0.01$) (b) stresses in the absence of self-propulsion ($\alpha_{0}=0$). 
Shown are the height profile (left) and the polarization profile (right). The transition from non-uniform to uniform polarization is accompanied by a strong transient motion of the droplet. In a frame moving with the droplet the domain wall moves towards the shrinking domain (here, to the left).
(a) For extensile stress, the droplet (in the laboratory frame) moves into the same direction as the domain wall (in the comoving frame), i.e., to the left.
(b) For contractile stress, the droplet (in the laboratory frame) moves into the opposite direction as the domain wall (in the comoving frame), i.e., to the right.
The droplet stops when the transformation into a uniformly polarized droplet is complete at $t\approx 2.2 \, 10^6$ and $t \approx 1.5\,  10^6$, respectively. Remaining parameters are as in Fig.~\ref{Fig:bif_ca}.}
\label{Fig:waterfall-ca}
\end{figure}
\begin{figure}[htbp]
\begin{center}
\includegraphics[width=0.5\textwidth]{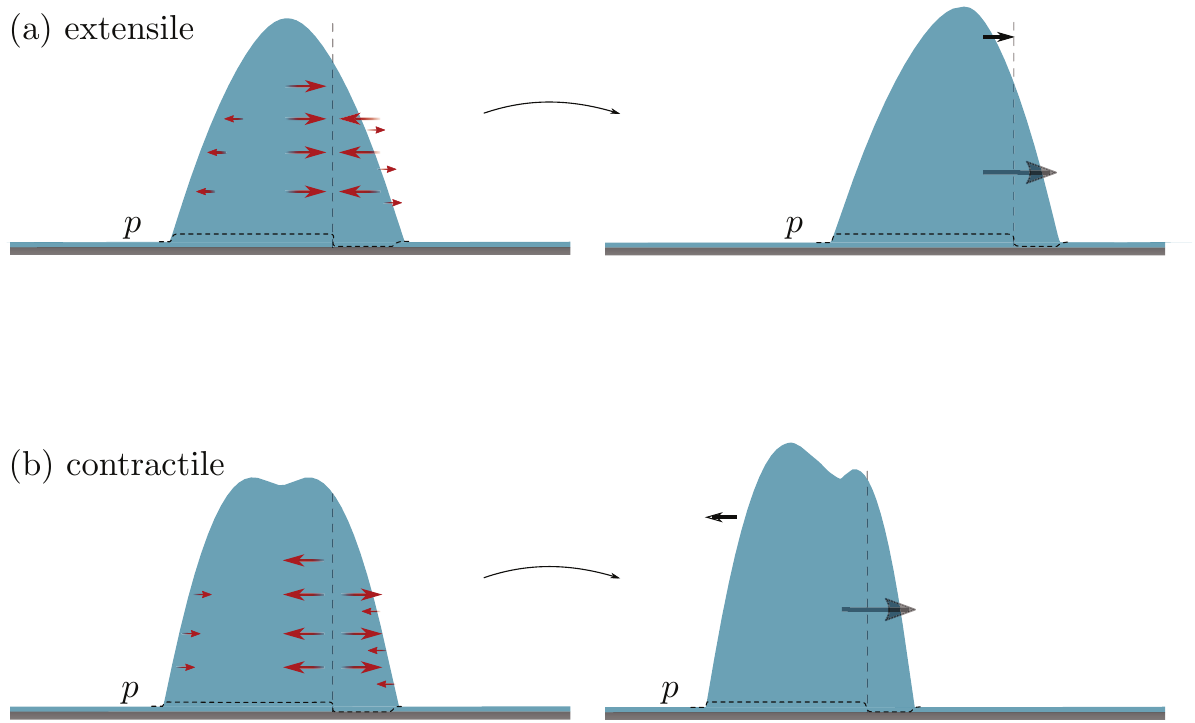}
\caption{Sketch of the droplet behavior during the transition from an unstable to a stable state characterized by a constantly moving domain wall, whose position is indicated by a vertical gray dashed line. 
The blue shaded drop profiles indicate (left) the initial symmetric resting state and (right) a later transient moving state. The black dashed lines indicate polarization profiles. 
(a) For extensile stress, the fluid in both domains is pushed towards the domain wall. 
As it moves off center, due to the local slope of the drop surface the net fluid flux around the wall is in the direction of its motion. 
Mass conservation implies that the entire drop moves into the same direction in the laboratory frame. (b) For contractile stress, the fluid is attracted into both domains resulting in a dip in the height profile at the domain wall.
However, as it moves off center, the net fluid flux around the wall and therefore the droplet motion in the laboratory frame is in the direction opposite to the motion of the domain wall in the comoving frame.} 
\label{Sketch:active-stress-transient} 
\end{center}
\end{figure}
Fluxes at the center of the droplet are stronger than at its periphery. This is due to the highly nonlinear $h$-dependence of the flux that is caused by the active stress [cf.~Eq.~\eqref{eq:fluxes}]. In consequence, mass conservation causes the droplet to become narrower, as fluid is more strongly attracted towards the domain wall at the droplet center. 
This behavior is sketched schematically in Fig.~\ref{Sketch:active-stress}~(a). The red arrows indicate the direction and strength (scaled with the gradient in $p$) of fluid flow 
for each domain. On the other hand, for contractile stresses, the two polarization domains compete to attract the fluid and the droplet's height profile develops a dip at the position of the domain wall [black dashed line in Fig.~\ref{Fig:bif_ca}~(c)]. 
\begin{figure*}[htbp]
\begin{center}
\includegraphics[width=1.0\textwidth]{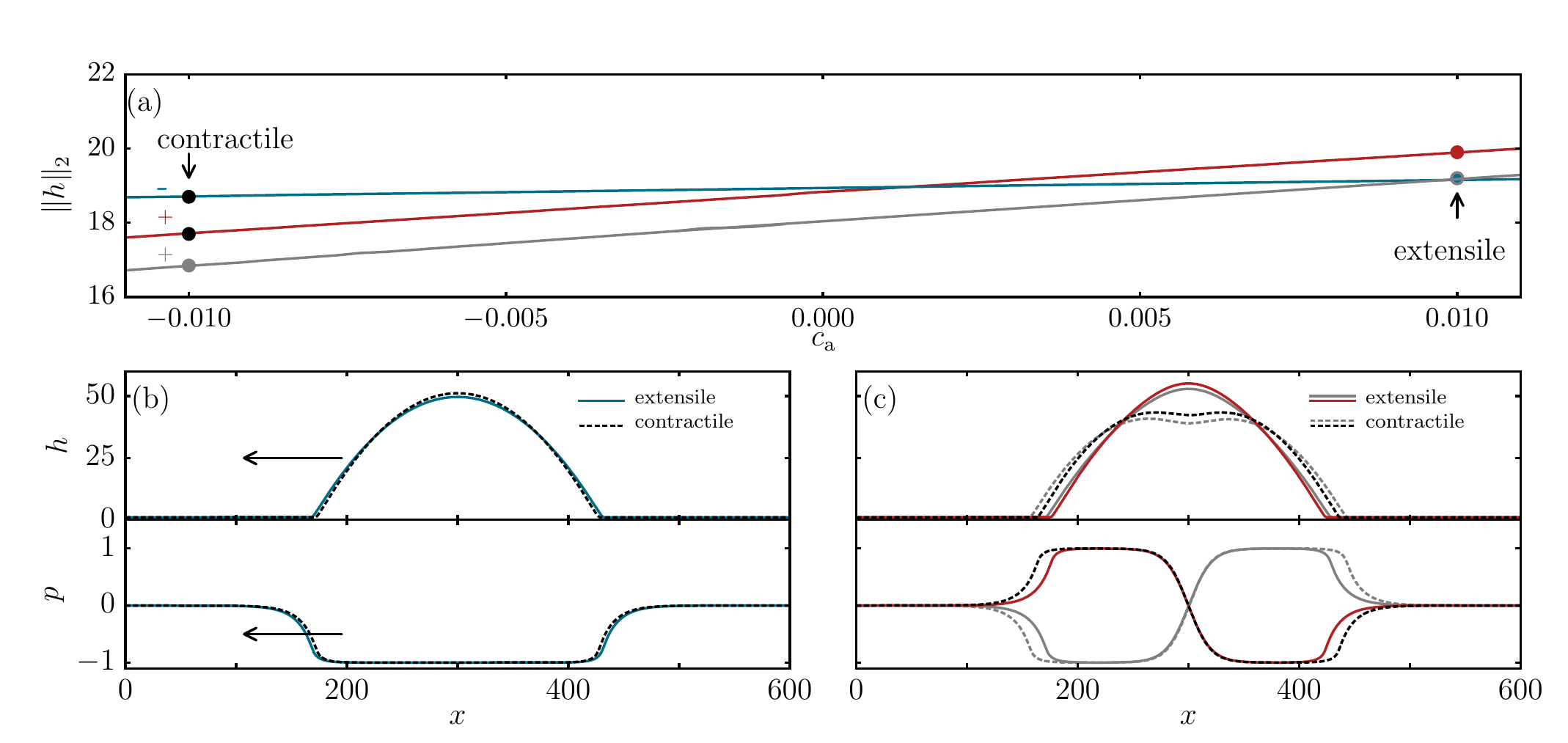}  
\caption{Dependence of the droplet shape and polarization profiles on the active stress parameter $c_\mathrm{a}$ for uniformly and non-uniformly polarized self-propelled ($\alpha_0=0.002$) droplets. 
(a) Bifurcation diagram showing the $L^2$-norm of $h$ for resting unstable (red and gray) and moving stable (blue) active droplets. The red [gray] branch corresponds to inward [outward]-pointing polarized droplets.
(b) Uniformly polarized moving (indicated by arrow) droplets for extensile (blue solid) and contractile (black dashed) active stress, corresponding to the stable states on the blue solid branch in (a) indicated by the respectively colored filled circles.
(c) Non-uniformly inward (red and black) and outward (gray) pointing polarized droplets for extensile (solid) and contractile (dashed) active stresses. The droplets correspond to the respectively colored filled circles on the red and gray solution branch in (a).
Due to symmetry breaking caused by self-propulsion, the inward (red and black) and outward (gray) pointing solutions show different behavior when varying the active stress.
Remaining parameters are as in Fig.~\ref{Fig:Simulation_active}.}
\label{Fig:bif_alpha_ca}
\end{center}
\end{figure*}
\begin{figure}[htbp]
\begin{center}
\includegraphics[width=0.5\textwidth]{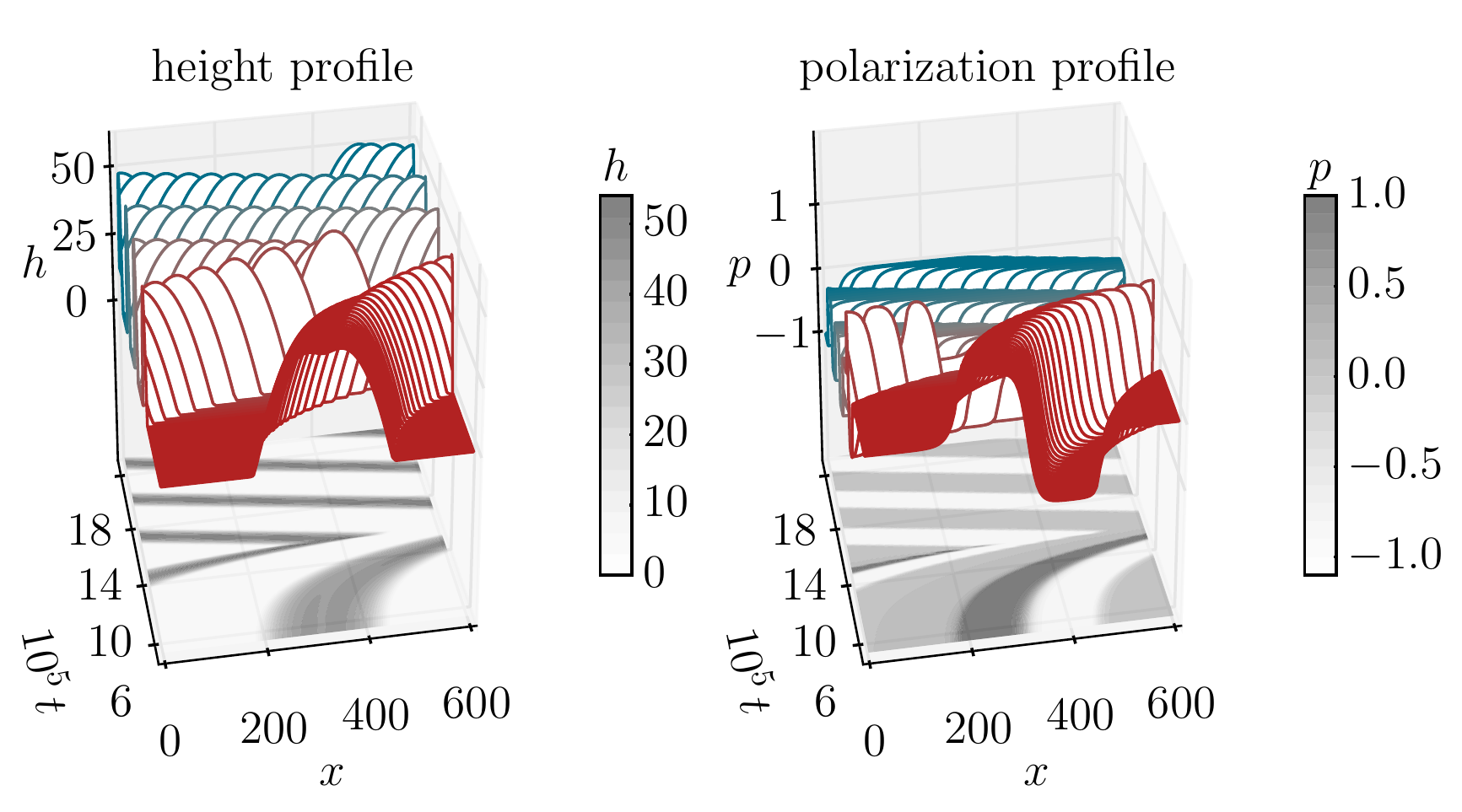}
\caption{Long time simulation for an initially non-uniformly polarized droplet with active contractile stresses and self-propulsion, i.e.,~$c_\mathrm{a}=-0.01$ and  $\alpha_{0}=0.002$. 
Shown is the height profile (left) and the polarization profile (right).
Initially, the droplet contains one domain wall between domains of inward pointing polarization and a dip in the height profile rapidly develops. 
At time $t\approx 10^6$ the drop transforms from the non-uniformly polarized to the uniformly polarized state (polarization into negative x-direction, i.e., $p=-1$). 
This transition is accompanied by a fast transient droplet motion into the positive x-direction. 
Eventually, the droplet starts to move to the left, consistent with its net polarization direction. 
The remaining parameters are as in Fig.~\ref{Fig:bif_alpha_ca}.}
\label{Fig:waterfall-appendix-alpha0-ca}
\end{center}
\end{figure}
Overall, due to mass conservation, the droplet becomes lower and wider as sketched in Fig.~\ref{Sketch:active-stress}~(b). Regarding the stability, we find that the uniformly polarized droplets are always stable in contrast to the non-uniformly polarized droplets, which are always unstable. The active stress does not influence stability in the given parameter range.
Direct numerical simulations show that in the long-time limit ($t\approx 10^6$), the polarization field for the unstable state transforms into a uniform one, see Fig.~\ref{Fig:waterfall-ca}. Interestingly, during the transient phase, the droplets spontaneously move even though there is no self-propulsion. In this transient, the droplets can cover distances corresponding to multiples of their own size. In the examples shown in Fig.~\ref{Fig:waterfall-ca} the droplet with extensile active stress [Fig.~\ref{Fig:waterfall-ca}~(a)] moves about seven times its own width while the droplet with contractile stress [Fig.~\ref{Fig:waterfall-ca}~(b)] covers three to four times its own width.
In any case, as soon as the polarization profile becomes uniform the droplets stop. Note, that the transition from an unstable non-uniformly polarized into a stable uniformly polarized state takes more time for extensile active stresses than for contractile stresses.
The question remains, what triggers the extensive transient droplet motion. During the transition from non-uniform to uniform polarization, the droplet undergoes a parity symmetry-breaking: One of the two polarization domains grows, i.e, the domain wall moves away from the droplet center. Because of the broken symmetry, active stresses induce a net fluid flux across the domain wall. 
Due to mass conservation this net flux results in a motion of the droplet. It is accompanied  by an increase [decrease] in the contact angle at the droplet edge in the direction of the fluid flux [opposite to it]. The motion of the domain wall within the droplet and the motion of the droplet itself continue until the polarization is uniform throughout the droplet and parity-symmetry is restored.
We illustrate this phenomenon in Fig.~\ref{Sketch:active-stress-transient}~(a) and (b) for extensile and contractile stress, respectively. For extensile stress the fluid in both domains is attracted towards the domain wall, analogously to Fig.~\ref{Sketch:active-stress}~(a). 
However, as the wall moves off center, due to the local slope of the drop surface, the net fluid flux around the wall is in the direction of the wall's motion. Mass conservation implies that the entire drop moves into the same direction.
For contractile stress, the fluid in both domains is pushed away from the domain wall resulting in a dip in the height profile at the domain wall, analogously to Fig.~\ref{Sketch:active-stress}~(b). 
The net fluid flux around the domain wall is in the direction opposite to the motion of the wall in the frame moving with the droplet. Therefore, in the laboratory frame the droplet moves into the same direction as the net fluid flux.
Thus, in both cases the interplay between droplet shape and the motion of a domain wall in polarization drives a transient motion of the droplet. Interestingly, the origin of motion 
lies in the relaxation of the polarization field which ultimately eliminates domain walls and establishes a uniform polarization. 
The nature of the active stress, contractile vs.\ extensile, determines the direction of the transient droplet motion, relative to the domain wall motion within a comoving frame. 
In the laboratory frame, for extensile active stress, the domain wall moves faster than the droplet, whereas for contractile active stress, the domain wall moves slower than the droplet itself.\\
In a final step we analyze the steady states (stationary in the lab frame or stationary in the co-moving frame) of active droplets in the presence of self-propulsion (sensitive to polar order) and active stresses (sensitive to nematic order).
To this end we use again parameter continuation: Starting from the self-propelled solutions marked in Fig.~\ref{Fig:bif_alpha}~(a) by the filled circles we increase the active stress and obtain the bifurcation diagram shown in Fig.~\ref{Fig:bif_alpha_ca}~(a) for moving stable (blue line) and resting unstable (red and gray lines) active droplets. 
For uniformly polarized droplets moving with constant shape and velocity, the addition of active stresses has only minor effects on drop shape and velocity.  For non-uniformly polarized droplets containing one domain wall, the picture is more differentiated. Self-propulsion breaks their symmetry, as it locally stretches the droplet 
with outward pointing polarization [gray profile in Fig.~\ref{Fig:bif_alpha_ca}~(c)] whereas it contracts drops with inward pointing polarization [red and black profiles in Fig.~\ref{Fig:bif_alpha_ca}~(c)]. Therefore, active stress has a different impact in the inward and outward pointing cases.
Stability does not change, namely, non-uniformly [uniformly] polarized states are still unstable [stable] in the considered parameter range.
Fig.~\ref{Fig:waterfall-appendix-alpha0-ca} shows a direct time simulation with parameters indicated by the black filled circle on the red branch in Fig.~\ref{Fig:bif_alpha_ca}~(a).
Initially, the droplet contains one central domain wall between domains with inward pointing polarization. 
The active stress is contractile, i.e., the initial drop profile contains a small dip at the center where the domain wall is located.
The simulation shown in Fig.~\ref{Fig:waterfall-appendix-alpha0-ca} demonstrates that the unstable states are long-time transients as a steadily moving droplet arises at $t\approx 10^6$.
The transition occurs via the growth of the domain of negative polarization, i.e., the domain wall moves to the left within the droplet. 
When, ultimately, the droplet is fully negatively polarized it moves steadily to the left. However, during the transition the droplet moves to the right, because the active stress causes local fluid flows close to the off center domain wall that push the droplet into the direction opposite to the relative motion of the wall, as illustrated in Fig.~\ref{Sketch:active-stress-transient}~(b). The direction of motion reverses when self-propulsion dominates. The transition occurs on the same time scale as for droplets without self-propulsion (cf.~Fig.~\ref{Fig:waterfall-ca}), i.e., $\alpha_0=0$.  Additional time simulations with different initial conditions for the polarization field for extensile and contractile active stress at otherwise identical parameters are provided in Appendix \ref{sec:append-act-drop}.

\subsection{Strong activity}
\label{sec:strong-activity}
\begin{figure*}
\begin{center}
\includegraphics[width=1.0\textwidth]{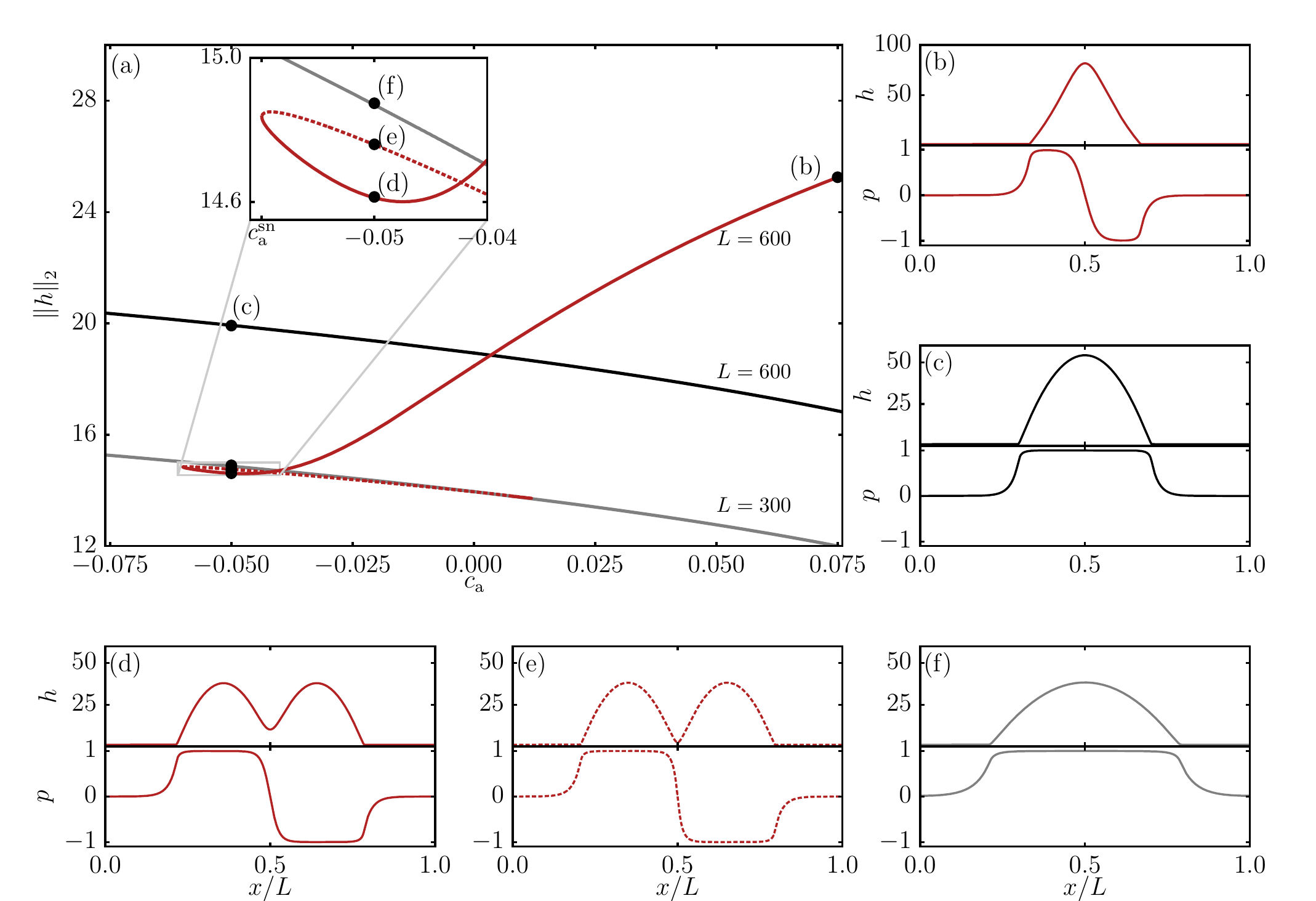}
\caption{Strong active stresses and droplet splitting. (a) Bifurcation diagram showing the L$^{2}$-norm of the film height $h$ depending on the active stress $c_\mathrm{a}$. All branches correspond to the same mean value $h\lowt{m}$. The black and red branch correspond to droplets with one and two polarization domains, respectively. For comparison, the gray branch corresponds to a uniformly polarized droplet for $L=300$, i.e., half the size of the system corresponding to the black and red branch with $L=600$.  At high contractile stresses ($c_\mathrm{a}<0$) a droplet with two polarization domains splits into two smaller droplet, each uniformly polarized with opposite polarizations. The respective red dashed branch coincides with the gray solid branch for $c\lowt{a}>0$, i.e., the corresponding solutions are compositions of two single uniformly polarized droplets on each half of the domain.
The inset shows a saddle-node bifurcation at $c\lowt{a}=c\lowt{a}^\mathrm{sn}$, where the drop starts to split (see profiles in (d) and (e)). We refrain from indicating stability. Note, that the crossing of branches does not imply the solutions to be equal, but to correspond to the same $\lVert h\rVert_{2}$-norm. Panels (b)-(f) show the different profiles in $h$ and $p$ corresponding to the parameters indicated by the respectively labeled black circles in (a). The remaining parameters are as in Fig.~\ref{Fig:bif_ca}.}
\label{Fig:strong-ca}
\end{center}
\end{figure*}
Up to here we have focused on a parameter range of relatively weak activity that is justified by estimates of activity in specific biophysical systems (cf.~appendix~\ref{sec:param-estim}).
However, as it is also of interest how the droplet behavior changes at strong activity and corresponding parameter ranges are considered in the literature \cite{LoEL2019prl,AuET2020sm}, we next explore this case.
\begin{figure}[htbp]
\begin{center}
\includegraphics[width=0.5\textwidth]{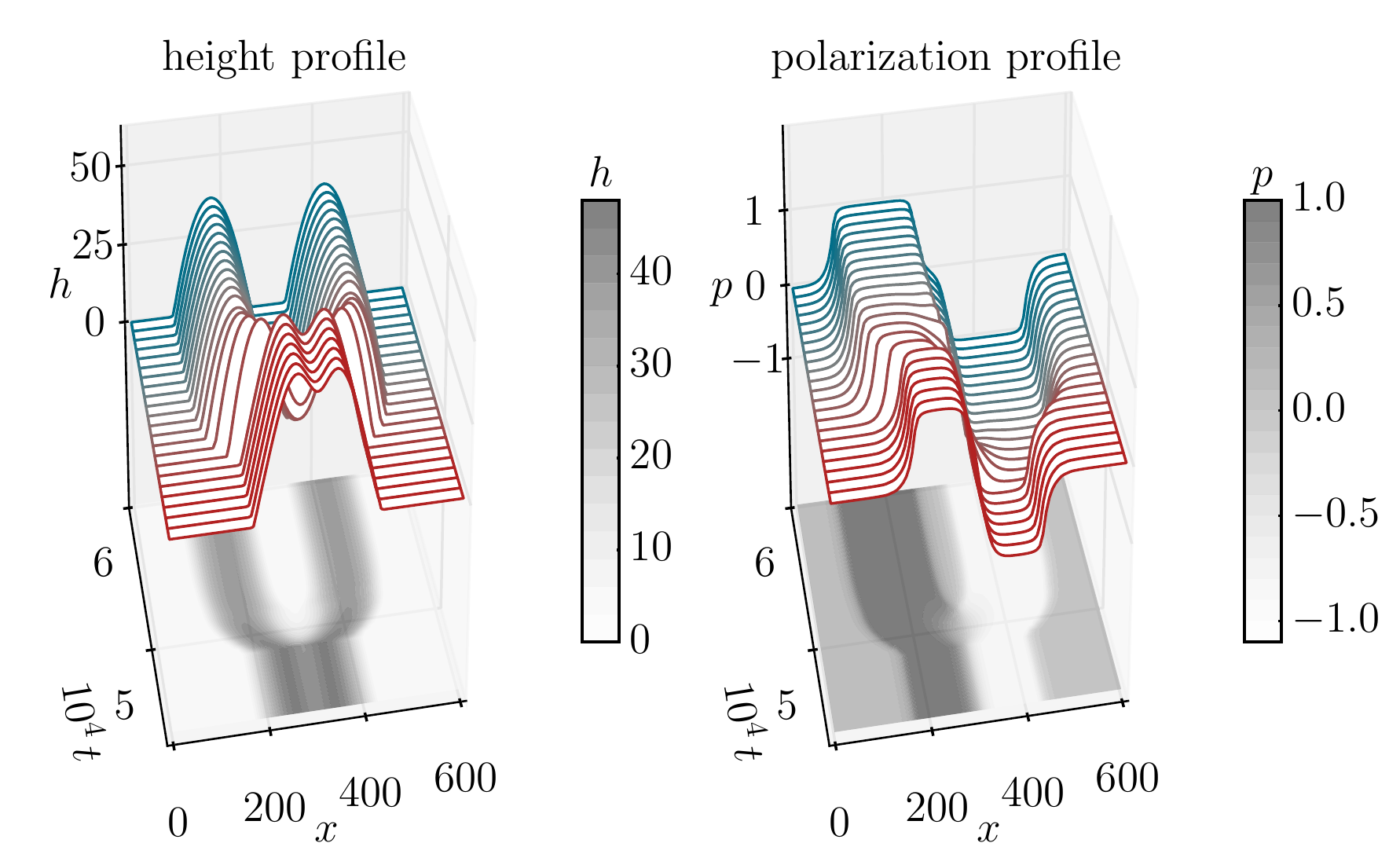}
\caption{Time simulation for an initially non-uniformly polarized droplet (state in Fig.~\ref{Fig:bif_csp2}~(c)) at active contractile stress $c_\mathrm{a}=-0.05$. The drop splits and evolves into a two-droplet steady state. Note, that this state is not equal to the one shown in Fig.~\ref{Fig:strong-ca}~(e) as the two droplets do not touch each other. Shown are (left) the height profile and (right) the polarization profile. The remaining parameters are as in Fig.~\ref{Fig:strong-ca}.}
\label{Fig:strong-alpha-time-sim}
\end{center}
\end{figure}
In Fig.~\ref{Fig:strong-ca}~(a) we show as red line the bifurcation diagram for strong active stresses for droplets with one defect in the polarization profile \footnote{
 Here, the red branch is obtained employing the continuation package \texttt{auto07p}, as we were not able to perform the continuation for $c_\mathrm{a}>0.02$ using \texttt{pde2path} due to finite size effects as we consider large system sizes to obtain large drop heights as compared to the precursor film. The results obtained for $c_\mathrm{a}<0.02$ with \texttt{pde2path} agree with the ones obtained with \texttt{auto07p}.}.
For large extensile active stress ($c_\mathrm{a}>0$) we observe a  drastic change in the droplet shape as it becomes more pointed and nearly doubles its maximum height, see the profile in Fig.~\ref{Fig:strong-ca}~(b). Direct time simulations show that this state survives for a long time (not shown). However, as this does not necessarily imply overall stability we refrain from indicating stabilities in the Fig.~\ref{Fig:strong-ca}~(a).
\begin{figure*}[htbp]
\begin{center}
\includegraphics[width=1.0\textwidth]{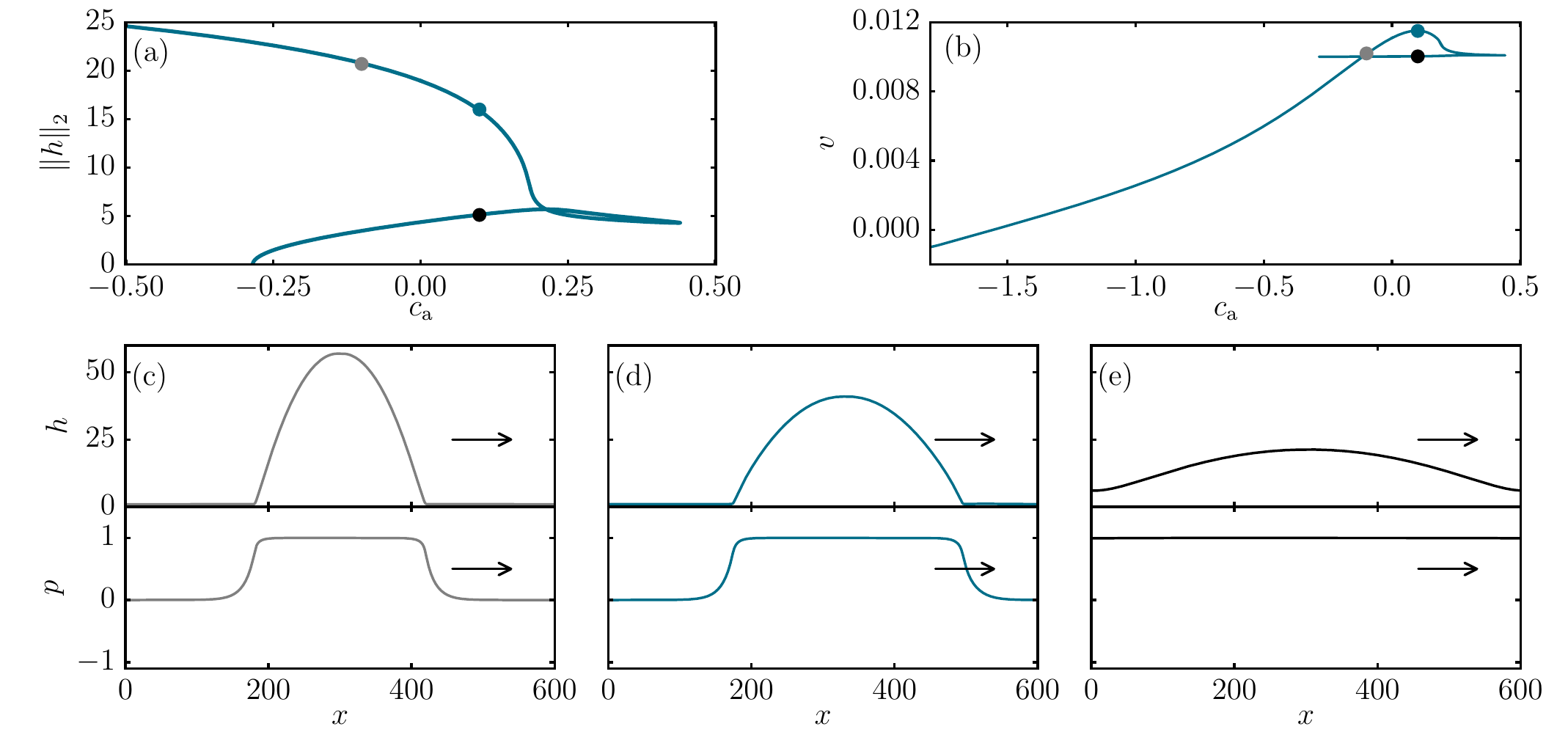} 
\caption{(a) Bifurcation diagram showing the $L^{2}$-norm of $h$ for a uniformly polarized droplet with strong self-propulsion $\alpha_{0}=0.01$. Panel (b) depicts the velocity $v$ depending on $c\lowt{a}$. For strong extensile stresses, we find a fold bifurcation to the polarized moving film solution, whereas the velocity monotonically decreases for  contractile stresses, where it may even become negative (see main text for details). The panels (c)-(e) depict solution profiles indicated by the respectively colored circles in (a) and (b). The remaining parameters are as in Fig.~\ref{Fig:bif_alpha_ca}~(a).}
\label{fig:bif-strong-ca-alpha}
\end{center}
\end{figure*}
For large contractile stresses ($c_\mathrm{a}<0$) we observe an interesting topological change in the drop shape as we follow the curve across the saddle-node bifurcation at $c_\mathrm{a} =c_\mathrm{a}^\mathrm{sn}$ focused on in the inset of Fig.~\ref{Fig:strong-ca}~(a). 
\begin{figure}[htbp]
\begin{center}
\includegraphics[width=0.5\textwidth]{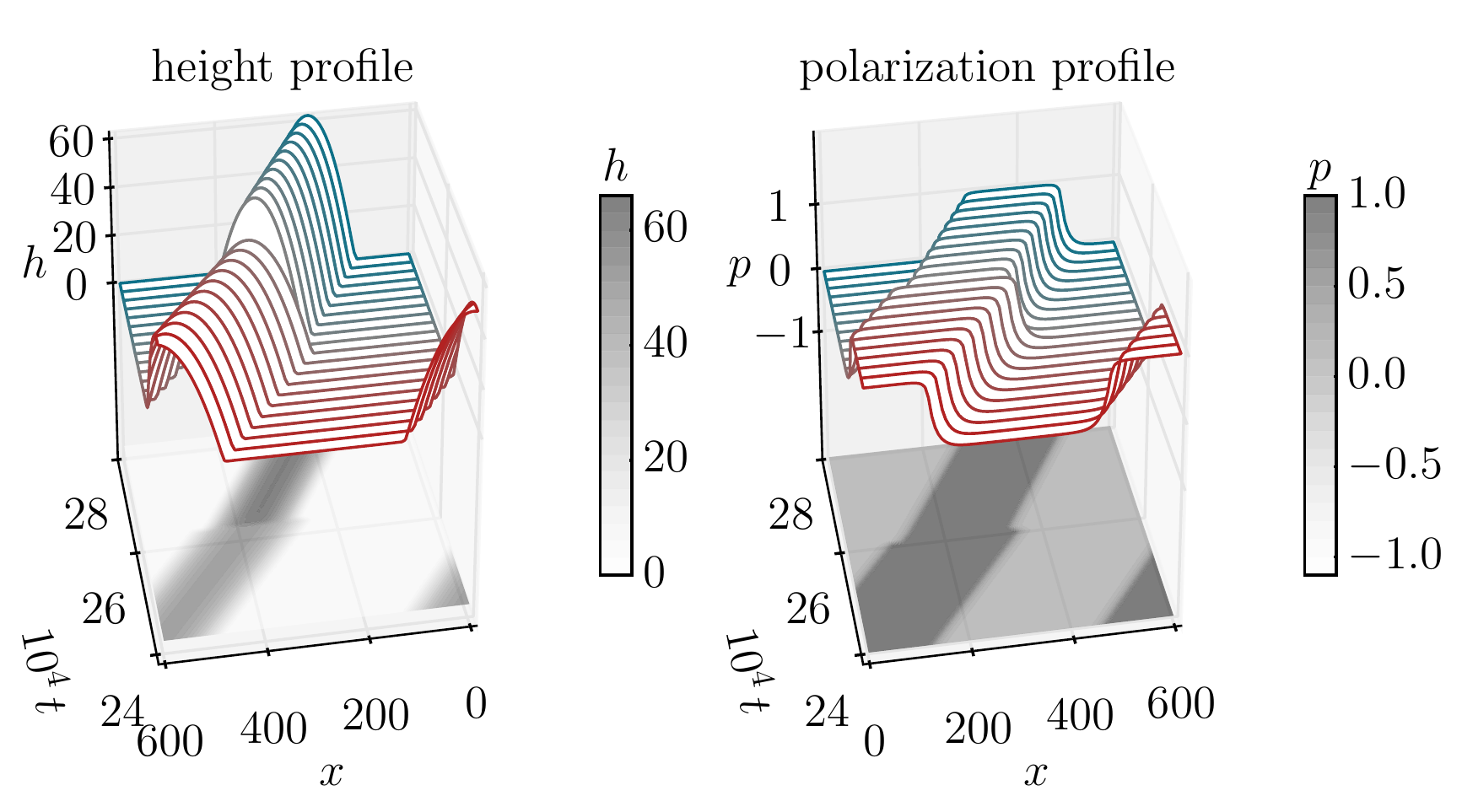} 
\caption{Time simulation for an initially uniformly polarized droplet with self-propulsion strength $\alpha_{0}=0.01$ and without active stress ($c\lowt{a}=0$). At $t=26\times10^{4}$ an active contractile stress $c\lowt{a}=-0.25$ is switched on inducing the droplet to contract and to move slower.
The remaining parameters are as in Fig.~\ref{Fig:strong-ca}.}
\label{Fig:strong-ca-alpha-vel}
\end{center}
\end{figure}
As one approaches the bifurcation on the solid red line the small central depression in the drop profile (described already at Fig.~\ref{Fig:bif_ca}~(c)) deepens (Fig.~\ref{Fig:strong-ca}~(d)) and then beyond the bifurcation the structure starts to resemble two drops of opposite uniform polarization (Fig.~\ref{Fig:strong-ca}~(e)). Following the red dashed branch towards larger $c_\mathrm{a}$ the distance between the two static split drops slightly increases and the drops become wider and lower. Moreover, the bifurcation diagram in Fig.~\ref{Fig:strong-ca}~(a) depicts the L$^{2}$-norm for uniformly polarized single droplets for system sizes $L=600$ (black) and $L=300$ (gray), respectively. For $c_\mathrm{a} > 0$ the red dashed branch closely matches the $L=300$ branch of single, uniformly polarized drops [gray solid line, example profile in Fig.~\ref{Fig:strong-ca}~(f)]. This reflects the fact that the $L=600$ split solution resembles a composition of two $L=300$ drops of respectively uniform but opposite polarization.
For $c\lowt{a}<0$ the distance between the two drops decreases, hence the corresponding L$^2$-norm slightly deviates from the L$^{2}$-norm representing the single droplet for $L=300$. Performing a direct time simulation at $c_\mathrm{a}^\mathrm{sn} < c_\mathrm{a}=-0.05$ starting with a non-uniformly inward pointing polarized single drop [Fig.~\ref{Fig:bif_csp2}~(c)] it spontaneously splits and develops into a steady state consisting of two completely separated droplets, as shown in Fig.~\ref{Fig:strong-alpha-time-sim}. However, this solution does not correspond to the state given in Fig.~\ref{Fig:strong-ca}~(e). Instead, the two droplets are much further apart and the state resembles a composition of two of the uniformly polarized states shown in Fig.~\ref{Fig:strong-ca}~(f) (again of opposite polarization). In consequence, in the course of the time evolution depicted in Fig.~\ref{Fig:strong-alpha-time-sim}, the norm approaches the value on the $L=300$ branch\footnote{As the interaction of the droplets exponentially weakens with increasing drop distance, it depends on numerical details at which exact distance the droplets stop.} in Fig.~\ref{Fig:strong-ca}~(a).
Finally, we investigate how strong active stresses affect self-propelling uniformly polarized droplets using the moving drop at self-propulsion strength $\alpha_{0}=0.01$ as reference state. The resulting bifurcation diagram\footnote{Again obtained employing \texttt{auto07p}, after having confirmed that for $c\lowt{a}<0.1$ all results are in good agreement with those obtained with \texttt{pde2path}.} in dependence of $c_\mathrm{a}$ in terms of the L$^{2}$-norm of $h$ and the drop velocity $v$ is shown in Fig.~\ref{fig:bif-strong-ca-alpha}~(a) and (b), respectively. Large active stresses have a strong impact on the shape and speed of the drop: Extensile stresses tend to spread the droplet out [Fig.~\ref{fig:bif-strong-ca-alpha}~(d)] until its edges reach the domain boundaries. Then a saddle-node bifurcation occurs (at $c\lowt{a}\approx0.45$) that connects the droplet states to a modulated film, i.e., a state of traveling waves [Fig.~\ref{fig:bif-strong-ca-alpha}~(e)]. Following this branch back towards smaller $c\lowt{a}$, the wave amplitude decreases (at approximately constant speed) before it ends at $c\lowt{a}\approx-0.2$ in a Hopf-bifurcation of the flat film state (there $\|h\|_2=0$). This transition is accompanied by a non-monotonous change in the velocity: First, extensile stresses make the droplet faster before the speed decreases again. The surface waves have a speed of $v\approx\alpha_0$.  
For large contractile active stresses, the drop contracts [Fig.~\ref{fig:bif-strong-ca-alpha}~(c)] and its velocity decreases. This is illustrated in the time simulation shown in Fig.~\ref{Fig:strong-ca-alpha-vel}. There, a droplet is initiated at $\alpha_{0}=0.01$ in the absence of active stress. It moves with speed $v\approx\alpha_0$. At $t=26\times10^{4}$ a contractile active stress $c\lowt{a}=-0.25$ is switched on, which results in a fast contraction (increase in the maximal height) and a marked slow down of the droplet.
For very strong contractile stresses, the velocity even becomes negative, i.e., the droplet changes its direction of motion. However, then the precursor film starts to polarize. As this is unphysical we stop the continuation there and do not further pursue the case of very large contractile active stresses.

\section{Summary and outlook}
\label{sec:conc}
We have presented a generic phenomenological model for free-surface thin films and shallow droplets of an active polar liquid on solid substrates. It couples evolution equations for the film height profile of the liquid and the local height-integrated polarization. The model consists of a passive part that forms a gradient dynamics on an underlying free energy functional and an active part that represents self-propulsion and active stresses. Here, the energy incorporates simple forms of capillarity, wettability, spontaneous polarization, elastic energy of the polarization field and a coupling between the polarization and free-surface shape.  We have shown that the gradient dynamics form can be translated into the usual hydrodynamic form of a thin-film model where the pressure-gradient driven liquid flux is determined by Laplace and Derjaguin (disjoining) pressure and elastic stress while polarization is transported by the same flux and additionally undergoes non-Fickean rotational and translational diffusion.  
Although the model has not been derived via a long-wave approximation from 3D bulk equations and appropriate boundary conditions, it has a number of features that to our knowledge no thin-film model in the literature combines: (i) it is a fully dynamical model where height profile \textit{and} polarization field can freely develop; (ii) it fully accounts for wettability and capillarity, allows for the motion of three-phase contact lines, and dynamic contact angles; (iii) it accounts for simple mechanisms of coupling between height and polarization; and (iv) active stress and self-propulsion are both included. In the future, the model can be extended and adapted in a number of ways. So it is straight forward to incorporate more complicated energies and energetic couplings as this does not change the general form of the equations (see, for example, the pertinent discussion for a surfactant-covered thin liquid film in Ref.~\cite{TAP2012pf}). Also the active terms may be easily adapted, e.g., incorporating the active stress term of Ref.~\cite{KA2015pre}. Ideally, it would be possible to derive a closed model in the form of two coupled partial differential equations, like the one presented here, via a long-wave approximation for films of active liquids as undertaken in \cite{KMW2018potrsa}. There, however, it was not possible to obtain such a closed form.  
After presenting our full thin-film model for 3D droplets, i.e., on 2D substrates, we have reduced the model to the description of 2D droplets on 1D substrates (i.e., transversally invariant liquid ridges), to allow for a first model analysis. Our study of this 1D geometry has mainly focused on basic phenomena: We have shown that the dewetting dynamics of a flat film of polar liquid is not solely determined by passive wetting forces, but also by the polarization field and activity.
In addition to the dewetting dynamics, the model is able to describe moving and resting drops of active liquids, with uniform and non-uniform polarization profiles.  
A parameter continuation has identified non-uniformly polarized solutions as linearly unstable. However, depending on the initial conditions they appear as long-lived transient states on the pathway to uniformly polarized droplets. This occurs in both, passive and active systems. During the transition phase, droplets start to move due to an interesting interplay of mass conservation and active stress.
We have also briefly explored the behavior of droplets over a larger range of active stresses. We have shown that strong contractile active stresses may result in drop splitting similar to morphological changes observed in 2D simulations of active nematics in \cite{WhHa2016njp}, and have provided a first insight into the underlying bifurcation structure. Moreover, we have shown that strong contractile active stresses slow down the motion of self-propelled drops. Strong extensile stresses also result in modulation of drop speeds and, ultimately, finite size effects result in a saddle-node bifurcation that connects the polarized moving drop state with traveling surface waves similar to the waves in thin films of living fluids described in Ref.~\cite{SR2009prl}.\\
Finally, we highlight specific features of the present model in comparison to literature models describing free surface droplets of active liquids on smooth solid substrates: The phase-field description of Ref.~\cite{TTMC2015nc} is a generalization of model-H \cite{AnMW1998arfm} for active liquids where the active driving is restricted to a finite thickness layer close to the substrate (in their quasi-2D simulations further restricted to the advancing part of the active drop). Otherwise, they consider similar physical ingredients in a full 2D/3D setting and provide full simulation results. In contrast, here we employ a thin-film description of reduced dimension to provide such simulation results and to track relevant states in parameter space to determine bifurcation diagrams. Our model is a precursor film model that naturally incorporates a wetting energy while in \cite{TTMC2015nc} seemingly a ninety degree equilibrium contact angle of the active phase with the substrate is imposed via a boundary condition. In our case, self-propulsion is not confined to a finite layer. In the case of a (quasi-)1D substrate, both models yield motile droplets with velocities close to the self-propulsion strength that are amended by additional active stresses. As expected, in the considered 1D case we have not observed non-transient droplet motion due to active stresses only, but could describe drop splitting not discussed in \cite{TTMC2015nc}. A detailed analysis of the presented model for a full 3D geometry, i.e., for 2D substrates, as done for the active model-H in \cite{TTMC2015nc} represents an important future challenge. Additionally, next to effects described here, one could then expect a spontaneous symmetry breaking to occur that results in a splay-induced motility in the presence of active stress (but without self-propulsion) as observed in \cite{TMC2012potnaos,WMV+2014tepje,MWP2015jotrsi,TTMC2015nc,WhHa2016njp}.\\
Note that in our model one may as in \cite{TTMC2015nc} restrict active driving to a finite thickness layer close to the substrate, e.g., using a Hill function with exponent two for the height dependence in the self-propulsion term in Eq.~(\ref{eq:fluxes}). Then one can observe steadily moving drops with a forward protrusion (not shown) as observed in layers of epithelial cells \cite{PAB+2019np}. The dynamical transition between moving drops with and without such a protrusion can have a continuous or discontinuous character as also seen in \cite{TTMC2015nc} in dependence of the employed slip strength. A thin-film model derived in the recent \cite{AuET2020sm} obtains a self-propulsion term that exponentially decays with increasing film height  which also results in a forward protrusion. In the thin-film active droplet model proposed in \cite{JoRa2012jfm}, nematic elasticity and active stresses are incorporated. It is used to determine steady fully spread shapes with zero microscopic contact angle as well as the scaling law of spreading dominated by active stresses alone. A microscopic contact angle different from zero can only be imposed in the case without elasticity and is independent of the active stress that only influences the drop profile away from the substrate. In the given version, contact lines would only be able to move if slip were incorporated.  In contrast, our thin-film model with a wetting energy accounting for partial wettability directly allows for fully dynamic considerations of, e.g., film dewetting, drop spreading, activity-driven surface waves. We have included elasticity resulting from polarization gradients parallel to the substrate, not considered in \cite{JoRa2012jfm}, but do not incorporate vertical contributions, i.e., consider a regime of very weak anchoring at the free surface.
A very recent thin-film model for a droplet of active nematics \cite{LoEL2019prl} (further investigated beside other models in \cite{AuET2020sm}) considers the case of strong anchoring parallel at the substrate and perpendicular at the free surface.
Assuming a strong elastic limit, the relaxation of the director field is very fast, adiabatically enslaving it to the film height profile. In consequence, a single thin-film evolution equation is derived where active stresses result in a directed driving term similar to the one obtained for a constant imposed shear stress, e.g., a Marangoni stress due to an imposed temperature gradient. Self-propulsion is not considered. Note that the considered director profile strongly differs from the one resulting for the strong parallel anchoring considered in \cite{JoRa2012jfm} or the very weak anchoring considered here. The calculations in the 1D case in \cite{LoEL2019prl} show that the drops move for any active stress and show shape transformations from spherical cap-like drops via drops with a backwards shoulder to long flat drops with a capillary ridge at the front very similar to transitions described for 1D drops driven by a body force \cite{TNBP2002csa}. Similar strong shape transformations related to the emergence of a backward protrusion can be obtained with our model if the self-propulsion term in Eq.~(\ref{eq:fluxes}) is chosen to depend quadratically on film height (not shown). It is a task for the future to develop an active liquid thin-film model that is able to capture all three anchoring modes individually considered in \cite{JoRa2012jfm}, \cite{LoEL2019prl} and the present work.
\appendix
\section{Initial conditions for droplet simulations}
\label{ap:init:cond}
For all direct numerical simulations of passive and active single droplets we use the initial conditions
\begin{align}\label{eq:initial}
 h(x) &= \text{max} \Big( h\lowt{max} -a (x-\tfrac{L_x}{2})^2 , 1\Big) \notag \\
 P(x) &= 0.01 \, \text{rand}(N_x) \,  \tfrac{h(x)-1}{h\lowt{max}} \, \text{Sym}(x) \notag\\
  \text{with} &\quad h\lowt{max} =50\quad L_x=600 \quad \notag \\
  \text{and}& \quad a = \tfrac{3}{20} \tfrac{A}{h\lowt{max}-1.}
\end{align}
where $\text{rand}(N_x)$ corresponds to a 1D array of random float numbers from the half-open interval $[0.0,1.0)$. The function $\text{Sym}(x)$ can be used to impose a slight asymmetry with respect to parity ($x\to-x$). 
Specifically, we use $\text{Sym}(x)=1$ to induce droplets with uniform polarization, $\text{Sym}(x)=\sin{(2\pi{x\over L_x})}$ to induce drops
with non-uniform inward polarization, and  $\text{Sym}(x)=\sin{(-2\pi{x\over L_x})}$ for non-uniform outward polarization. This corresponds to the scenarios shown in Fig.~\ref{Fig:Simulation_active}~(a-c).
\section{Further time simulation for active droplets}
\label{sec:append-act-drop}
\begin{figure}[!h]
\begin{minipage}[t]{1.0\hsize}
(a) extensile stress\\
\includegraphics[width=1.0\textwidth]{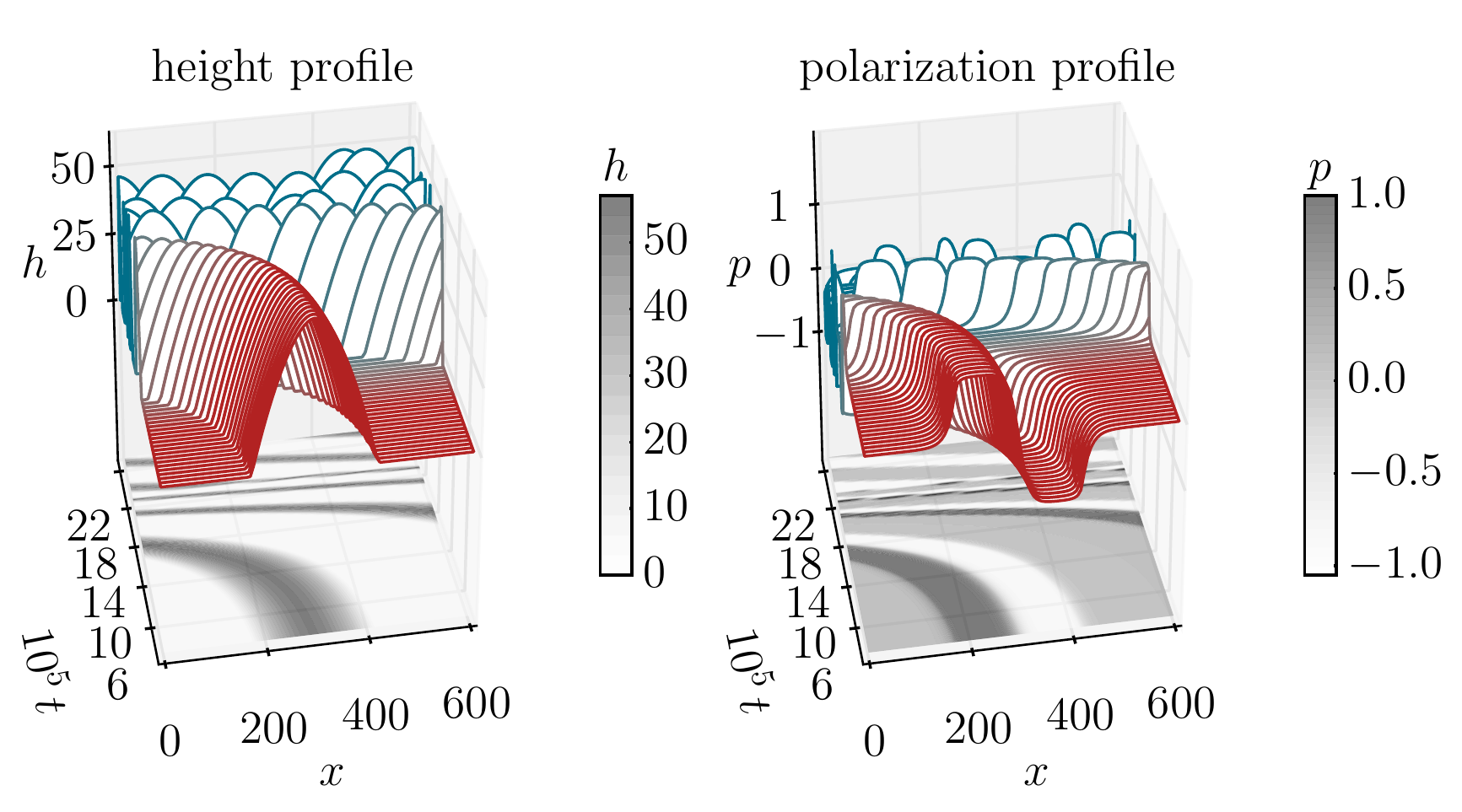} \\
\end{minipage}\hfill
\begin{minipage}[t]{1.0\hsize}
(b) extensile stress\\
\includegraphics[width=1.0\textwidth]{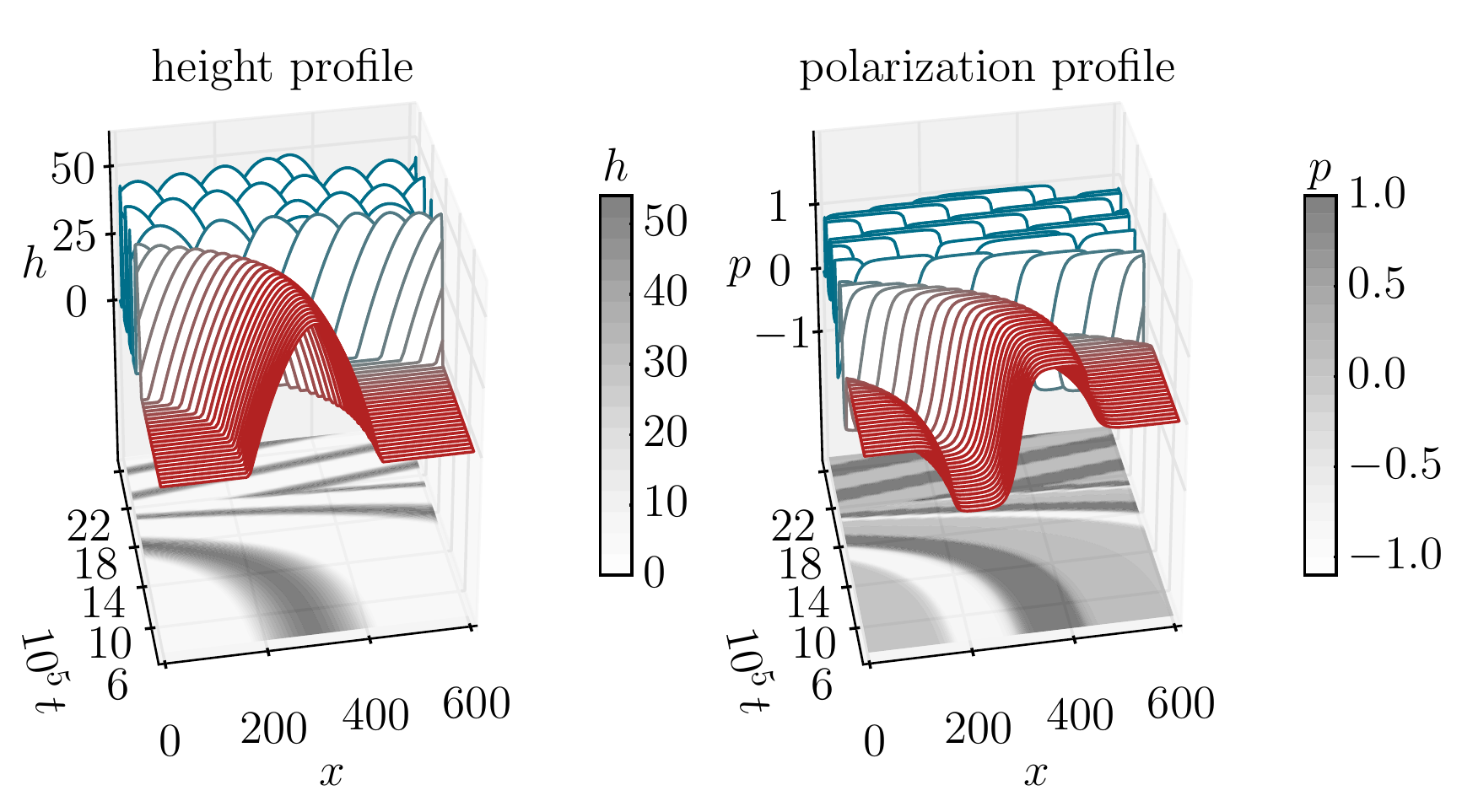} \\
\end{minipage}\hfill
\begin{minipage}[t]{1.0\hsize}
(c) contractile stress\\
\includegraphics[width=1.0\textwidth]{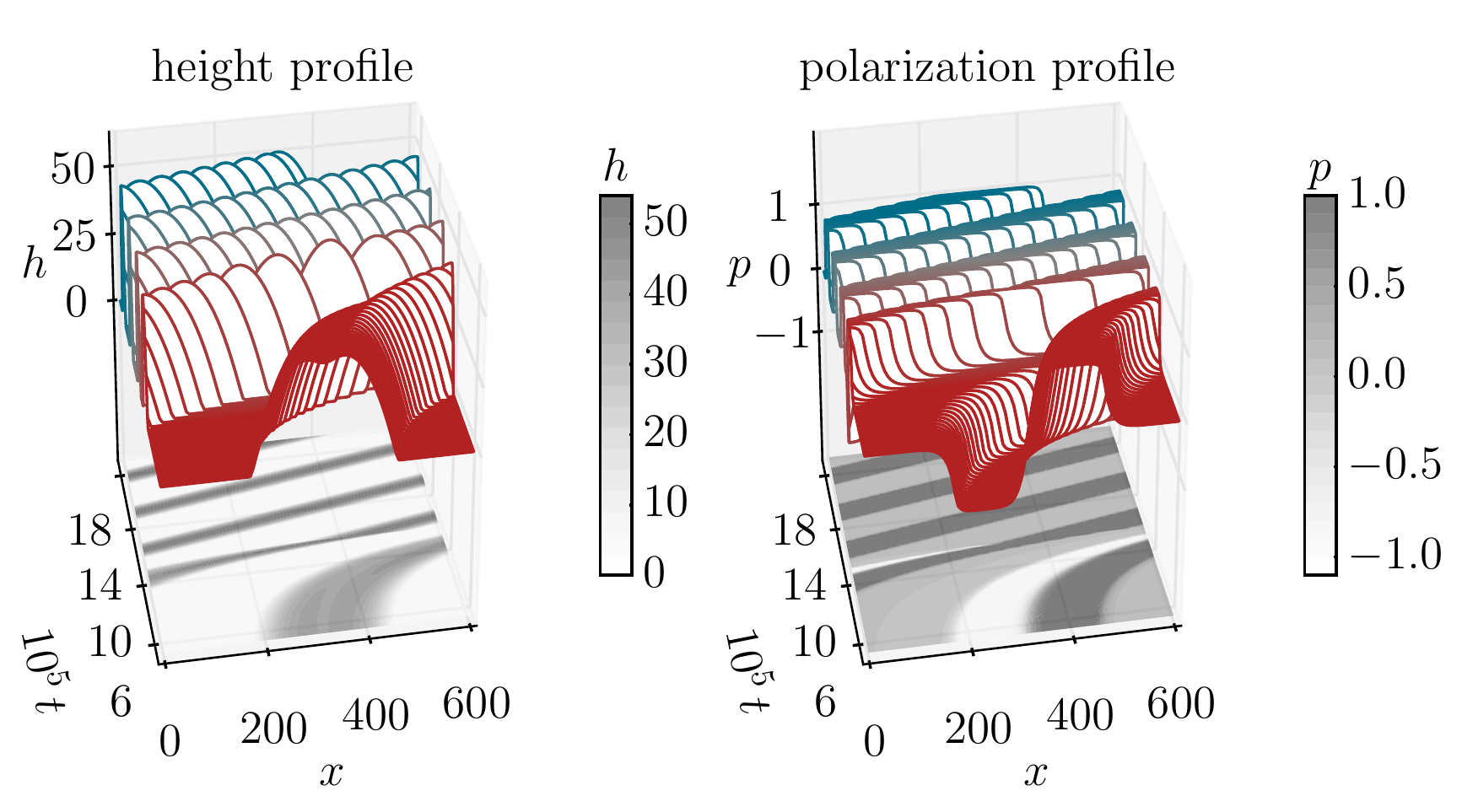} \\
\end{minipage}\hfill
\caption{Long time simulation for an initially non-uniformly polarized droplet with active (a,b) extensile ($ c_\mathrm{a}=0.01$) and (c) contractile ($ c_\mathrm{a}=-0.01$) stresses and self-propulsion ($\alpha_{0}=0.002$). 
Shown is (left) the height profile and (right) the polarization profile. The transient droplet motion is caused by the moving domain wall in the polarization which either 
moves into the same (extensile) or into the opposite (contractile) direction in the laboratory than in the comoving frame. (a)A drop with initially inward pointing polarization eventually evolves into a uniformly polarized droplet ($p=-1$) moving to the left. 
(b) and (c) show that drops with initially outward pointing polarization eventually evolve into uniformly polarized droplets ($p=1$) moving to the right. Note that during the transient they move into different directions.
Remaining parameters are as in Fig.~\ref{Fig:bif_ca}.}
\label{Fig:waterfall-appendix-ca-alpha}
\end{figure}
For completeness we show here time simulations for self-propelled ($\alpha_0=0.002$) non-uniformly polarized droplets under active stress [Fig.~\ref{Fig:waterfall-appendix-ca-alpha}]. 
The parameters are identical to the parameters used for the simulations in Fig.~\ref{Fig:waterfall-appendix-alpha0-ca}, except for the nature of the active stress, i.e., contractile vs extensile, or the initial polarization profile, i.e., inward vs. outward pointing polarization.

\section{Parameter Estimate}
\label{sec:param-estim}
For a water droplet with surface tension $\gamma=70\,$mN\,m$^{-1}$
  and viscosity $\eta=1\,$mPa\,s containing a high concentration of
  self-propelled particles (e.g., swimming bacteria or treadmilling
  filaments with a typical swimming/treadmilling speed of
  $\alpha_0 \sim$ 1-100\,$\mu$m\,s$^{-1}$) we find the dimensionless
  self-propulsion parameter
\begin{eqnarray}
\tilde\alpha_0 &=&\alpha_0 \eta \sqrt{\gamma}
\left({h\lowt{a}^2\over A}\right)^{3/2} \nonumber\\
&=&{\alpha_0\eta\over \gamma} {1\over
  \left[{3\over 10}(1-\cos{\theta_0})\right]^{3/2}}\,\nonumber
\end{eqnarray}
where $\theta_0$ denotes the equilibrium contact angle of the passive
unpolarized droplet, which is related to the wetting energy at the
precursor film thickness $f_w(h\lowt{a})$ via the
well-known relation $\cos{\theta_0}={\gamma+f_w(h\lowt{a})\over \gamma}$
(see, e.g., \cite{TST+2018l}). Using a small contact angle of
5\,$^\circ$ gives values for $\tilde a_0$ of about $10^{-5}\ldots
10^{-3}$, which is well below the range of $\tilde\alpha_0$ which we
have investigated (Fig.~\ref{Fig:bif_alpha}). For the non-dimensional active stress
parameter $\tilde c_\mathrm{a}$ we find
\begin{eqnarray}
\tilde c_\mathrm{a} & =& {c_\mathrm{a} h\lowt{a}^3\over A}\nonumber\\
 &=& {c_\mathrm{a} h\lowt{a} \over \gamma} {1\over
  \left[{3\over 10}(1-\cos{\theta_0})\right]^{1/2}}\,.\nonumber
\end{eqnarray} 
Assuming a precursor film thickness of $h\lowt{a}=$1\,nm and an active stress
comparable to the elastic modulus of Arp2/3 cross-linked actin
networks of $c_\mathrm{a}=$1\,kPa \cite{Chau2007nat} we find the
non-dimensional active stress of $\tilde c_\mathrm{a}=0.001$
which is also below our tested parameter range (see Figs.~\ref{Fig:bif_ca} and \ref{Fig:bif_alpha_ca}).

\section{Author contribution}
S.T., U.T., and K.J. developed the presented model. S.T. performed the simulations in section~\ref{sec:flat-film}. F.S. performed the simulations in section~\ref{sec:res-drop}. All authors together developed the interpretation and progression of modeling and wrote the manuscript.\\
\section{Data supplement}
The data that support the findings of this study are openly available in www.zenodo.org at https://doi.org/10.5281/zenodo.3813574 in Ref. \cite{TSJTzenodo2020}.

%\bibliography{Literature_abbr}

\begin{thebibliography}{73}%
\makeatletter
\providecommand \@ifxundefined [1]{%
 \@ifx{#1\undefined}
}%
\providecommand \@ifnum [1]{%
 \ifnum #1\expandafter \@firstoftwo
 \else \expandafter \@secondoftwo
 \fi
}%
\providecommand \@ifx [1]{%
 \ifx #1\expandafter \@firstoftwo
 \else \expandafter \@secondoftwo
 \fi
}%
\providecommand \natexlab [1]{#1}%
\providecommand \enquote  [1]{``#1''}%
\providecommand \bibnamefont  [1]{#1}%
\providecommand \bibfnamefont [1]{#1}%
\providecommand \citenamefont [1]{#1}%
\providecommand \href@noop [0]{\@secondoftwo}%
\providecommand \href [0]{\begingroup \@sanitize@url \@href}%
\providecommand \@href[1]{\@@startlink{#1}\@@href}%
\providecommand \@@href[1]{\endgroup#1\@@endlink}%
\providecommand \@sanitize@url [0]{\catcode `\\12\catcode `\$12\catcode
  `\&12\catcode `\#12\catcode `\^12\catcode `\_12\catcode `\%12\relax}%
\providecommand \@@startlink[1]{}%
\providecommand \@@endlink[0]{}%
\providecommand \url  [0]{\begingroup\@sanitize@url \@url }%
\providecommand \@url [1]{\endgroup\@href {#1}{\urlprefix }}%
\providecommand \urlprefix  [0]{URL }%
\providecommand \Eprint [0]{\href }%
\providecommand \doibase [0]{http://dx.doi.org/}%
\providecommand \selectlanguage [0]{\@gobble}%
\providecommand \bibinfo  [0]{\@secondoftwo}%
\providecommand \bibfield  [0]{\@secondoftwo}%
\providecommand \translation [1]{[#1]}%
\providecommand \BibitemOpen [0]{}%
\providecommand \bibitemStop [0]{}%
\providecommand \bibitemNoStop [0]{.\EOS\space}%
\providecommand \EOS [0]{\spacefactor3000\relax}%
\providecommand \BibitemShut  [1]{\csname bibitem#1\endcsname}%
\let\auto@bib@innerbib\@empty
%</preamble>
\bibitem [{\citenamefont {Wensink}\ \emph {et~al.}(2012)\citenamefont
  {Wensink}, \citenamefont {Dunkel}, \citenamefont {Heidenreich}, \citenamefont
  {Drescher}, \citenamefont {Goldstein}, \citenamefont {L{\"o}wen},\ and\
  \citenamefont {Yeomans}}]{WDH+2012pnasusa}%
  \BibitemOpen
  \bibfield  {author} {\bibinfo {author} {\bibfnamefont {H.}~\bibnamefont
  {Wensink}}, \bibinfo {author} {\bibfnamefont {J.}~\bibnamefont {Dunkel}},
  \bibinfo {author} {\bibfnamefont {S.}~\bibnamefont {Heidenreich}}, \bibinfo
  {author} {\bibfnamefont {K.}~\bibnamefont {Drescher}}, \bibinfo {author}
  {\bibfnamefont {R.}~\bibnamefont {Goldstein}}, \bibinfo {author}
  {\bibfnamefont {H.}~\bibnamefont {L{\"o}wen}}, \ and\ \bibinfo {author}
  {\bibfnamefont {J.}~\bibnamefont {Yeomans}},\ }\href {\doibase
  10.1073/pnas.1202032109} {\bibfield  {journal} {\bibinfo  {journal} {Proc.
  Natl. Acad. Sci. U. S. A.}\ }\textbf {\bibinfo {volume} {109}},\ \bibinfo
  {pages} {14308} (\bibinfo {year} {2012})}\BibitemShut {NoStop}%
\bibitem [{\citenamefont {N{\'e}d{\'e}lec}\ \emph {et~al.}(1997)\citenamefont
  {N{\'e}d{\'e}lec}, \citenamefont {Surrey}, \citenamefont {Maggs},\ and\
  \citenamefont {Leibler}}]{NSM+1997n}%
  \BibitemOpen
  \bibfield  {author} {\bibinfo {author} {\bibfnamefont {F.}~\bibnamefont
  {N{\'e}d{\'e}lec}}, \bibinfo {author} {\bibfnamefont {T.}~\bibnamefont
  {Surrey}}, \bibinfo {author} {\bibfnamefont {A.~C.}\ \bibnamefont {Maggs}}, \
  and\ \bibinfo {author} {\bibfnamefont {S.}~\bibnamefont {Leibler}},\
  }\href@noop {} {\bibfield  {journal} {\bibinfo  {journal} {Nature}\ }\textbf
  {\bibinfo {volume} {389}},\ \bibinfo {pages} {305} (\bibinfo {year}
  {1997})}\BibitemShut {NoStop}%
\bibitem [{\citenamefont {Surrey}\ \emph {et~al.}(2001)\citenamefont {Surrey},
  \citenamefont {N{\'e}d{\'e}lec}, \citenamefont {Leibler},\ and\ \citenamefont
  {Karsenti}}]{SNL+2001s}%
  \BibitemOpen
  \bibfield  {author} {\bibinfo {author} {\bibfnamefont {T.}~\bibnamefont
  {Surrey}}, \bibinfo {author} {\bibfnamefont {F.}~\bibnamefont
  {N{\'e}d{\'e}lec}}, \bibinfo {author} {\bibfnamefont {S.}~\bibnamefont
  {Leibler}}, \ and\ \bibinfo {author} {\bibfnamefont {E.}~\bibnamefont
  {Karsenti}},\ }\href@noop {} {\bibfield  {journal} {\bibinfo  {journal}
  {Science}\ }\textbf {\bibinfo {volume} {292}},\ \bibinfo {pages} {1167}
  (\bibinfo {year} {2001})}\BibitemShut {NoStop}%
\bibitem [{\citenamefont {Sumino}\ \emph {et~al.}(2012)\citenamefont {Sumino},
  \citenamefont {Nagai}, \citenamefont {Shitaka}, \citenamefont {Tanaka},
  \citenamefont {Yoshikawa}, \citenamefont {Chat{\'e}},\ and\ \citenamefont
  {Oiwa}}]{SNS+2012n}%
  \BibitemOpen
  \bibfield  {author} {\bibinfo {author} {\bibfnamefont {Y.}~\bibnamefont
  {Sumino}}, \bibinfo {author} {\bibfnamefont {K.~H.}\ \bibnamefont {Nagai}},
  \bibinfo {author} {\bibfnamefont {Y.}~\bibnamefont {Shitaka}}, \bibinfo
  {author} {\bibfnamefont {D.}~\bibnamefont {Tanaka}}, \bibinfo {author}
  {\bibfnamefont {K.}~\bibnamefont {Yoshikawa}}, \bibinfo {author}
  {\bibfnamefont {H.}~\bibnamefont {Chat{\'e}}}, \ and\ \bibinfo {author}
  {\bibfnamefont {K.}~\bibnamefont {Oiwa}},\ }\href@noop {} {\bibfield
  {journal} {\bibinfo  {journal} {Nature}\ }\textbf {\bibinfo {volume} {483}},\
  \bibinfo {pages} {448} (\bibinfo {year} {2012})}\BibitemShut {NoStop}%
\bibitem [{\citenamefont {Peruani}\ \emph {et~al.}(2012)\citenamefont
  {Peruani}, \citenamefont {Starru{\ss}}, \citenamefont {Jakovljevic},
  \citenamefont {S{\o}gaard-Andersen}, \citenamefont {Deutsch},\ and\
  \citenamefont {B{\"a}r}}]{PSJ+2012prl}%
  \BibitemOpen
  \bibfield  {author} {\bibinfo {author} {\bibfnamefont {F.}~\bibnamefont
  {Peruani}}, \bibinfo {author} {\bibfnamefont {J.}~\bibnamefont
  {Starru{\ss}}}, \bibinfo {author} {\bibfnamefont {V.}~\bibnamefont
  {Jakovljevic}}, \bibinfo {author} {\bibfnamefont {L.}~\bibnamefont
  {S{\o}gaard-Andersen}}, \bibinfo {author} {\bibfnamefont {A.}~\bibnamefont
  {Deutsch}}, \ and\ \bibinfo {author} {\bibfnamefont {M.}~\bibnamefont
  {B{\"a}r}},\ }\href@noop {} {\bibfield  {journal} {\bibinfo  {journal} {Phys.
  Rev. Lett.}\ }\textbf {\bibinfo {volume} {108}},\ \bibinfo {pages} {098102}
  (\bibinfo {year} {2012})}\BibitemShut {NoStop}%
\bibitem [{\citenamefont {Zhang}\ \emph {et~al.}(2010)\citenamefont {Zhang},
  \citenamefont {Be'er}, \citenamefont {Florin},\ and\ \citenamefont
  {Swinney}}]{ZBF+2010potnaos}%
  \BibitemOpen
  \bibfield  {author} {\bibinfo {author} {\bibfnamefont {H.-P.}\ \bibnamefont
  {Zhang}}, \bibinfo {author} {\bibfnamefont {A.}~\bibnamefont {Be'er}},
  \bibinfo {author} {\bibfnamefont {E.-L.}\ \bibnamefont {Florin}}, \ and\
  \bibinfo {author} {\bibfnamefont {H.~L.}\ \bibnamefont {Swinney}},\
  }\href@noop {} {\bibfield  {journal} {\bibinfo  {journal} {Proc. Natl. Acad.
  Sci. U. S. A}\ }\textbf {\bibinfo {volume} {107}},\ \bibinfo {pages} {13626}
  (\bibinfo {year} {2010})}\BibitemShut {NoStop}%
\bibitem [{\citenamefont {Buttinoni}\ \emph {et~al.}(2013)\citenamefont
  {Buttinoni}, \citenamefont {Bialk{\'e}}, \citenamefont {K{\"u}mmel},
  \citenamefont {L{\"o}wen}, \citenamefont {Bechinger},\ and\ \citenamefont
  {Speck}}]{BBK+2013prl}%
  \BibitemOpen
  \bibfield  {author} {\bibinfo {author} {\bibfnamefont {I.}~\bibnamefont
  {Buttinoni}}, \bibinfo {author} {\bibfnamefont {J.}~\bibnamefont
  {Bialk{\'e}}}, \bibinfo {author} {\bibfnamefont {F.}~\bibnamefont
  {K{\"u}mmel}}, \bibinfo {author} {\bibfnamefont {H.}~\bibnamefont
  {L{\"o}wen}}, \bibinfo {author} {\bibfnamefont {C.}~\bibnamefont
  {Bechinger}}, \ and\ \bibinfo {author} {\bibfnamefont {T.}~\bibnamefont
  {Speck}},\ }\href@noop {} {\bibfield  {journal} {\bibinfo  {journal} {Phys.
  Rev. Lett.}\ }\textbf {\bibinfo {volume} {110}},\ \bibinfo {pages} {238301}
  (\bibinfo {year} {2013})}\BibitemShut {NoStop}%
\bibitem [{\citenamefont {Wioland}\ \emph {et~al.}(2013)\citenamefont
  {Wioland}, \citenamefont {Woodhouse}, \citenamefont {Dunkel}, \citenamefont
  {Kessler},\ and\ \citenamefont {Goldstein}}]{WWD+2013prl}%
  \BibitemOpen
  \bibfield  {author} {\bibinfo {author} {\bibfnamefont {H.}~\bibnamefont
  {Wioland}}, \bibinfo {author} {\bibfnamefont {F.~G.}\ \bibnamefont
  {Woodhouse}}, \bibinfo {author} {\bibfnamefont {J.}~\bibnamefont {Dunkel}},
  \bibinfo {author} {\bibfnamefont {J.~O.}\ \bibnamefont {Kessler}}, \ and\
  \bibinfo {author} {\bibfnamefont {R.~E.}\ \bibnamefont {Goldstein}},\
  }\href@noop {} {\bibfield  {journal} {\bibinfo  {journal} {Phys. Rev. Lett.}\
  }\textbf {\bibinfo {volume} {110}},\ \bibinfo {pages} {268102} (\bibinfo
  {year} {2013})}\BibitemShut {NoStop}%
\bibitem [{\citenamefont {Shah}\ and\ \citenamefont {Keren}(2014)}]{SK2014e}%
  \BibitemOpen
  \bibfield  {author} {\bibinfo {author} {\bibfnamefont {E.~A.}\ \bibnamefont
  {Shah}}\ and\ \bibinfo {author} {\bibfnamefont {K.}~\bibnamefont {Keren}},\
  }\href@noop {} {\bibfield  {journal} {\bibinfo  {journal} {Elife}\ }\textbf
  {\bibinfo {volume} {3}},\ \bibinfo {pages} {e01433} (\bibinfo {year}
  {2014})}\BibitemShut {NoStop}%
\bibitem [{\citenamefont {Sanchez}\ \emph {et~al.}(2012)\citenamefont
  {Sanchez}, \citenamefont {Chen}, \citenamefont {DeCamp}, \citenamefont
  {Heymann},\ and\ \citenamefont {Dogic}}]{SCD+2012n}%
  \BibitemOpen
  \bibfield  {author} {\bibinfo {author} {\bibfnamefont {T.}~\bibnamefont
  {Sanchez}}, \bibinfo {author} {\bibfnamefont {D.~T.~N.}\ \bibnamefont
  {Chen}}, \bibinfo {author} {\bibfnamefont {S.~J.}\ \bibnamefont {DeCamp}},
  \bibinfo {author} {\bibfnamefont {M.}~\bibnamefont {Heymann}}, \ and\
  \bibinfo {author} {\bibfnamefont {Z.}~\bibnamefont {Dogic}},\ }\href@noop {}
  {\bibfield  {journal} {\bibinfo  {journal} {Nature}\ }\textbf {\bibinfo
  {volume} {491}},\ \bibinfo {pages} {431} (\bibinfo {year}
  {2012})}\BibitemShut {NoStop}%
\bibitem [{\citenamefont {De~Gennes}(1985)}]{DeGennes1985}%
  \BibitemOpen
  \bibfield  {author} {\bibinfo {author} {\bibfnamefont {P.-G.}\ \bibnamefont
  {De~Gennes}},\ }\href@noop {} {\bibfield  {journal} {\bibinfo  {journal}
  {Rev. Mod. Phys.}\ }\textbf {\bibinfo {volume} {57}},\ \bibinfo {pages} {827}
  (\bibinfo {year} {1985})}\BibitemShut {NoStop}%
\bibitem [{\citenamefont {Wallmeyer}\ \emph {et~al.}(2018)\citenamefont
  {Wallmeyer}, \citenamefont {Trinschek}, \citenamefont {Yigit}, \citenamefont
  {Thiele},\ and\ \citenamefont {Betz}}]{WTY+2018bj}%
  \BibitemOpen
  \bibfield  {author} {\bibinfo {author} {\bibfnamefont {B.}~\bibnamefont
  {Wallmeyer}}, \bibinfo {author} {\bibfnamefont {S.}~\bibnamefont
  {Trinschek}}, \bibinfo {author} {\bibfnamefont {S.}~\bibnamefont {Yigit}},
  \bibinfo {author} {\bibfnamefont {U.}~\bibnamefont {Thiele}}, \ and\ \bibinfo
  {author} {\bibfnamefont {T.}~\bibnamefont {Betz}},\ }\href@noop {} {\bibfield
   {journal} {\bibinfo  {journal} {Biophys. J.}\ }\textbf {\bibinfo {volume}
  {114}},\ \bibinfo {pages} {213} (\bibinfo {year} {2018})}\BibitemShut
  {NoStop}%
\bibitem [{\citenamefont {Douezan}\ \emph {et~al.}(2011)\citenamefont
  {Douezan}, \citenamefont {Guevorkian}, \citenamefont {Naouar}, \citenamefont
  {Dufour}, \citenamefont {Cuvelier},\ and\ \citenamefont
  {Brochard-Wyart}}]{DGN+2011potnaos}%
  \BibitemOpen
  \bibfield  {author} {\bibinfo {author} {\bibfnamefont {S.}~\bibnamefont
  {Douezan}}, \bibinfo {author} {\bibfnamefont {K.}~\bibnamefont {Guevorkian}},
  \bibinfo {author} {\bibfnamefont {R.}~\bibnamefont {Naouar}}, \bibinfo
  {author} {\bibfnamefont {S.}~\bibnamefont {Dufour}}, \bibinfo {author}
  {\bibfnamefont {D.}~\bibnamefont {Cuvelier}}, \ and\ \bibinfo {author}
  {\bibfnamefont {F.}~\bibnamefont {Brochard-Wyart}},\ }\href@noop {}
  {\bibfield  {journal} {\bibinfo  {journal} {Proc. Natl. Acad. Sci. U.S.A}\
  }\textbf {\bibinfo {volume} {108}},\ \bibinfo {pages} {7315} (\bibinfo {year}
  {2011})}\BibitemShut {NoStop}%
\bibitem [{\citenamefont {Douezan}\ \emph {et~al.}(2012)\citenamefont
  {Douezan}, \citenamefont {Dumond},\ and\ \citenamefont
  {Brochard-Wyart}}]{DDB2012sm}%
  \BibitemOpen
  \bibfield  {author} {\bibinfo {author} {\bibfnamefont {S.}~\bibnamefont
  {Douezan}}, \bibinfo {author} {\bibfnamefont {J.}~\bibnamefont {Dumond}}, \
  and\ \bibinfo {author} {\bibfnamefont {F.}~\bibnamefont {Brochard-Wyart}},\
  }\href@noop {} {\bibfield  {journal} {\bibinfo  {journal} {Soft Matter}\
  }\textbf {\bibinfo {volume} {8}},\ \bibinfo {pages} {4578} (\bibinfo {year}
  {2012})}\BibitemShut {NoStop}%
\bibitem [{\citenamefont {P{\'e}rez-Gonz{\'a}lez}\ \emph
  {et~al.}(2019)\citenamefont {P{\'e}rez-Gonz{\'a}lez}, \citenamefont {Alert},
  \citenamefont {Blanch-Mercader}, \citenamefont {G{\'o}mez-Gonz{\'a}lez},
  \citenamefont {Kolodziej}, \citenamefont {Bazellieres}, \citenamefont
  {Casademunt},\ and\ \citenamefont {Trepat}}]{PAB+2019np}%
  \BibitemOpen
  \bibfield  {author} {\bibinfo {author} {\bibfnamefont {C.}~\bibnamefont
  {P{\'e}rez-Gonz{\'a}lez}}, \bibinfo {author} {\bibfnamefont {R.}~\bibnamefont
  {Alert}}, \bibinfo {author} {\bibfnamefont {C.}~\bibnamefont
  {Blanch-Mercader}}, \bibinfo {author} {\bibfnamefont {M.}~\bibnamefont
  {G{\'o}mez-Gonz{\'a}lez}}, \bibinfo {author} {\bibfnamefont {T.}~\bibnamefont
  {Kolodziej}}, \bibinfo {author} {\bibfnamefont {E.}~\bibnamefont
  {Bazellieres}}, \bibinfo {author} {\bibfnamefont {J.}~\bibnamefont
  {Casademunt}}, \ and\ \bibinfo {author} {\bibfnamefont {X.}~\bibnamefont
  {Trepat}},\ }\href@noop {} {\bibfield  {journal} {\bibinfo  {journal} {Nat.
  Phys.}\ }\textbf {\bibinfo {volume} {15}},\ \bibinfo {pages} {79} (\bibinfo
  {year} {2019})}\BibitemShut {NoStop}%
\bibitem [{\citenamefont {Morita}\ \emph {et~al.}(2017)\citenamefont {Morita},
  \citenamefont {Grigolon}, \citenamefont {Bock}, \citenamefont {Krens},
  \citenamefont {Salbreux},\ and\ \citenamefont {Heisenberg}}]{MGB+2017dc}%
  \BibitemOpen
  \bibfield  {author} {\bibinfo {author} {\bibfnamefont {H.}~\bibnamefont
  {Morita}}, \bibinfo {author} {\bibfnamefont {S.}~\bibnamefont {Grigolon}},
  \bibinfo {author} {\bibfnamefont {M.}~\bibnamefont {Bock}}, \bibinfo {author}
  {\bibfnamefont {S.~G.}\ \bibnamefont {Krens}}, \bibinfo {author}
  {\bibfnamefont {G.}~\bibnamefont {Salbreux}}, \ and\ \bibinfo {author}
  {\bibfnamefont {C.-P.}\ \bibnamefont {Heisenberg}},\ }\href@noop {}
  {\bibfield  {journal} {\bibinfo  {journal} {Dev. Cell}\ }\textbf {\bibinfo
  {volume} {40}},\ \bibinfo {pages} {354} (\bibinfo {year} {2017})}\BibitemShut
  {NoStop}%
\bibitem [{\citenamefont {Marchetti}\ \emph {et~al.}(2013)\citenamefont
  {Marchetti}, \citenamefont {Joanny}, \citenamefont {Ramaswamy}, \citenamefont
  {Liverpool}, \citenamefont {Prost}, \citenamefont {Rao},\ and\ \citenamefont
  {Simha}}]{MJR+2013rmp}%
  \BibitemOpen
  \bibfield  {author} {\bibinfo {author} {\bibfnamefont {M.~C.}\ \bibnamefont
  {Marchetti}}, \bibinfo {author} {\bibfnamefont {J.-F.}\ \bibnamefont
  {Joanny}}, \bibinfo {author} {\bibfnamefont {S.}~\bibnamefont {Ramaswamy}},
  \bibinfo {author} {\bibfnamefont {T.~B.}\ \bibnamefont {Liverpool}}, \bibinfo
  {author} {\bibfnamefont {J.}~\bibnamefont {Prost}}, \bibinfo {author}
  {\bibfnamefont {M.}~\bibnamefont {Rao}}, \ and\ \bibinfo {author}
  {\bibfnamefont {R.~A.}\ \bibnamefont {Simha}},\ }\href {\doibase
  10.1103/RevModPhys.85.1143} {\bibfield  {journal} {\bibinfo  {journal} {Rev.
  Mod. Phys.}\ }\textbf {\bibinfo {volume} {85}},\ \bibinfo {pages} {1143}
  (\bibinfo {year} {2013})}\BibitemShut {NoStop}%
\bibitem [{\citenamefont {Menzel}(2015)}]{Menzel2015pr}%
  \BibitemOpen
  \bibfield  {author} {\bibinfo {author} {\bibfnamefont {A.~M.}\ \bibnamefont
  {Menzel}},\ }\href {\doibase http://dx.doi.org/10.1016/j.physrep.2014.10.001}
  {\bibfield  {journal} {\bibinfo  {journal} {Phys. Rep.}\ }\textbf {\bibinfo
  {volume} {554}},\ \bibinfo {pages} {1 } (\bibinfo {year} {2015})}\BibitemShut
  {NoStop}%
\bibitem [{\citenamefont {Ramaswamy}(2010)}]{Ramaswamy2010}%
  \BibitemOpen
  \bibfield  {author} {\bibinfo {author} {\bibfnamefont {S.}~\bibnamefont
  {Ramaswamy}},\ }\href {\doibase 10.1146/annurev-conmatphys-070909-104101}
  {\bibfield  {journal} {\bibinfo  {journal} {Annu. Rev. Condens. Matter Phys}\
  }\textbf {\bibinfo {volume} {1}},\ \bibinfo {pages} {323} (\bibinfo {year}
  {2010})}\BibitemShut {NoStop}%
\bibitem [{\citenamefont {Chandrasekhar}(1992)}]{Chandrasekhar1992}%
  \BibitemOpen
  \bibfield  {author} {\bibinfo {author} {\bibfnamefont {S.}~\bibnamefont
  {Chandrasekhar}},\ }\href@noop {} {\emph {\bibinfo {title} {Liquid
  Crystals}}},\ \bibinfo {edition} {2nd}\ ed.\ (\bibinfo  {publisher}
  {Cambridge University Press},\ \bibinfo {year} {1992})\BibitemShut {NoStop}%
\bibitem [{\citenamefont {Prost}(1995)}]{Prost1995}%
  \BibitemOpen
  \bibfield  {author} {\bibinfo {author} {\bibfnamefont {J.}~\bibnamefont
  {Prost}},\ }\href@noop {} {\emph {\bibinfo {title} {The Physics of Liquid
  Crystals}}},\ Vol.~\bibinfo {volume} {83}\ (\bibinfo  {publisher} {Oxford
  University Press},\ \bibinfo {year} {1995})\BibitemShut {NoStop}%
\bibitem [{\citenamefont {Baskaran}\ and\ \citenamefont
  {Marchetti}(2009)}]{BM2009potnaos}%
  \BibitemOpen
  \bibfield  {author} {\bibinfo {author} {\bibfnamefont {A.}~\bibnamefont
  {Baskaran}}\ and\ \bibinfo {author} {\bibfnamefont {M.~C.}\ \bibnamefont
  {Marchetti}},\ }\href@noop {} {\bibfield  {journal} {\bibinfo  {journal}
  {Proc. Natl. Acad. Sci. U.S.A}\ }\textbf {\bibinfo {volume} {106}},\ \bibinfo
  {pages} {15567} (\bibinfo {year} {2009})}\BibitemShut {NoStop}%
\bibitem [{\citenamefont {Liverpool}\ and\ \citenamefont
  {Marchetti}(2006)}]{LM2006prl}%
  \BibitemOpen
  \bibfield  {author} {\bibinfo {author} {\bibfnamefont {T.~B.}\ \bibnamefont
  {Liverpool}}\ and\ \bibinfo {author} {\bibfnamefont {M.~C.}\ \bibnamefont
  {Marchetti}},\ }\href@noop {} {\bibfield  {journal} {\bibinfo  {journal}
  {Phys. Rev. Lett.}\ }\textbf {\bibinfo {volume} {97}},\ \bibinfo {pages}
  {268101} (\bibinfo {year} {2006})}\BibitemShut {NoStop}%
\bibitem [{\citenamefont {Reinken}\ \emph {et~al.}(2018)\citenamefont
  {Reinken}, \citenamefont {Klapp}, \citenamefont {B\"ar},\ and\ \citenamefont
  {Heidenreich}}]{RKBH2018pre}%
  \BibitemOpen
  \bibfield  {author} {\bibinfo {author} {\bibfnamefont {H.}~\bibnamefont
  {Reinken}}, \bibinfo {author} {\bibfnamefont {S.~H.~L.}\ \bibnamefont
  {Klapp}}, \bibinfo {author} {\bibfnamefont {M.}~\bibnamefont {B\"ar}}, \ and\
  \bibinfo {author} {\bibfnamefont {S.}~\bibnamefont {Heidenreich}},\ }\href
  {\doibase 10.1103/PhysRevE.97.022613} {\bibfield  {journal} {\bibinfo
  {journal} {Phys. Rev. E}\ }\textbf {\bibinfo {volume} {97}},\ \bibinfo
  {pages} {022613} (\bibinfo {year} {2018})}\BibitemShut {NoStop}%
\bibitem [{\citenamefont {Hatwalne}\ \emph {et~al.}(2004)\citenamefont
  {Hatwalne}, \citenamefont {Ramaswamy}, \citenamefont {Rao},\ and\
  \citenamefont {Simha}}]{HRR+2004prl}%
  \BibitemOpen
  \bibfield  {author} {\bibinfo {author} {\bibfnamefont {Y.}~\bibnamefont
  {Hatwalne}}, \bibinfo {author} {\bibfnamefont {S.}~\bibnamefont {Ramaswamy}},
  \bibinfo {author} {\bibfnamefont {M.}~\bibnamefont {Rao}}, \ and\ \bibinfo
  {author} {\bibfnamefont {R.~A.}\ \bibnamefont {Simha}},\ }\href@noop {}
  {\bibfield  {journal} {\bibinfo  {journal} {Phys. Rev. Lett.}\ }\textbf
  {\bibinfo {volume} {92}},\ \bibinfo {pages} {118101} (\bibinfo {year}
  {2004})}\BibitemShut {NoStop}%
\bibitem [{\citenamefont {Kruse}\ \emph {et~al.}(2005)\citenamefont {Kruse},
  \citenamefont {Joanny}, \citenamefont {J{\"u}licher}, \citenamefont {Prost},\
  and\ \citenamefont {Sekimoto}}]{KJJ+2005tepje}%
  \BibitemOpen
  \bibfield  {author} {\bibinfo {author} {\bibfnamefont {K.}~\bibnamefont
  {Kruse}}, \bibinfo {author} {\bibfnamefont {J.-F.}\ \bibnamefont {Joanny}},
  \bibinfo {author} {\bibfnamefont {F.}~\bibnamefont {J{\"u}licher}}, \bibinfo
  {author} {\bibfnamefont {J.}~\bibnamefont {Prost}}, \ and\ \bibinfo {author}
  {\bibfnamefont {K.}~\bibnamefont {Sekimoto}},\ }\href@noop {} {\bibfield
  {journal} {\bibinfo  {journal} {Eur. Phys. J. E}\ }\textbf {\bibinfo {volume}
  {16}},\ \bibinfo {pages} {5} (\bibinfo {year} {2005})}\BibitemShut {NoStop}%
\bibitem [{\citenamefont {J\"{u}licher}\ \emph {et~al.}(2007)\citenamefont
  {J\"{u}licher}, \citenamefont {Kruse}, \citenamefont {Prost},\ and\
  \citenamefont {Joanny}}]{JKP+2007pr}%
  \BibitemOpen
  \bibfield  {author} {\bibinfo {author} {\bibfnamefont {F.}~\bibnamefont
  {J\"{u}licher}}, \bibinfo {author} {\bibfnamefont {K.}~\bibnamefont {Kruse}},
  \bibinfo {author} {\bibfnamefont {J.}~\bibnamefont {Prost}}, \ and\ \bibinfo
  {author} {\bibfnamefont {J.-F.}\ \bibnamefont {Joanny}},\ }\href@noop {}
  {\bibfield  {journal} {\bibinfo  {journal} {Phys. Rep}\ }\textbf {\bibinfo
  {volume} {449}},\ \bibinfo {pages} {3} (\bibinfo {year} {2007})}\BibitemShut
  {NoStop}%
\bibitem [{\citenamefont {Joanny}\ and\ \citenamefont
  {Prost}(2009)}]{JP2009hj}%
  \BibitemOpen
  \bibfield  {author} {\bibinfo {author} {\bibfnamefont {J.-F.}\ \bibnamefont
  {Joanny}}\ and\ \bibinfo {author} {\bibfnamefont {J.}~\bibnamefont {Prost}},\
  }\href@noop {} {\bibfield  {journal} {\bibinfo  {journal} {HFSP}\ }\textbf
  {\bibinfo {volume} {3}},\ \bibinfo {pages} {94} (\bibinfo {year}
  {2009})}\BibitemShut {NoStop}%
\bibitem [{\citenamefont {Prost}\ \emph {et~al.}(2015)\citenamefont {Prost},
  \citenamefont {J{\"u}licher},\ and\ \citenamefont {Joanny}}]{PJJ2015np}%
  \BibitemOpen
  \bibfield  {author} {\bibinfo {author} {\bibfnamefont {J.}~\bibnamefont
  {Prost}}, \bibinfo {author} {\bibfnamefont {F.}~\bibnamefont {J{\"u}licher}},
  \ and\ \bibinfo {author} {\bibfnamefont {J.-F.}\ \bibnamefont {Joanny}},\
  }\href@noop {} {\bibfield  {journal} {\bibinfo  {journal} {‎Nat. Phys.}\
  }\textbf {\bibinfo {volume} {11}},\ \bibinfo {pages} {111} (\bibinfo {year}
  {2015})}\BibitemShut {NoStop}%
\bibitem [{\citenamefont {Kruse}\ \emph {et~al.}(2004)\citenamefont {Kruse},
  \citenamefont {Joanny}, \citenamefont {J{\"u}licher}, \citenamefont {Prost},\
  and\ \citenamefont {Sekimoto}}]{KJJ+2004prl}%
  \BibitemOpen
  \bibfield  {author} {\bibinfo {author} {\bibfnamefont {K.}~\bibnamefont
  {Kruse}}, \bibinfo {author} {\bibfnamefont {J.-F.}\ \bibnamefont {Joanny}},
  \bibinfo {author} {\bibfnamefont {F.}~\bibnamefont {J{\"u}licher}}, \bibinfo
  {author} {\bibfnamefont {J.}~\bibnamefont {Prost}}, \ and\ \bibinfo {author}
  {\bibfnamefont {K.}~\bibnamefont {Sekimoto}},\ }\href@noop {} {\bibfield
  {journal} {\bibinfo  {journal} {Phys. Rev. Lett.}\ }\textbf {\bibinfo
  {volume} {92}},\ \bibinfo {pages} {078101} (\bibinfo {year}
  {2004})}\BibitemShut {NoStop}%
\bibitem [{\citenamefont {Tjhung}\ \emph {et~al.}(2011)\citenamefont {Tjhung},
  \citenamefont {Cates},\ and\ \citenamefont {Marenduzzo}}]{TCM2011sm}%
  \BibitemOpen
  \bibfield  {author} {\bibinfo {author} {\bibfnamefont {E.}~\bibnamefont
  {Tjhung}}, \bibinfo {author} {\bibfnamefont {M.~E.}\ \bibnamefont {Cates}}, \
  and\ \bibinfo {author} {\bibfnamefont {D.}~\bibnamefont {Marenduzzo}},\
  }\href@noop {} {\bibfield  {journal} {\bibinfo  {journal} {Soft Matter}\
  }\textbf {\bibinfo {volume} {7}},\ \bibinfo {pages} {7453} (\bibinfo {year}
  {2011})}\BibitemShut {NoStop}%
\bibitem [{\citenamefont {Giomi}\ \emph {et~al.}(2008)\citenamefont {Giomi},
  \citenamefont {Marchetti},\ and\ \citenamefont {Liverpool}}]{GML2008prl}%
  \BibitemOpen
  \bibfield  {author} {\bibinfo {author} {\bibfnamefont {L.}~\bibnamefont
  {Giomi}}, \bibinfo {author} {\bibfnamefont {M.~C.}\ \bibnamefont
  {Marchetti}}, \ and\ \bibinfo {author} {\bibfnamefont {T.~B.}\ \bibnamefont
  {Liverpool}},\ }\href@noop {} {\bibfield  {journal} {\bibinfo  {journal}
  {Phys. Rev. Lett.}\ }\textbf {\bibinfo {volume} {101}},\ \bibinfo {pages}
  {198101} (\bibinfo {year} {2008})}\BibitemShut {NoStop}%
\bibitem [{\citenamefont {Joanny}\ \emph {et~al.}(2007)\citenamefont {Joanny},
  \citenamefont {J{\"u}licher}, \citenamefont {Kruse},\ and\ \citenamefont
  {Prost}}]{JJK+2007NJoP}%
  \BibitemOpen
  \bibfield  {author} {\bibinfo {author} {\bibfnamefont {J.-F.}\ \bibnamefont
  {Joanny}}, \bibinfo {author} {\bibfnamefont {F.}~\bibnamefont
  {J{\"u}licher}}, \bibinfo {author} {\bibfnamefont {K.}~\bibnamefont {Kruse}},
  \ and\ \bibinfo {author} {\bibfnamefont {J.}~\bibnamefont {Prost}},\
  }\href@noop {} {\bibfield  {journal} {\bibinfo  {journal} {New J. Phys.}\
  }\textbf {\bibinfo {volume} {9}},\ \bibinfo {pages} {422} (\bibinfo {year}
  {2007})}\BibitemShut {NoStop}%
\bibitem [{\citenamefont {Voituriez}\ \emph {et~al.}(2006)\citenamefont
  {Voituriez}, \citenamefont {Joanny},\ and\ \citenamefont
  {Prost}}]{VJP2006prl}%
  \BibitemOpen
  \bibfield  {author} {\bibinfo {author} {\bibfnamefont {R.}~\bibnamefont
  {Voituriez}}, \bibinfo {author} {\bibfnamefont {J.-F.}\ \bibnamefont
  {Joanny}}, \ and\ \bibinfo {author} {\bibfnamefont {J.}~\bibnamefont
  {Prost}},\ }\href@noop {} {\bibfield  {journal} {\bibinfo  {journal} {Phys.
  Rev. Lett.}\ }\textbf {\bibinfo {volume} {96}},\ \bibinfo {pages} {028102}
  (\bibinfo {year} {2006})}\BibitemShut {NoStop}%
\bibitem [{\citenamefont {Cortese}\ \emph {et~al.}(2016)\citenamefont
  {Cortese}, \citenamefont {Eggers},\ and\ \citenamefont
  {Liverpool}}]{CoEL2016el}%
  \BibitemOpen
  \bibfield  {author} {\bibinfo {author} {\bibfnamefont {D.}~\bibnamefont
  {Cortese}}, \bibinfo {author} {\bibfnamefont {J.}~\bibnamefont {Eggers}}, \
  and\ \bibinfo {author} {\bibfnamefont {T.~B.}\ \bibnamefont {Liverpool}},\
  }\href {\doibase 10.1209/0295-5075/115/28002} {\bibfield  {journal} {\bibinfo
   {journal} {Europhys. Lett.}\ }\textbf {\bibinfo {volume} {115}},\ \bibinfo
  {pages} {28002} (\bibinfo {year} {2016})}\BibitemShut {NoStop}%
\bibitem [{\citenamefont {Loisy}\ \emph {et~al.}(2018)\citenamefont {Loisy},
  \citenamefont {Eggers},\ and\ \citenamefont {Liverpool}}]{LoEL2018prl}%
  \BibitemOpen
  \bibfield  {author} {\bibinfo {author} {\bibfnamefont {A.}~\bibnamefont
  {Loisy}}, \bibinfo {author} {\bibfnamefont {J.}~\bibnamefont {Eggers}}, \
  and\ \bibinfo {author} {\bibfnamefont {T.~B.}\ \bibnamefont {Liverpool}},\
  }\href@noop {} {\bibfield  {journal} {\bibinfo  {journal} {Phys. Rev. Lett.}\
  }\textbf {\bibinfo {volume} {121}} (\bibinfo {year} {2018})}\BibitemShut
  {NoStop}%
\bibitem [{\citenamefont {Loisy}\ \emph
  {et~al.}(2019{\natexlab{a}})\citenamefont {Loisy}, \citenamefont {Thompson},
  \citenamefont {Eggers},\ and\ \citenamefont {Liverpool}}]{LTEL2019jcp}%
  \BibitemOpen
  \bibfield  {author} {\bibinfo {author} {\bibfnamefont {A.}~\bibnamefont
  {Loisy}}, \bibinfo {author} {\bibfnamefont {A.~P.}\ \bibnamefont {Thompson}},
  \bibinfo {author} {\bibfnamefont {J.}~\bibnamefont {Eggers}}, \ and\ \bibinfo
  {author} {\bibfnamefont {T.~B.}\ \bibnamefont {Liverpool}},\ }\href {\doibase
  10.1063/1.5080343} {\bibfield  {journal} {\bibinfo  {journal} {J. Chem.
  Phys.}\ }\textbf {\bibinfo {volume} {150}},\ \bibinfo {pages} {104902}
  (\bibinfo {year} {2019}{\natexlab{a}})}\BibitemShut {NoStop}%
\bibitem [{\citenamefont {Tjhung}\ \emph {et~al.}(2012)\citenamefont {Tjhung},
  \citenamefont {Marenduzzo},\ and\ \citenamefont {Cates}}]{TMC2012potnaos}%
  \BibitemOpen
  \bibfield  {author} {\bibinfo {author} {\bibfnamefont {E.}~\bibnamefont
  {Tjhung}}, \bibinfo {author} {\bibfnamefont {D.}~\bibnamefont {Marenduzzo}},
  \ and\ \bibinfo {author} {\bibfnamefont {M.~E.}\ \bibnamefont {Cates}},\
  }\href@noop {} {\bibfield  {journal} {\bibinfo  {journal} {Proc. Natl. Acad.
  Sci. U. S. A}\ }\textbf {\bibinfo {volume} {109}},\ \bibinfo {pages} {12381}
  (\bibinfo {year} {2012})}\BibitemShut {NoStop}%
\bibitem [{\citenamefont {Whitfield}\ \emph {et~al.}(2014)\citenamefont
  {Whitfield}, \citenamefont {Marenduzzo}, \citenamefont {Voituriez},\ and\
  \citenamefont {Hawkins}}]{WMV+2014tepje}%
  \BibitemOpen
  \bibfield  {author} {\bibinfo {author} {\bibfnamefont {C.~A.}\ \bibnamefont
  {Whitfield}}, \bibinfo {author} {\bibfnamefont {D.}~\bibnamefont
  {Marenduzzo}}, \bibinfo {author} {\bibfnamefont {R.}~\bibnamefont
  {Voituriez}}, \ and\ \bibinfo {author} {\bibfnamefont {R.~J.}\ \bibnamefont
  {Hawkins}},\ }\href@noop {} {\bibfield  {journal} {\bibinfo  {journal} {Eur.
  Phys. J. E}\ }\textbf {\bibinfo {volume} {37}},\ \bibinfo {pages} {8}
  (\bibinfo {year} {2014})}\BibitemShut {NoStop}%
\bibitem [{\citenamefont {Marth}\ \emph {et~al.}(2015)\citenamefont {Marth},
  \citenamefont {Wieland},\ and\ \citenamefont {Praetorius}}]{MWP2015jotrsi}%
  \BibitemOpen
  \bibfield  {author} {\bibinfo {author} {\bibfnamefont {S.}~\bibnamefont
  {Marth}}, \bibinfo {author} {\bibfnamefont {V.}~\bibnamefont {Wieland}}, \
  and\ \bibinfo {author} {\bibfnamefont {A.}~\bibnamefont {Praetorius}},\
  }\href@noop {} {\bibfield  {journal} {\bibinfo  {journal} {J. Royal Soc.
  Interface}\ }\textbf {\bibinfo {volume} {12}},\ \bibinfo {pages} {20150161}
  (\bibinfo {year} {2015})}\BibitemShut {NoStop}%
\bibitem [{\citenamefont {Whitfield}\ and\ \citenamefont
  {Hawkins}(2016)}]{WhHa2016njp}%
  \BibitemOpen
  \bibfield  {author} {\bibinfo {author} {\bibfnamefont {C.~A.}\ \bibnamefont
  {Whitfield}}\ and\ \bibinfo {author} {\bibfnamefont {R.~J.}\ \bibnamefont
  {Hawkins}},\ }\href {\doibase 10.1088/1367-2630/18/12/123016} {\bibfield
  {journal} {\bibinfo  {journal} {New J. Phys.}\ }\textbf {\bibinfo {volume}
  {18}},\ \bibinfo {pages} {123016} (\bibinfo {year} {2016})}\BibitemShut
  {NoStop}%
\bibitem [{\citenamefont {Ziebert}\ \emph {et~al.}(2012)\citenamefont
  {Ziebert}, \citenamefont {Swaminathan},\ and\ \citenamefont
  {Aranson}}]{ZiSA2012jrsi}%
  \BibitemOpen
  \bibfield  {author} {\bibinfo {author} {\bibfnamefont {F.}~\bibnamefont
  {Ziebert}}, \bibinfo {author} {\bibfnamefont {S.}~\bibnamefont
  {Swaminathan}}, \ and\ \bibinfo {author} {\bibfnamefont {I.~S.}\ \bibnamefont
  {Aranson}},\ }\href {\doibase 10.1098/rsif.2011.0433} {\bibfield  {journal}
  {\bibinfo  {journal} {J. R. Soc. Interface}\ }\textbf {\bibinfo {volume}
  {9}},\ \bibinfo {pages} {1084} (\bibinfo {year} {2012})}\BibitemShut
  {NoStop}%
\bibitem [{\citenamefont {Giomi}\ and\ \citenamefont
  {DeSimone}(2014)}]{GD2014prl}%
  \BibitemOpen
  \bibfield  {author} {\bibinfo {author} {\bibfnamefont {L.}~\bibnamefont
  {Giomi}}\ and\ \bibinfo {author} {\bibfnamefont {A.}~\bibnamefont
  {DeSimone}},\ }\href@noop {} {\bibfield  {journal} {\bibinfo  {journal}
  {Phys. Rev. Lett.}\ }\textbf {\bibinfo {volume} {112}},\ \bibinfo {pages}
  {147802} (\bibinfo {year} {2014})}\BibitemShut {NoStop}%
\bibitem [{\citenamefont {Ziebert}\ and\ \citenamefont
  {Aranson}(2016)}]{ZiAr2016ncm}%
  \BibitemOpen
  \bibfield  {author} {\bibinfo {author} {\bibfnamefont {F.}~\bibnamefont
  {Ziebert}}\ and\ \bibinfo {author} {\bibfnamefont {I.~S.}\ \bibnamefont
  {Aranson}},\ }\href {\doibase 10.1038/npjcompumats.2016.19} {\bibfield
  {journal} {\bibinfo  {journal} {npj Comput. Mater.}\ }\textbf {\bibinfo
  {volume} {2}},\ \bibinfo {pages} {16019} (\bibinfo {year}
  {2016})}\BibitemShut {NoStop}%
\bibitem [{\citenamefont {Tjhung}\ \emph {et~al.}(2015)\citenamefont {Tjhung},
  \citenamefont {Tiribocchi}, \citenamefont {Marenduzzo},\ and\ \citenamefont
  {Cates}}]{TTMC2015nc}%
  \BibitemOpen
  \bibfield  {author} {\bibinfo {author} {\bibfnamefont {E.}~\bibnamefont
  {Tjhung}}, \bibinfo {author} {\bibfnamefont {A.}~\bibnamefont {Tiribocchi}},
  \bibinfo {author} {\bibfnamefont {D.}~\bibnamefont {Marenduzzo}}, \ and\
  \bibinfo {author} {\bibfnamefont {M.~E.}\ \bibnamefont {Cates}},\ }\href@noop
  {} {\bibfield  {journal} {\bibinfo  {journal} {Nat. Commun.}\ }\textbf
  {\bibinfo {volume} {6}},\ \bibinfo {pages} {5420} (\bibinfo {year}
  {2015})}\BibitemShut {NoStop}%
\bibitem [{\citenamefont {Anderson}\ \emph {et~al.}(1998)\citenamefont
  {Anderson}, \citenamefont {McFadden},\ and\ \citenamefont
  {Wheeler}}]{AnMW1998arfm}%
  \BibitemOpen
  \bibfield  {author} {\bibinfo {author} {\bibfnamefont {D.~M.}\ \bibnamefont
  {Anderson}}, \bibinfo {author} {\bibfnamefont {G.~B.}\ \bibnamefont
  {McFadden}}, \ and\ \bibinfo {author} {\bibfnamefont {A.~A.}\ \bibnamefont
  {Wheeler}},\ }\href {\doibase 10.1146/annurev.fluid.30.1.139} {\bibfield
  {journal} {\bibinfo  {journal} {Ann. Rev. Fluid Mech.}\ }\textbf {\bibinfo
  {volume} {30}},\ \bibinfo {pages} {139} (\bibinfo {year} {1998})}\BibitemShut
  {NoStop}%
\bibitem [{\citenamefont {Oron}\ \emph {et~al.}(1997)\citenamefont {Oron},
  \citenamefont {Davis},\ and\ \citenamefont {Bankoff}}]{ODB1997rmp}%
  \BibitemOpen
  \bibfield  {author} {\bibinfo {author} {\bibfnamefont {A.}~\bibnamefont
  {Oron}}, \bibinfo {author} {\bibfnamefont {S.}~\bibnamefont {Davis}}, \ and\
  \bibinfo {author} {\bibfnamefont {S.}~\bibnamefont {Bankoff}},\ }\href
  {\doibase 10.1103/RevModPhys.69.931} {\bibfield  {journal} {\bibinfo
  {journal} {Rev. Mod. Phys.}\ }\textbf {\bibinfo {volume} {69}},\ \bibinfo
  {pages} {931} (\bibinfo {year} {1997})}\BibitemShut {NoStop}%
\bibitem [{\citenamefont {Thiele}(2007)}]{Thiele2007}%
  \BibitemOpen
  \bibfield  {author} {\bibinfo {author} {\bibfnamefont {U.}~\bibnamefont
  {Thiele}},\ }in\ \href {\doibase 10.1007/978-3-211-69808-2\_2} {\emph
  {\bibinfo {booktitle} {Thin Films of Soft Matter}}},\ \bibinfo {editor}
  {edited by\ \bibinfo {editor} {\bibfnamefont {S.}~\bibnamefont
  {Kalliadasis}}\ and\ \bibinfo {editor} {\bibfnamefont {U.}~\bibnamefont
  {Thiele}}}\ (\bibinfo  {publisher} {Springer},\ \bibinfo {address} {Wien},\
  \bibinfo {year} {2007})\ pp.\ \bibinfo {pages} {25--93}\BibitemShut {NoStop}%
\bibitem [{\citenamefont {Amar}\ and\ \citenamefont
  {Cummings}(2001)}]{BeCu2001pf}%
  \BibitemOpen
  \bibfield  {author} {\bibinfo {author} {\bibfnamefont {B.~M.}\ \bibnamefont
  {Amar}}\ and\ \bibinfo {author} {\bibfnamefont {L.~J.}\ \bibnamefont
  {Cummings}},\ }\href {\doibase 10.1063/1.1359748} {\bibfield  {journal}
  {\bibinfo  {journal} {Phys. Fluids}\ }\textbf {\bibinfo {volume} {13}},\
  \bibinfo {pages} {1160} (\bibinfo {year} {2001})}\BibitemShut {NoStop}%
\bibitem [{\citenamefont {Lin}\ \emph {et~al.}(2013{\natexlab{a}})\citenamefont
  {Lin}, \citenamefont {Cummings}, \citenamefont {Archer}, \citenamefont
  {Kondic},\ and\ \citenamefont {Thiele}}]{LCA+2013pf}%
  \BibitemOpen
  \bibfield  {author} {\bibinfo {author} {\bibfnamefont {T.-S.}\ \bibnamefont
  {Lin}}, \bibinfo {author} {\bibfnamefont {L.~J.}\ \bibnamefont {Cummings}},
  \bibinfo {author} {\bibfnamefont {A.~J.}\ \bibnamefont {Archer}}, \bibinfo
  {author} {\bibfnamefont {L.}~\bibnamefont {Kondic}}, \ and\ \bibinfo {author}
  {\bibfnamefont {U.}~\bibnamefont {Thiele}},\ }\href {\doibase
  10.1063/1.4816508} {\bibfield  {journal} {\bibinfo  {journal} {Phys. Fluids}\
  }\textbf {\bibinfo {volume} {25}},\ \bibinfo {pages} {082102} (\bibinfo
  {year} {2013}{\natexlab{a}})}\BibitemShut {NoStop}%
\bibitem [{\citenamefont {Lin}\ \emph {et~al.}(2013{\natexlab{b}})\citenamefont
  {Lin}, \citenamefont {Kondic}, \citenamefont {Thiele},\ and\ \citenamefont
  {Cummings}}]{LKT+2013jofm}%
  \BibitemOpen
  \bibfield  {author} {\bibinfo {author} {\bibfnamefont {T.-S.}\ \bibnamefont
  {Lin}}, \bibinfo {author} {\bibfnamefont {L.}~\bibnamefont {Kondic}},
  \bibinfo {author} {\bibfnamefont {U.}~\bibnamefont {Thiele}}, \ and\ \bibinfo
  {author} {\bibfnamefont {L.~J.}\ \bibnamefont {Cummings}},\ }\href@noop {}
  {\bibfield  {journal} {\bibinfo  {journal} {J. Fluid Mech.}\ }\textbf
  {\bibinfo {volume} {729}},\ \bibinfo {pages} {214} (\bibinfo {year}
  {2013}{\natexlab{b}})}\BibitemShut {NoStop}%
\bibitem [{\citenamefont {Sankararaman}\ and\ \citenamefont
  {Ramaswamy}(2009)}]{SR2009prl}%
  \BibitemOpen
  \bibfield  {author} {\bibinfo {author} {\bibfnamefont {S.}~\bibnamefont
  {Sankararaman}}\ and\ \bibinfo {author} {\bibfnamefont {S.}~\bibnamefont
  {Ramaswamy}},\ }\href {\doibase 10.1103/PhysRevLett.102.118107} {\bibfield
  {journal} {\bibinfo  {journal} {Phys. Rev. Lett.}\ }\textbf {\bibinfo
  {volume} {102}},\ \bibinfo {pages} {118107} (\bibinfo {year}
  {2009})}\BibitemShut {NoStop}%
\bibitem [{\citenamefont {Joanny}\ and\ \citenamefont
  {Ramaswamy}(2012)}]{JoRa2012jfm}%
  \BibitemOpen
  \bibfield  {author} {\bibinfo {author} {\bibfnamefont {J.~F.}\ \bibnamefont
  {Joanny}}\ and\ \bibinfo {author} {\bibfnamefont {S.}~\bibnamefont
  {Ramaswamy}},\ }\href {\doibase 10.1017/jfm.2012.131} {\bibfield  {journal}
  {\bibinfo  {journal} {J. Fluid Mech.}\ }\textbf {\bibinfo {volume} {705}},\
  \bibinfo {pages} {46} (\bibinfo {year} {2012})}\BibitemShut {NoStop}%
\bibitem [{\citenamefont {Khoromskaia}\ and\ \citenamefont
  {Alexander}(2015)}]{KA2015pre}%
  \BibitemOpen
  \bibfield  {author} {\bibinfo {author} {\bibfnamefont {D.}~\bibnamefont
  {Khoromskaia}}\ and\ \bibinfo {author} {\bibfnamefont {G.~P.}\ \bibnamefont
  {Alexander}},\ }\href {\doibase 10.1103/PhysRevE.92.062311} {\bibfield
  {journal} {\bibinfo  {journal} {Phys. Rev. E}\ }\textbf {\bibinfo {volume}
  {92}},\ \bibinfo {pages} {062311} (\bibinfo {year} {2015})}\BibitemShut
  {NoStop}%
\bibitem [{\citenamefont {Kitavtsev}\ \emph {et~al.}(2018)\citenamefont
  {Kitavtsev}, \citenamefont {M{\"u}nch},\ and\ \citenamefont
  {Wagner}}]{KMW2018potrsa}%
  \BibitemOpen
  \bibfield  {author} {\bibinfo {author} {\bibfnamefont {G.}~\bibnamefont
  {Kitavtsev}}, \bibinfo {author} {\bibfnamefont {A.}~\bibnamefont
  {M{\"u}nch}}, \ and\ \bibinfo {author} {\bibfnamefont {B.}~\bibnamefont
  {Wagner}},\ }\href@noop {} {\bibfield  {journal} {\bibinfo  {journal} {Proc.
  R. Soc. A.}\ }\textbf {\bibinfo {volume} {474}},\ \bibinfo {pages} {20170828}
  (\bibinfo {year} {2018})}\BibitemShut {NoStop}%
\bibitem [{\citenamefont {Loisy}\ \emph
  {et~al.}(2019{\natexlab{b}})\citenamefont {Loisy}, \citenamefont {Eggers},\
  and\ \citenamefont {Liverpool}}]{LoEL2019prl}%
  \BibitemOpen
  \bibfield  {author} {\bibinfo {author} {\bibfnamefont {A.}~\bibnamefont
  {Loisy}}, \bibinfo {author} {\bibfnamefont {J.}~\bibnamefont {Eggers}}, \
  and\ \bibinfo {author} {\bibfnamefont {T.~B.}\ \bibnamefont {Liverpool}},\
  }\href {\doibase 10.1103/PhysRevLett.123.248006} {\bibfield  {journal}
  {\bibinfo  {journal} {Phys. Rev. Lett.}\ }\textbf {\bibinfo {volume} {123}},\
  \bibinfo {pages} {248006} (\bibinfo {year} {2019}{\natexlab{b}})}\BibitemShut
  {NoStop}%
\bibitem [{\citenamefont {Loisy}\ \emph {et~al.}(2020)\citenamefont {Loisy},
  \citenamefont {Eggers},\ and\ \citenamefont {Liverpool}}]{AuET2020sm}%
  \BibitemOpen
  \bibfield  {author} {\bibinfo {author} {\bibfnamefont {A.}~\bibnamefont
  {Loisy}}, \bibinfo {author} {\bibfnamefont {J.}~\bibnamefont {Eggers}}, \
  and\ \bibinfo {author} {\bibfnamefont {T.~B.}\ \bibnamefont {Liverpool}},\
  }\href {\doibase 10.1039/D0SM00070A} {\bibfield  {journal} {\bibinfo
  {journal} {Soft Matter}\ }\textbf {\bibinfo {volume} {16}},\ \bibinfo {pages}
  {3106} (\bibinfo {year} {2020})}\BibitemShut {NoStop}%
\bibitem [{\citenamefont {Xu}\ \emph {et~al.}(2015)\citenamefont {Xu},
  \citenamefont {Thiele},\ and\ \citenamefont {Qian}}]{XTQ2015JPCM}%
  \BibitemOpen
  \bibfield  {author} {\bibinfo {author} {\bibfnamefont {X.}~\bibnamefont
  {Xu}}, \bibinfo {author} {\bibfnamefont {U.}~\bibnamefont {Thiele}}, \ and\
  \bibinfo {author} {\bibfnamefont {T.}~\bibnamefont {Qian}},\ }\href@noop {}
  {\bibfield  {journal} {\bibinfo  {journal} {J. Phys.: Condens. Matter}\
  }\textbf {\bibinfo {volume} {27}},\ \bibinfo {pages} {085005} (\bibinfo
  {year} {2015})}\BibitemShut {NoStop}%
\bibitem [{\citenamefont {Bonn}\ \emph {et~al.}(2009)\citenamefont {Bonn},
  \citenamefont {Eggers}, \citenamefont {Indekeu}, \citenamefont {Meunier},\
  and\ \citenamefont {Rolley}}]{BEI+2009rmp}%
  \BibitemOpen
  \bibfield  {author} {\bibinfo {author} {\bibfnamefont {D.}~\bibnamefont
  {Bonn}}, \bibinfo {author} {\bibfnamefont {J.}~\bibnamefont {Eggers}},
  \bibinfo {author} {\bibfnamefont {J.}~\bibnamefont {Indekeu}}, \bibinfo
  {author} {\bibfnamefont {J.}~\bibnamefont {Meunier}}, \ and\ \bibinfo
  {author} {\bibfnamefont {E.}~\bibnamefont {Rolley}},\ }\href {\doibase
  10.1103/RevModPhys.81.739} {\bibfield  {journal} {\bibinfo  {journal} {Rev.
  Mod. Phys.}\ }\textbf {\bibinfo {volume} {81}},\ \bibinfo {pages} {739}
  (\bibinfo {year} {2009})}\BibitemShut {NoStop}%
\bibitem [{\citenamefont {Thiele}(2010)}]{Thiele2010jpcm}%
  \BibitemOpen
  \bibfield  {author} {\bibinfo {author} {\bibfnamefont {U.}~\bibnamefont
  {Thiele}},\ }\href {\doibase 10.1088/0953-8984/22/8/084019} {\bibfield
  {journal} {\bibinfo  {journal} {J. Phys.: Condens. Matter}\ }\textbf
  {\bibinfo {volume} {22}},\ \bibinfo {pages} {084019} (\bibinfo {year}
  {2010})}\BibitemShut {NoStop}%
\bibitem [{\citenamefont {Thiele}\ \emph {et~al.}(2013)\citenamefont {Thiele},
  \citenamefont {Todorova},\ and\ \citenamefont {Lopez}}]{TTL2013prl}%
  \BibitemOpen
  \bibfield  {author} {\bibinfo {author} {\bibfnamefont {U.}~\bibnamefont
  {Thiele}}, \bibinfo {author} {\bibfnamefont {D.~V.}\ \bibnamefont
  {Todorova}}, \ and\ \bibinfo {author} {\bibfnamefont {H.}~\bibnamefont
  {Lopez}},\ }\href {\doibase 10.1103/PhysRevLett.111.117801} {\bibfield
  {journal} {\bibinfo  {journal} {Phys. Rev. Lett.}\ }\textbf {\bibinfo
  {volume} {111}},\ \bibinfo {pages} {117801} (\bibinfo {year}
  {2013})}\BibitemShut {NoStop}%
\bibitem [{\citenamefont {Bastian}\ \emph {et~al.}(2008)\citenamefont
  {Bastian}, \citenamefont {Blatt}, \citenamefont {Dedner}, \citenamefont
  {Engwer}, \citenamefont {Kl{\"o}fkorn}, \citenamefont {Kornhuber},
  \citenamefont {Ohlberger},\ and\ \citenamefont {Sander}}]{BBD+2008c}%
  \BibitemOpen
  \bibfield  {author} {\bibinfo {author} {\bibfnamefont {P.}~\bibnamefont
  {Bastian}}, \bibinfo {author} {\bibfnamefont {M.}~\bibnamefont {Blatt}},
  \bibinfo {author} {\bibfnamefont {A.}~\bibnamefont {Dedner}}, \bibinfo
  {author} {\bibfnamefont {C.}~\bibnamefont {Engwer}}, \bibinfo {author}
  {\bibfnamefont {R.}~\bibnamefont {Kl{\"o}fkorn}}, \bibinfo {author}
  {\bibfnamefont {R.}~\bibnamefont {Kornhuber}}, \bibinfo {author}
  {\bibfnamefont {M.}~\bibnamefont {Ohlberger}}, \ and\ \bibinfo {author}
  {\bibfnamefont {O.}~\bibnamefont {Sander}},\ }\href@noop {} {\bibfield
  {journal} {\bibinfo  {journal} {Computing}\ }\textbf {\bibinfo {volume}
  {82}},\ \bibinfo {pages} {103} (\bibinfo {year} {2008})}\BibitemShut
  {NoStop}%
\bibitem [{\citenamefont {Heil}\ and\ \citenamefont {Hazel}(2006)}]{HH2006}%
  \BibitemOpen
  \bibfield  {author} {\bibinfo {author} {\bibfnamefont {M.}~\bibnamefont
  {Heil}}\ and\ \bibinfo {author} {\bibfnamefont {A.~L.}\ \bibnamefont
  {Hazel}},\ }in\ \href@noop {} {\emph {\bibinfo {booktitle} {Fluid-structure
  interaction}}}\ (\bibinfo  {publisher} {Springer},\ \bibinfo {year} {2006})\
  pp.\ \bibinfo {pages} {19--49}\BibitemShut {NoStop}%
\bibitem [{\citenamefont {Doedel}\ \emph {et~al.}(1991)\citenamefont {Doedel},
  \citenamefont {Keller},\ and\ \citenamefont {Kernevez}}]{DoKK1991ijbc}%
  \BibitemOpen
  \bibfield  {author} {\bibinfo {author} {\bibfnamefont {E.}~\bibnamefont
  {Doedel}}, \bibinfo {author} {\bibfnamefont {H.~B.}\ \bibnamefont {Keller}},
  \ and\ \bibinfo {author} {\bibfnamefont {J.~P.}\ \bibnamefont {Kernevez}},\
  }\href {\doibase 10.1142/S0218127491000397} {\bibfield  {journal} {\bibinfo
  {journal} {Int. J. Bifurcation Chaos}\ }\textbf {\bibinfo {volume} {1}},\
  \bibinfo {pages} {493} (\bibinfo {year} {1991})}\BibitemShut {NoStop}%
\bibitem [{\citenamefont {Dijkstra}\ \emph {et~al.}(2014)\citenamefont
  {Dijkstra}, \citenamefont {Wubs}, \citenamefont {Cliffe}, \citenamefont
  {Doedel}, \citenamefont {Dragomirescu}, \citenamefont {Eckhardt},
  \citenamefont {Gelfgat}, \citenamefont {Hazel}, \citenamefont {Lucarini},
  \citenamefont {Salinger}, \citenamefont {Phipps}, \citenamefont
  {Sanchez-Umbria}, \citenamefont {Schuttelaars}, \citenamefont {Tuckerman},\
  and\ \citenamefont {Thiele}}]{DWCD2014ccp}%
  \BibitemOpen
  \bibfield  {author} {\bibinfo {author} {\bibfnamefont {H.~A.}\ \bibnamefont
  {Dijkstra}}, \bibinfo {author} {\bibfnamefont {F.~W.}\ \bibnamefont {Wubs}},
  \bibinfo {author} {\bibfnamefont {A.~K.}\ \bibnamefont {Cliffe}}, \bibinfo
  {author} {\bibfnamefont {E.}~\bibnamefont {Doedel}}, \bibinfo {author}
  {\bibfnamefont {I.~F.}\ \bibnamefont {Dragomirescu}}, \bibinfo {author}
  {\bibfnamefont {B.}~\bibnamefont {Eckhardt}}, \bibinfo {author}
  {\bibfnamefont {A.~Y.}\ \bibnamefont {Gelfgat}}, \bibinfo {author}
  {\bibfnamefont {A.}~\bibnamefont {Hazel}}, \bibinfo {author} {\bibfnamefont
  {V.}~\bibnamefont {Lucarini}}, \bibinfo {author} {\bibfnamefont {A.~G.}\
  \bibnamefont {Salinger}}, \bibinfo {author} {\bibfnamefont {E.~T.}\
  \bibnamefont {Phipps}}, \bibinfo {author} {\bibfnamefont {J.}~\bibnamefont
  {Sanchez-Umbria}}, \bibinfo {author} {\bibfnamefont {H.}~\bibnamefont
  {Schuttelaars}}, \bibinfo {author} {\bibfnamefont {L.~S.}\ \bibnamefont
  {Tuckerman}}, \ and\ \bibinfo {author} {\bibfnamefont {U.}~\bibnamefont
  {Thiele}},\ }\href {\doibase 10.4208/cicp.240912.180613a} {\bibfield
  {journal} {\bibinfo  {journal} {Commun. Comput. Phys.}\ }\textbf {\bibinfo
  {volume} {15}},\ \bibinfo {pages} {1} (\bibinfo {year} {2014})}\BibitemShut
  {NoStop}%
\bibitem [{\citenamefont {Engelnkemper}\ \emph {et~al.}(2019)\citenamefont
  {Engelnkemper}, \citenamefont {Gurevich}, \citenamefont {Uecker},
  \citenamefont {Wetzel},\ and\ \citenamefont {Thiele}}]{EGUW2019springer}%
  \BibitemOpen
  \bibfield  {author} {\bibinfo {author} {\bibfnamefont {S.}~\bibnamefont
  {Engelnkemper}}, \bibinfo {author} {\bibfnamefont {S.}~\bibnamefont
  {Gurevich}}, \bibinfo {author} {\bibfnamefont {H.}~\bibnamefont {Uecker}},
  \bibinfo {author} {\bibfnamefont {D.}~\bibnamefont {Wetzel}}, \ and\ \bibinfo
  {author} {\bibfnamefont {U.}~\bibnamefont {Thiele}},\ }in\ \href {\doibase
  10.1007/978-3-319-91494-7_13} {\emph {\bibinfo {booktitle} {Computational
  Modeling of Bifurcations and Instabilities in Fluid Mechanics}}},\ \bibinfo
  {series and number} {Computational Methods in Applied Sciences, vol 50}\
  (\bibinfo  {publisher} {Springer},\ \bibinfo {year} {2019})\ pp.\ \bibinfo
  {pages} {459--501}\BibitemShut {NoStop}%
\bibitem [{\citenamefont {Uecker}\ \emph {et~al.}(2014)\citenamefont {Uecker},
  \citenamefont {Wetzel},\ and\ \citenamefont {Rademacher}}]{UeWR2014nmma}%
  \BibitemOpen
  \bibfield  {author} {\bibinfo {author} {\bibfnamefont {H.}~\bibnamefont
  {Uecker}}, \bibinfo {author} {\bibfnamefont {D.}~\bibnamefont {Wetzel}}, \
  and\ \bibinfo {author} {\bibfnamefont {J.}~\bibnamefont {Rademacher}},\
  }\href {\doibase 10.4208/nmtma.2014.1231nm} {\bibfield  {journal} {\bibinfo
  {journal} {Numer. Math.-Theory Methods Appl.}\ }\textbf {\bibinfo {volume}
  {7}},\ \bibinfo {pages} {58} (\bibinfo {year} {2014})}\BibitemShut {NoStop}%
\bibitem [{\citenamefont {Uecker}\ and\ \citenamefont
  {Wetzel}(2014)}]{UeWe2014sjads}%
  \BibitemOpen
  \bibfield  {author} {\bibinfo {author} {\bibfnamefont {H.}~\bibnamefont
  {Uecker}}\ and\ \bibinfo {author} {\bibfnamefont {D.}~\bibnamefont
  {Wetzel}},\ }\href {\doibase 10.1137/130918484} {\bibfield  {journal}
  {\bibinfo  {journal} {SIAM J. Appl. Dyn. Syst.}\ }\textbf {\bibinfo {volume}
  {13}},\ \bibinfo {pages} {94} (\bibinfo {year} {2014})}\BibitemShut {NoStop}%
\bibitem [{\citenamefont {Thiele}\ \emph {et~al.}(2012)\citenamefont {Thiele},
  \citenamefont {Archer},\ and\ \citenamefont {Plapp}}]{TAP2012pf}%
  \BibitemOpen
  \bibfield  {author} {\bibinfo {author} {\bibfnamefont {U.}~\bibnamefont
  {Thiele}}, \bibinfo {author} {\bibfnamefont {A.~J.}\ \bibnamefont {Archer}},
  \ and\ \bibinfo {author} {\bibfnamefont {M.}~\bibnamefont {Plapp}},\ }\href
  {\doibase 10.1063/1.4758476} {\bibfield  {journal} {\bibinfo  {journal}
  {Phys. Fluids}\ }\textbf {\bibinfo {volume} {24}},\ \bibinfo {pages} {102107}
  (\bibinfo {year} {2012})}\BibitemShut {NoStop}%
\bibitem [{\citenamefont {Thiele}\ \emph {et~al.}(2002)\citenamefont {Thiele},
  \citenamefont {Neuffer}, \citenamefont {Bestehorn}, \citenamefont {Pomeau},\
  and\ \citenamefont {Velarde}}]{TNBP2002csa}%
  \BibitemOpen
  \bibfield  {author} {\bibinfo {author} {\bibfnamefont {U.}~\bibnamefont
  {Thiele}}, \bibinfo {author} {\bibfnamefont {K.}~\bibnamefont {Neuffer}},
  \bibinfo {author} {\bibfnamefont {M.}~\bibnamefont {Bestehorn}}, \bibinfo
  {author} {\bibfnamefont {Y.}~\bibnamefont {Pomeau}}, \ and\ \bibinfo {author}
  {\bibfnamefont {M.~G.}\ \bibnamefont {Velarde}},\ }\href@noop {} {\bibfield
  {journal} {\bibinfo  {journal} {Colloid Surf. A}\ }\textbf {\bibinfo {volume}
  {206}},\ \bibinfo {pages} {87} (\bibinfo {year} {2002})}\BibitemShut
  {NoStop}%
\bibitem [{\citenamefont {Thiele}\ \emph {et~al.}(2018)\citenamefont {Thiele},
  \citenamefont {Snoeijer}, \citenamefont {Trinschek},\ and\ \citenamefont
  {John}}]{TST+2018l}%
  \BibitemOpen
  \bibfield  {author} {\bibinfo {author} {\bibfnamefont {U.}~\bibnamefont
  {Thiele}}, \bibinfo {author} {\bibfnamefont {J.~H.}\ \bibnamefont
  {Snoeijer}}, \bibinfo {author} {\bibfnamefont {S.}~\bibnamefont {Trinschek}},
  \ and\ \bibinfo {author} {\bibfnamefont {K.}~\bibnamefont {John}},\
  }\href@noop {} {\bibfield  {journal} {\bibinfo  {journal} {Langmuir}\
  }\textbf {\bibinfo {volume} {34}},\ \bibinfo {pages} {7210} (\bibinfo {year}
  {2018})}\BibitemShut {NoStop}%
\bibitem [{\citenamefont {Chaudhuri}\ \emph {et~al.}(2007)\citenamefont
  {Chaudhuri}, \citenamefont {Parekh},\ and\ \citenamefont
  {Fletcher}}]{Chau2007nat}%
  \BibitemOpen
  \bibfield  {author} {\bibinfo {author} {\bibfnamefont {O.}~\bibnamefont
  {Chaudhuri}}, \bibinfo {author} {\bibfnamefont {S.~H.}\ \bibnamefont
  {Parekh}}, \ and\ \bibinfo {author} {\bibfnamefont {D.~A.}\ \bibnamefont
  {Fletcher}},\ }\href {\doibase 10.1038/nature05459} {\bibfield  {journal}
  {\bibinfo  {journal} {Nature}\ }\textbf {\bibinfo {volume} {445}},\ \bibinfo
  {pages} {295} (\bibinfo {year} {2007})}\BibitemShut {NoStop}%
\bibitem [{\citenamefont {Trinschek}\ \emph {et~al.}(2020)\citenamefont
  {Trinschek}, \citenamefont {Stegemerten}, \citenamefont {John},\ and\
  \citenamefont {Thiele}}]{TSJTzenodo2020}%
  \BibitemOpen
  \bibfield  {author} {\bibinfo {author} {\bibfnamefont {S.}~\bibnamefont
  {Trinschek}}, \bibinfo {author} {\bibfnamefont {F.}~\bibnamefont
  {Stegemerten}}, \bibinfo {author} {\bibfnamefont {K.}~\bibnamefont {John}}, \
  and\ \bibinfo {author} {\bibfnamefont {U.}~\bibnamefont {Thiele}},\ }\href
  {\doibase 10.5281/zenodo.3813574} {\enquote {\bibinfo {title} {{Data
  supplement for "Thin-Film Modeling of Resting and Moving Active
  Droplets"}},}\ } (\bibinfo {year} {2020})\BibitemShut {NoStop}%
\end{thebibliography}%

\end{document}